\documentclass[10pt,a4paper]{article}

\usepackage{booktabs}
\usepackage{multirow}
\usepackage{epsfig}
\usepackage{lscape}
\usepackage{rotating}
\usepackage{epsf}
\usepackage{slashed}
\usepackage{bm}
\usepackage{amsmath}
\usepackage{amssymb}
\usepackage{mathtools}
\usepackage{booktabs}
\usepackage{array}
\usepackage{caption}
\usepackage{subcaption}
\usepackage[utf8]{inputenc}
\usepackage{float}
\usepackage[affil-it]{authblk}
\usepackage{cite}
 
\usepackage{color}

\usepackage{hyperref}

\newcommand*{\email}[1]{\href{mailto:#1}{\nolinkurl{#1}} } 

\newcolumntype{L}[1]{>{\raggedright\let\newline\\\arraybackslash\hspace{0pt}}m{#1}}
\newcolumntype{C}[1]{>{\centering\let\newline\\\arraybackslash\hspace{0pt}}m{#1}}
\newcolumntype{R}[1]{>{\raggedleft\let\newline\\\arraybackslash\hspace{0pt}}m{#1}}

\allowdisplaybreaks

\newcommand{\beq}{\begin{align}}
\newcommand{\eeq}{\end{align}}

\begin{document}

\title{Interplay of the LHC and non-LHC Dark Matter searches in the Effective Field Theory approach}

\author[a,b]{Alexander Belyaev\footnote{\email{ a.belyaev@soton.ac.uk}}}
\author[c]{Enrico Bertuzzo\footnote{\email{bertuzzo@if.usp.br}}}
\author[c]{Cristian Caniu Barros\footnote{\email{caniu@if.usp.br}}}
\author[c]{Oscar Eboli\footnote{\email{eboli@if.usp.br}}}
\author[c]{Giovanni Grilli di Cortona\footnote{\email{ggrilli@if.usp.br}}}
\author[d]{Fabio Iocco\footnote{\email{iocco@ift.unesp.br}}}
\author[e]{Alexander Pukhov\footnote{\email{pukhov@lapp.in2p3.fr}}}

\affil[a]{School of Physics and Astronomy, University of Southampton, Southampton SO17 1BJ, UK}
\affil[b]{Particle Physics Department, Rutherford Appleton Laboratory, Didcot, Oxon OX11 0QX, UK}
\affil[c]{Instituto de F\'isica, Universidade de S\~ao Paulo, C.P. 66.318, 05315-970 S\~ao Paulo, Brazil}
\affil[d]{ICTP South American Institute for Fundamental Research, and Instituto de F\'isica Te\'orica - Universidade Estadual Paulista (UNESP), Rua Dr. Bento Teobaldo Ferraz 271, 01140-070 S\~ao Paulo, SP Brazil}
\affil[e]{Skobeltsyn Inst. of Nuclear Physics, Moscow State Univ., Moscow 119992, Russia}

\maketitle

\begin{abstract}
We present accurate and up-to-date constraints on the complete
set of dimension five and six operators with scalar, fermion and vector Dark Matter (DM).
We find  limits using LHC mono-jet data, spin independent and spin dependent direct searches, relic density and CMB, and show the interplay between high and low energy data in setting bounds on the parameter space. In order to properly compare data taken at different energies, we take into account the effect of the running and mixing of operators. We also take into account the local density uncertainties affecting direct detection data, and apply EFT validity criteria related to the cut on the invariant mass of DM pair production at the LHC, which turns out to be especially important for the case of vector DM. Finally, we estimate the potential of the future LHC runs to probe
DM parameter space.

\end{abstract}

\newcommand{\tcb}[1]{\textcolor{blue}{#1}}
\newcommand{\tco}[1]{ {\textcolor{orange}{#1}} }
\newcommand{\tcm}[1]{\textcolor{magenta}{#1}}
\newcommand{\tcr}[1]{\textcolor{red}{#1}}

\newpage
\tableofcontents
\newpage

\section{Introduction}\label{sec:intro}

Understanding the nature of Dark Matter (DM) is one of the greatest puzzles of modern particle physics and cosmology. Although overwhelming observational evidences from galactic to cosmological scales point to the existence of DM~\cite{Ade:2015xua,Blumenthal:1984bp,Bullock:1999he}, after decades of experimental effort only its gravitational interaction has been experimentally confirmed.
Currently, no information is available on the DM properties, such as its spin, mass, interactions other than gravitational, symmetry responsible for its stability, number of states associated to it, and possible particles that would mediate the interactions between DM and the standard model (SM) particles.\smallskip

If DM is light enough and interacts with SM particles directly or via some mediators with a strength beyond the gravitational one, its elusive nature can be detected or constrained in different ways:
\begin{itemize}

\item from direct production at colliders, resulting in a signature exhibiting an observed SM object, such as jet, Higgs, $Z$, or photon, that recoils against the missing energy from the DM pair~\cite{Aaboud:2017phn,CMS:2017tbk,ATLAS-CONF-2018-005,Aaboud:2017dor};

\item via the relic density constraint obtained through the observations of cosmic microwave background (CMB) anisotropies, such as those of WMAP and PLANCK collaborations~\cite{Hinshaw:2012aka,Ade:2015xua};

\item from DM direct detection (DD) experiments, which are sensitive to elastic spin independent (SI) or spin dependent (SD) DM scattering off nuclei~\cite{Goodman:1984dc,Aprile:2017iyp,Akerib:2016vxi,Fu:2016ega};

\item from DM indirect detection searches, that look for SM particles produced in the decay or annihilation of DM present in the cosmos, both with high energies observables (gamma-rays, neutrinos, charge cosmic rays) produced in the local Universe~\cite{Slatyer:2017sev,Ackermann:2015zua,Zitzer:2015eqa,Ahnen:2016qkx,Abdallah:2016ygi,Abramowski:2013ax}, and by studying the effects of 
energy produced by DM annihilation in the early universe on the properties of the CMB spectrum~\cite{Galli:2009zc, Galli:2011rz, Ade:2015xua}.

\end{itemize}

In this work we obtain the present constraints on three scenarios for the DM particles: complex scalars ($\phi$), Dirac fermions ($\chi$) and complex vectors ($V_\mu$).  In order to describe the interactions of the new states, we parametrize the DM interactions with the SM quarks and gluons as an effective field theory (EFT) that contains a complete set of operators of dimension six or less.
Notice the presence of the coupling $g_*$ in the definition of the effective operators, which we insert according to the Naive Dimensional Analysis~\cite{Contino:2016jqw}.
Moreover, for the vector DM case we choose the parametrisation suggested in Ref.~\cite{Belyaev:2016pxe} that takes into account the high energy behaviour of the scattering amplitudes that are enhanced by an energy factor $(E/m_{\mathrm{DM}})$  for every longitudinal vector DM polarisation.

\begin{table}[htp]
  \centering
  \begin{minipage}[t]{.4\textwidth}
    \begin{align*}
      \begin{array}{l@{\quad}l}
        \toprule
        \multicolumn{2}{c}{\text{Complex Scalar DM}}\\
        \midrule
        \frac{g_*^2}{\Lambda} \phi^{\dagger}\phi \bar{q}  q		    				& [C1] \\[2pt]
        \frac{g_*^2}{\Lambda} \phi^{\dagger}\phi \bar{q} i \gamma^5 q    					& [C2] \\[2pt]
        \frac{g_*^2}{\Lambda^2} \phi^{\dagger} i \overleftrightarrow{\partial_\mu} \phi \bar{q}\gamma^\mu q		& [C3] \\[2pt]
        \frac{g_*^2}{\Lambda^2} \phi^{\dagger} i \overleftrightarrow{\partial_\mu} \phi \bar{q}\gamma^\mu\gamma^5 q	& [C4] \\[2pt]
        \midrule
        \frac{g_*^2}{\Lambda^2} \phi^{\dagger}\phi G^{\mu\nu} G_{\mu\nu}						& [C5] \\[2pt]
        \frac{g_*^2}{\Lambda^2} \phi^{\dagger}\phi \tilde{G}^{\mu\nu} G_{\mu\nu}					& [C6]\\
        \bottomrule
      \end{array}
    \end{align*}
    \begin{align*}
      \begin{array}{l@{\quad}l}
        \toprule
        \multicolumn{2}{c}{\text{Dirac Fermion DM}}\\
        \midrule
        \frac{g_*^2}{\Lambda^2} \bar{\chi}\chi \bar{q} q 						& [\textrm{D1}] \\[2pt]
        \frac{g_*^2}{\Lambda^2} \bar{\chi} i \gamma^5\chi \bar{q} q 			       	& [\textrm{D2}] \\[2pt]
        \frac{g_*^2}{\Lambda^2} \bar{\chi}\chi \bar{q} i \gamma^5q 					& [\textrm{D3}] \\[2pt]
        \frac{g_*^2}{\Lambda^2} \bar{\chi}\gamma^5\chi \bar{q} \gamma^5q 				& [\textrm{D4}] \\[2pt]
        \frac{g_*^2}{\Lambda^2} \bar{\chi}\gamma^\mu\chi \bar{q}\gamma_\mu q			& [\textrm{D5}] \\[2pt]
        \frac{g_*^2}{\Lambda^2} \bar{\chi}\gamma^\mu\gamma^5\chi \bar{q}\gamma_\mu q 		& [\textrm{D6}] \\[2pt]
        \frac{g_*^2}{\Lambda^2} \bar{\chi}\gamma^\mu\chi \bar{q}\gamma_\mu \gamma^5 q		& [\textrm{D7}] \\[2pt]
        \frac{g_*^2}{\Lambda^2} \bar{\chi}\gamma^\mu \gamma^5\chi \bar{q}\gamma_\mu \gamma^5 q	& [\textrm{D8}] \\[2pt]
        \frac{g_*^2}{\Lambda^2} \bar{\chi}\sigma^{\mu\nu}\chi \bar{q}\sigma_{\mu\nu} q		& [\textrm{D9}] \\[2pt]
        \frac{g_*^2}{\Lambda^2} \bar{\chi}\sigma^{\mu\nu}i\gamma^5\chi \bar{q}\sigma_{\mu\nu} q	& [\textrm{D10}] \\
        \bottomrule
      \end{array}
    \end{align*}
  \end{minipage}\hfill
  \begin{minipage}[t]{0.6\textwidth}
    \begin{align*}
      \begin{array}{l@{\quad}l}
        \toprule
        \multicolumn{2}{c}{\text{Complex Vector DM}}\\
        \midrule
        \frac{g_*^2\, m_{\mathrm{DM}}^2}{\Lambda^3} V^\dagger_\mu V^\mu \bar{q}  q													& [\textrm{V1}] \\[2pt]
        \frac{g_*^2\,m_{\mathrm{DM}}^2}{\Lambda^3} V^\dagger_\mu V^\mu \bar{q} i \gamma^5 q												& [\textrm{V2}] \\[2pt]
         \frac{g_*^2\,m_{\mathrm{DM}}^2}{2\Lambda^4} i(V^\dagger_\nu\partial_\mu V^\nu - V^\nu \partial_\mu V^\dagger_\nu ) \bar{q} \gamma^\mu q 						& [\textrm{V3}] \\[2pt]
        \frac{g_*^2\,m_{\mathrm{DM}}^2}{2\Lambda^4} (V^\dagger_\nu\partial_\mu V^\nu - V^\nu \partial_\mu V^\dagger_\nu ) \bar{q} i \gamma^\mu \gamma^5 q					& [\textrm{V4}] \\[2pt]
         \frac{g_*^2\,m_{\mathrm{DM}}^2}{\Lambda^3} V^\dagger_\mu V_\nu \bar{q} i \sigma^{\mu\nu} q											& [\textrm{V5}] \\[2pt]
         \frac{g_*^2\,m_{\mathrm{DM}}^2}{\Lambda^3} V^\dagger_\mu V_\nu \bar{q} \sigma^{\mu\nu} \gamma^5 q										& [\textrm{V6}] \\[2pt]
         \frac{g_*^2\,m_{\mathrm{DM}}}{2\Lambda^3} (V^\dagger_\nu\partial^\nu V_\mu + V_\nu \partial^\nu V^\dagger_\mu ) \bar{q} \gamma^\mu q							& [\textrm{V7P}] \\[2pt]
         \frac{g_*^2\,m_{\mathrm{DM}}^2}{2\Lambda^4} (V^\dagger_\nu\partial^\nu V_\mu - V_\nu \partial^\nu V^\dagger_\mu ) \bar{q} i \gamma^\mu q						& [\textrm{V7M}] \\[2pt]
        \frac{g_*^2\,m_{\mathrm{DM}}}{2\Lambda^3} (V^\dagger_\nu\partial^\nu V_\mu + V_\nu \partial^\nu V^\dagger_\mu ) \bar{q} \gamma^\mu \gamma^5 q					& [\textrm{V8P}] \\[2pt]
         \frac{g_*^2\,m_{\mathrm{DM}}^2}{2\Lambda^4} (V^\dagger_\nu\partial^\nu V_\mu - V_\nu \partial^\nu V^\dagger_\mu ) \bar{q} i \gamma^\mu \gamma^5 q					& [\textrm{V8M}] \\[2pt]
         \frac{g_*^2\,m_{\mathrm{DM}}}{2\Lambda^3} \epsilon^{\mu\nu\rho\sigma} (V^\dagger_\nu\partial_\rho V_\sigma + V_\nu \partial_\rho V^\dagger_\sigma ) \bar{q} \gamma_\mu q		& [\textrm{V9P}] \\[2pt]
         \frac{g_*^2\,m_{\mathrm{DM}}}{2\Lambda^3} \epsilon^{\mu\nu\rho\sigma} (V^\dagger_\nu\partial_\rho V_\sigma - V_\nu \partial_\rho V^\dagger_\sigma  ) \bar{q} i \gamma_\mu q			& [\textrm{V9M}] \\[2pt]
        \frac{g_*^2\,m_{\mathrm{DM}}}{2\Lambda^3} \epsilon^{\mu\nu\rho\sigma} (V^\dagger_\nu\partial_\rho V_\sigma + V_\nu \partial_\rho V^\dagger_\sigma ) \bar{q} \gamma_\mu \gamma^5 q	& [\textrm{V10P}] \\[2pt]
        \frac{g_*^2\,m_{\mathrm{DM}}}{2\Lambda^3} \epsilon^{\mu\nu\rho\sigma} (V^\dagger_\nu\partial_\rho V_\sigma - V_\nu \partial_\rho V^\dagger_\sigma  ) \bar{q} i \gamma_\mu \gamma^5 q		& [\textrm{V10M}] \\[2pt]
        \frac{g_*^2\,m_{\mathrm{DM}}^2}{\Lambda^4} V^\dagger_\mu V^\mu G^{\rho\sigma} G_{\rho\sigma}			                                                    			& [V11] \\[2pt]
        \frac{g_*^2\,m_{\mathrm{DM}}^2}{\Lambda^4} V^\dagger_\mu V^\mu \tilde{G}^{\rho\sigma} G_{\rho\sigma}			                                                		& [V12]\\[2pt]
        \bottomrule
      \end{array}
    \end{align*}
  \end{minipage}
  \caption{
    Minimal basis of  operators of dimension six or less  involving only complex scalar DM ($\phi$), Dirac fermion DM ($\chi$) or complex vector DM ($V^\mu$) interacting with SM quarks ($q$) or gluons. Here we denote the field strength tensor of the gluons as $G^{\mu\nu}$ and its dual as $\tilde G^{\mu\nu}$.}
  \label{tab:EFToperators}
\end{table}

Here, our goal is to explore the complementarity of the collider and non-collider experiments mentioned above for all DM EFT operators and DM spin listed in Table~\ref{tab:EFToperators}~\cite{Goodman:2010ku,Kumar:2015wya,Belyaev:2016pxe} for DM masses in the GeV-TeV range; for lighter DM masses see for example~\cite{Bertuzzo:2017lwt}.
For the sake of consistency of our analyses, we obtain the present constraints from the LHC data taking into account the validity of the EFT \cite{Busoni:2013lha, Busoni:2014sya, Busoni:2014haa} using the prescription of Ref.~\cite{Racco:2015dxa}; see Section~\ref{sec:analysis-setup} for details.
Furthermore, the correct comparison between the LHC and non-collider bounds requires that we consider in our analyses the running of the EFT operators from the TeV scale down to the GeV one.  This is important because the running of operators leads to mixing between them at low energy which can give rise to stronger DM DD limits~\cite{Hill:2011be, Frandsen:2012db, Vecchi:2013iza, Crivellin:2014qxa, DEramo:2014nmf, DEramo:2016gos, Bishara:2017pfq}.

As it is well known, the DM DD searches are plagued by the uncertainty on the local DM density~\cite{Iocco:2015xga, Pato:2015dua, Read:2014qva, Green:2010gw, Green:2011bv, Bozorgnia:2016ogo} which propagates to the limits on the DM DD cross sections reported by the experiments, inducing variations up to one order of magnitude. Here we estimate the impact of this uncertainty on the DM bounds, presenting three scenarios that range from a conservative to a more optimistic one.
On the other hand, indirect constraints from CMB are unaffected by the usual unknowns related to the DM density profile within structures. For this reason, we also include the CMB data in our analyses that lead to more robust limits.

This paper is organized as follows: in Section~\ref{sec:eft-run} we study the running of the EFT DM operators from the TeV to the GeV scale and the effect of their mixing,
while we present our analyses framework and available constraints in Section~\ref{sec:analysis-setup}.
Section~\ref{sec:results} contains our main results for all operators in Table~\ref{tab:EFToperators} that show the complementarity of the collider and non-collider constraints.  Finally, we draw our conclusions in Section~\ref{sec:conclusions}.


\section{Direct detection and  Running effect of the EFT operators}\label{sec:eft-run}

In this section we demonstrate the importance of the running of the operators for the DM DD constraints.
As we know, in  Quantum Field Theory radiative corrections may be important to properly assess the phenomenological implications of a model. In the case of DM EFT this is even more so, because (i) the bounds on the parameter space involve experiments with very different energy scales, and (ii) the Wilson coefficients can vary substantially between the typical LHC energies (a few TeV) and the typical energies of DM DD experiments (below the GeV). The first radiative effects to be considered were the QCD ones~\cite{Hill:2011be,Frandsen:2012db,Vecchi:2013iza}, while only later it has been realized that EW loops could be important as well~\cite{Crivellin:2014qxa,DEramo:2014nmf,DEramo:2016gos,Bishara:2017pfq,Brod:2018ust}.

In our discussion we will make use of the renormalisation of currents~\cite{Vecchi:2013iza}, which we found fits best the purpose of our study.
As can be seen from Table~\ref{tab:EFToperators}, all the operators are written as a product
\begin{align}
	\mathcal{O} = J_\chi^A J_{SM}^A\, ,
\end{align}
where $A$ denotes some combination of Lorentz indexes and $J_\chi$ and $J_{SM}$ are currents constructed out of DM and SM fields only. Since we suppose $\chi$ (and hence $J_\chi$) to be a gauge singlet, only the renormalisation of the SM currents has to be computed. 
The lowest dimensional currents considered in this work are the dimension 3 and 4 operators\footnote{There are two dimension 2 currents which are gauge singlets, $B_{\mu\nu}$ and $H^\dag H$, which however do not mix under renormalisation.}
\begin{align}\label{eq:dim_3_currents}
	J_{d=3} = \left\{ \overline{q} \gamma^\mu q,\; \overline{q} \gamma^\mu \gamma_5 q,\; \overline{q}q,\; \overline{q} \gamma_5 q \right\}\, ,~~~~ J_{d=4} = \left\{  G_{\mu\nu} G^{\mu\nu} ,\; \tilde{G}_{\mu\nu} G^{\mu\nu} \right\} \, ,
\end{align}
where $q$ denotes any of the SM quarks and $G$ is the gluon field. At the low energies involved in the DM DD experiments, the relevant degrees of freedom are nucleons and nuclei, rather than quarks and gluons. For this reason, we should match the amplitudes involving quarks and gluons with matrix elements involving nucleons (see for example \cite{Belanger:2008sj, DelNobile:2013sia} and Table \ref{operatorsSI} for some of the operators).
\begin{table*}[t]
\centering
\begin{tabular}{|c|c|c|c|}
\hline
&WIMP & Even & Odd \\
&spin&  operators      & operators   \\
\hline
&&&\\
& 0  & $2m_{DM} \phi  \phi^\dagger \overline{N} N $ & $
i (\partial_{\mu} \phi \phi^\dagger - \phi
\partial_{\mu}\phi^\dagger) \overline{N} \gamma^\mu N $ \\
\rule{0pt}{3.5ex}
 SI &1/2 & $\overline{\chi} \chi  \overline{N} N $  &
$\overline{\chi}\gamma_\mu \chi
 \overline{N} \gamma^\mu N   $ \\
\rule{0pt}{3.5ex}
& 1  & $ 2 m_{DM} V^*_{\mu} V^{\mu}
  \overline{N} N$ & $i(V^{\dagger \alpha}\partial_\mu
V_{\alpha}
-V^\alpha \partial_\mu V_{\alpha}^\dagger)\overline{N} \gamma^\mu N$      \\
\hline
\end{tabular}
\caption{ Operators for WIMP-nucleon SI interactions. Even and odd refer to the properties with respect to quark/anti-quark exchange. \label{operatorsSI}} \vspace{.3cm}
\end{table*}

Let us start with the scalar quark current. The matrix elements at zero momentum transfer are  \cite{Shifman:1978zn,Drees:1993bu}
\begin{equation}\label{eq:scalar_current_matrix_element}
    \langle N| m_q(\mu) \bar{q} q |N\rangle =  m_N  f_{Tq}^{(N)} \, , ~~~~ \langle N| m_Q(\mu) \bar{Q} Q |N\rangle =  {2 \over 27} m_N  f_{TG}^{(N)}\, ,
\end{equation} 
for light ($q=u,d,s$) and heavy quarks ($Q=c,b,t$), respectively. In the above equation, $N$ stands either for proton $p$ or neutron $n$. The quantity $f_{Tq}^{(N)}$ amounts to the light quark contribution to the nucleon mass $m_N$, while $f_{TG}^{(N)} = 1- \sum_{q} f_{Tq}^{(N)}$. Notice that both the quark condensate and the running mass depend on the scale $\mu$, in such a way that their product is scale independent. As a consequence, the form factors are also scale independent. The light quarks form factors are known from hadron spectroscopy and lattice calculations. In our numerical analysis we use the quark scalar form factors presented in micrOMEGAs~\cite{Belanger:2013oya}. 
Using the matrix elements of Eq.~\eqref{eq:scalar_current_matrix_element} the DM-nucleon spin independent (SI) cross section can be written as 
\begin{equation}\label{eq:SI_xsec}
  \sigma^{SI} = \frac{4\lambda^2}{\pi} \left(\frac{ m_N m_{DM}}{m_N+m_{DM}}\right)^2,
\end{equation} 
where $\lambda$ is an operator dependent coefficient that will be defined below.

To discuss the importance of the running, let us discuss the example of the C1 operator. We will suppose that such operator is generated at the scale $\Lambda \sim \mu_{LHC} \sim \mathrm{TeV}$ with Wilson coefficient $g_*^2$. As already explained, the $\overline{q}q$ operator runs as the inverse of the quark mass, in such a way that the C1 operator at an arbitrary scale $\mu$ looks like 
\begin{equation}
{m_q(\mu)\, \over m_q(\mu_{LHC})} {g_*^2 \over \Lambda} (\phi^\dagger \phi) \left(\sum_q  (\bar{q} q) + \sum_Q \overline{Q} Q \right)\, .
\label{eq:C1}
\end{equation}
The same is true for the D1 and V1 operators, and the $\lambda$ coefficient of Eq.~\eqref{eq:SI_xsec} reads
\begin{align}\begin{aligned}
\lambda_{C1}^{(N)} &= \frac{m_N g_*^2}{2 m_{DM} \, \Lambda} \left[ \sum_{q} \frac{f_{Tq}^{(N)}}{m_q(\mu_{LHC})} + {2 \over 27 } f_{TG}^{(N)} \sum_{Q} \frac{1}{m_Q(\mu_{LHC})} \right]\, ,\\
\lambda_{D1}^{(N)} &= \frac{m_N g_*^2}{ \, \Lambda^2} \left[ \sum_{q} \frac{f_{Tq}^{(N)}}{m_q(\mu_{LHC})} + {2 \over 27 } f_{TG}^{(N)} \sum_{Q} \frac{1}{m_Q(\mu_{LHC})} \right]\, ,\\
\lambda_{V1}^{(N)} &= \frac{m_N \, m_{DM}^2 \, g_*^2}{2 \, \Lambda^3} \left[ \sum_{q} \frac{f_{Tq}^{(N)}}{m_q(\mu_{LHC})} + {2 \over 27 } f_{TG}^{(N)} \sum_{Q} \frac{1}{m_Q(\mu_{LHC})} \right]\, ,
\end{aligned}\end{align}
where $N=n,p$.\\

A similar effect is present for the operators involving the $G^{\mu\nu}G_{\mu\nu}$ current (C5 and V11). In this case, the matrix element of the gluon current is  \cite{Shifman:1978zn,Drees:1993bu}
\begin{equation}
    \langle N| \alpha_s(\mu) G_{\mu\nu} G^{\mu\nu} |N\rangle = -{8 \pi \over 9} m_N  f_{TG}^{(N)},   
\end{equation} 
and the combination $\alpha_s(\mu) G_{\mu\nu} G^{\mu\nu}$ (and hence the form factor) is scale independent. Suppose now that the C5 operator is generated at the $\Lambda$ scale with Wilson coefficient $g_*^2$. At an arbitrary scale $\mu$, the operator looks like
\begin{align}
{\alpha_s(\mu)  \over \alpha_s(\mu_{LHC})} {g_*^2\over \Lambda^2} (\phi^\dagger \phi) (G_{\mu\nu} G^{\mu\nu})\, .
\label{eq:C5}
\end{align}
The same result applies to the V11 operator. The $\lambda$ coefficients of Eq.~\eqref{eq:SI_xsec} reads
 \begin{align}\begin{aligned}\label{eq:lambda_GG}
\lambda_{C5}^{(N)} &=- {4 \pi \over 9} g_* f_{TG}^{(N)} {m_N \over  \alpha_s(\mu_{LHC})\,\Lambda^2\,m_{\mathrm{DM}}}\, , \\
\lambda_{V11}^{(N)} &=-{4 \pi \over 9} g_* f_{TG}^{(N)} {m_N \, m_{\mathrm{DM}} \over  \alpha_s(\mu_{LHC}) \, \Lambda^4}.
 \end{aligned}\end{align}
In principle, in addition to the running of the individual Wilson coefficients, a mixing between operators is generated~\cite{Hill:2011be,Frandsen:2012db,Vecchi:2013iza,Crivellin:2014qxa}. However, for operators involving the $\left\{m_q \overline{q}q,\;  G_{\mu\nu} G^{\mu\nu} \right\}$ currents the mixing was found to be numerically unimportant~\cite{Crivellin:2014qxa}.\\

Let us now turn to the analysis of the $\left\{ \overline{q} \gamma^\mu q,\; \overline{q} \gamma^\mu \gamma_5 q \right\}$ currents. In this case, the mixing between operators can be numerically important. Let us consider for instance the operators involving the $\overline{q} \gamma^\mu \gamma_5 q$ axial vector current, \textit{i.e.} the operators C4, D7-D8, V4. These operators are responsible for spin dependent DM scattering at DD experiments, with bounds much weaker than those of spin independent experiments. Nonetheless, the $\overline{q} \gamma^\mu \gamma_5 q$ axial vector current mixes with the $\overline{q}\gamma^\mu q$ vector current during the running, and a SI cross section is generated~\cite{Crivellin:2014qxa,DEramo:2014nmf,DEramo:2016gos}. Suppose for instance the C4 operator is generated at the $\Lambda$ scale above the top mass 
\begin{align}\label{eq:CP_odd_Lambda}
	\mathcal{L} = {g_*^2 \over \Lambda^2}(\phi^{\dagger} i \overleftrightarrow{\partial_\mu} \phi)  \left[\sum_q  (\bar{q}\gamma^\mu\gamma^5 q) + \sum_Q  (\bar{Q}\gamma^\mu\gamma^5 Q) \right]\, ,
\end{align}
where the sum is taken over all light ($q$) and heavy ($Q$) quark flavors. Then, using the results of References~\cite{Crivellin:2014qxa,DEramo:2014nmf,DEramo:2016gos}, the operators present in the Lagrangian at the DD scale are
\begin{align}\label{eq:running_example}
	\begin{aligned}
	\mathcal{L} &\simeq \left[ 1- \frac{3 \alpha_t}{2\pi} \log\left( \frac{\mu_{LHC}}{m_t}\right) + ... \right] {g_*^2 \over \Lambda^2} (\phi^{\dagger} i \overleftrightarrow{\partial_\mu} \phi) (\overline{u}  \gamma^\mu \gamma_5 u) \\
	& + \left[ 1 + \frac{3 \alpha_t}{2\pi} \log\left( \frac{\mu_{LHC}}{m_t}\right) + ...  \right] {g_*^2 \over \Lambda^2} (\phi^{\dagger} i \overleftrightarrow{\partial_\mu} \phi)  \left( \overline{d}  \gamma^\mu \gamma_5 d +  \overline{s}  \gamma^\mu \gamma_5 s\right) \\
	& + \left( 3 - 8 s_w^2 \right)\left[ \frac{ \alpha_t}{2\pi} \log\left(  \frac{\mu_{LHC}}{m_t}\right) + ...   \right] {g_*^2 \over \Lambda^2} (\phi^{\dagger} i \overleftrightarrow{\partial_\mu} \phi)  (\overline{u}  \gamma^\mu  u) \\
	&+ \left( 3 - 4 s_w^2 \right)\left[ -\frac{ \alpha_t}{2\pi} \log\left(  \frac{\mu_{LHC}}{m_t}\right) + ...   \right] {g_*^2 \over \Lambda^2} (\phi^{\dagger} i \overleftrightarrow{\partial_\mu} \phi)  (\overline{d}  \gamma^\mu  d + \overline{s} \gamma^\mu s) \, ,
\end{aligned}
\end{align}
where we show only the most relevant contribution, coming from top loops.
In the previous expression, $\alpha_{t} \equiv y_{t}^2/4\pi$, with $y_{t}$ the top Yukawa coupling. Notice that the top contribution is present only down to the top scale, where the top quark is integrated out. As already stressed, in addition to the SD operators of the first two lines, the running has generated the SI operators of the last two lines. Even though the Wilson coefficients of the SI operators are smaller than those of the SD ones, it has been shown in Reference~\cite{DEramo:2016gos} that they are sufficient to put bounds on $g_*^2 /\Lambda^2$ which are up to a factor of 100 stronger with respect to the typical bounds that can be obtained from SD experiments (\textit{i.e.} considering only the operators in the first two lines). This shows clearly the importance of the running in setting consistent bounds on the parameter space of the DM EFT. Let us point out however that for the effect to be numerically relevant, it is instrumental for the coupling between the DM and the top currents to be switched on in Eq.~\eqref{eq:running_example}. If this is not the case, the SI operators are still generated in the running, but with much smaller Wilson coefficients and weaker bounds (see Reference~\cite{DEramo:2016gos} for more details). In the analysis of Sec.~\ref{sec:results} we will numerically implement the running and the mixing of the currents using the runDM code~\cite{Crivellin:2014qxa,DEramo:2014nmf,DEramo:2016gos,runDM}, which takes into account all the contributions.


\section{Analysis setup and constraints}
\label{sec:analysis-setup}

In this section we describe the analysis setup and constraints used in this study. In particular we delineate the limits originating from CMB, direct detection experiments and collider searches.

Direct and indirect detection constraints are affected by uncertainties of astrophysical nature. On one hand, the scattering of DM off nuclei on the Earth depends on the DM local density and velocity distributions around Earth. On the other hand, the DM self-annihilation rate in our galaxy depends on its particle density distribution therein. For what is concern of this paper, whenever possible we make the conservative choice to select targets that can reduce as much as possible the uncertainties, and thoroughly account for the remaining ones.  In practice, this means that: (i) for indirect searches we adopt CMB limits, as the energy injection of DM annihilation is unaffected by the usual unknowns related to DM density profile within structures; (ii) for direct searches we explicitly take into account the systematic effects generated by the astrophysical uncertainties in the determination of the local DM density $\rho_0$.

\subsection{CMB constraints}

The observation of byproducts of DM annihilation (or decay) in astrophysical targets can be used to determine (or constrain, in case of missing observations) relevant DM properties such as its mass and annihilation cross section. Such bounds depend on the unknown DM distribution within the astrophysical objects chosen as targets. A detailed analysis has shown that, choosing the CMB as target, the leading signal of DM annihilation is produced around redshift $z\sim 600$~\cite{Finkbeiner:2011dx}. 
This makes the CMB a quantitatively competitive channel for indirect searches~\cite{Galli:2009zc, Galli:2011rz, Ade:2015xua}, since at $z \sim 600$ the DM has not fallen into structures yet, and the observation is free from the usual astrophysical sources of uncertainties (density profile within a halo, distribution and density of subhalos, mass of the smallest bound halo). Moreover, additional  sources of systematics affecting the CMB constraints have also been thoroughly examined~\cite{Galli:2013dna, Weniger:2013hja}, and shown to affect the results below the sensitivity needed for this paper. In the following, we will neglect them leaving our conclusions unaltered.\\

In order to set the CMB bounds on the quantity of our interest $\Lambda$, we first obtain the observational bound on the thermally averaged annihilation cross-section at redshift 600:
\begin{equation}
p_{\mathrm{ann}}= \sum_j f_j(600, m_{\mathrm{DM}}) \frac{\langle \sigma v  \rangle_j(600)}{m_{\mathrm{DM}}} \;,
\end{equation}
where $\langle \sigma v  \rangle_j$(600) is the thermally averaged partial annihilation cross section for the $j$-th channel at redshift 600 and $f_j(z,m_{\mathrm{DM}})$ is the fraction of annihilation energy that is absorbed by the plasma at redshift $z$. The quantity $p_{\mathrm{ann}}$ is constrained by Planck TT, TE, EE and lowP  data~\cite{Ade:2015xua}:
\begin{equation}
p_{\mathrm{ann}} < 4.1\times 10^{-28} \,\,\frac{\mathrm{cm^3}}{\mathrm{s\,\,GeV}}\,\,\,\mathrm{at}\,\,\,95\%\,\,\mathrm{C.L.}, 
\end{equation}
and the values of the variable $f_j(z,m_{\mathrm{DM}})$ are taken form Ref. \cite{Slatyer:2009yq}. 

In order to bound the new physics scale $\Lambda$, we numerically compute the velocity dependent annihilation cross section with micrOMEGAs~\cite{Belanger:2008sj}, obtaining the relationship between $\langle \sigma v \rangle_j$(600) and $\Lambda$ for each effective operator and each final state. It may be noted that for those operators for which the s--wave process is dominating, the thermally averaged cross section is constant~\footnote{We remind the reader that the thermally averaged cross section is given by $$\langle \sigma v \rangle = \langle a + b v^2 + \dots \rangle = a + \frac{3}{2} b\frac{ T}{m_{DM}} + \dots$$ with $T$ the temperature at which the process is computed.} and $\langle \sigma v \rangle_j$(600)=$\langle \sigma v \rangle_j$(0). On the other hand, for those operators in which the p--wave contribution dominates the annihilation cross section, the CMB bound is almost ineffective, since it is suppressed by the low $T_{CMB}$ temperature. A possible bound in this case can be obtained from the Big Bang Nucleosynthesis (BBN), since this process happens at $T_{BBN} \sim \mathrm{MeV}$. As shown in References~\cite{Henning:2012rm,Kawasaki:2015yya}, the bounds on $\langle \sigma v \rangle$ are generically weaker than those obtained from the CMB, and we have explicitly checked that in all the cases in which the CMB bound is ineffective, the BBN bound is also not relevant.

\subsection{DM direct detection constraints \label{sec:setup-dm-dd}}

The determination of the DM mass and elastic scattering cross section in a DM DD experiments is affected by uncertainties associated to the flux of DM particles crossing the Earth at any given time. Although the uncertainties on the Sun's relative motion with respect to the Galactic Center, or the exact morphology of the Galactic bulge, do not affect the conclusion of the presence of a sizable component of DM at the Sun's location~\cite{Iocco:2015xga}, they impact the reconstruction of the DM profile throughout the Milky Way~\cite{Pato:2015dua}. More specifically, two sources of uncertainties are relevant for DM DD experiments:  the local DM distribution~\footnote{See Ref.~\cite{Read:2014qva} for a recent and thorough review.} and its velocity structure~\cite{Green:2010gw, Green:2011bv, Bozorgnia:2016ogo,Ibarra:2018yxq}. Although relevant, their effect is often overlooked in putting DD bounds on the parameter space of a model. Recently, Ref.~\cite{Benito:2016kyp} has shown that the effect of known Galactic uncertainties on the reconstruction of particle physics parameters may overcome those on the velocity structure. We thus decide to follow closely the analysis therein and take into account the different determinations of the local DM density due to the variation of the galactic parameters. In particular, we consider three possible values $\rho_{0} = 0.06\;\mathrm{GeV/cm^3}$, $\rho_{0} = 0.3\;\mathrm{GeV/cm^3}$ and $\rho_{0} = 1.8\;\mathrm{GeV/cm^3}$ (corresponding to the lowest and largest possible $\rho_0$, and to the one used by the experimental collaborations), and apply the experimental bounds stemming from Xenon1T SI searches~\cite{Aprile:2017iyp} and from PandaX-II SD analyses~\cite{Cui:2017nnn}.

In addition, when we obtain DD bounds, we also take into account the relic density constraint. Direct detection experiments constrain DM particles assuming that their relic density matches the one of the cold DM component. The simplest way to compute the detection rates is to rescale the DM density distribution according to the prescription
\begin{equation}
\rho_{i} \equiv \rho_0 \min\left(1, \frac{\Omega_{i}}{\Omega_{\mathrm{DM}}} \right)\, , ~~~~~~ i = \phi, \, \chi, \, V\, ,
\label{eq:rescale-rho}
\end{equation}
where $\rho_0$ is the local DM density, $\Omega_{\chi}$ is the theoretical relic density obtained via micrOMEGAs for every operator listed in Table~\ref{tab:EFToperators} and $\Omega_{\mathrm{DM}} \simeq 0.12$  from Ref.~\cite{Ade:2015xua}.

\subsection{Collider constraints}
\label{sec:collider-setup}

In the upcoming sections, we also present bounds coming from collider searches. In order to perform the required simulations, we implemented the different effective operators independently in FeynRules~\cite{Alloul:2013bka} and LanHep~\cite{Semenov:2010qt} and generated the signal using MadGraph5\_aMC@NLO \cite{Alwall:2011uj}. The hadronization and parton showering was done using PYTHIA 6.4~\cite{Sjostrand2006}, with subsequent detector simulation performed using MadAnalysis5~\cite{Conte:2012fm} and Delphes~\cite{deFavereau:2013fsa}. 

In order to obtain the limits on the scale $\Lambda$ we consider the CMS analysis of final state presenting jets and missing transverse energy~\cite{CMS:2017tbk} based on data obtained at 13 TeV with an integrated luminosity of 35.9 fb$^{-1}$. This analysis was performed 
as  a counting experiment in 22 independent signal regions characterized by (i) $E_T^{\mathrm{miss}}>250$ GeV, (ii) one jet with $p_T^{\mathrm{jet}}>100$ GeV and (iii) $|\eta_j|< 4.5$.  In order to simulate the DM contribution to this process we studied
\begin{equation}
p p \to j + ( \phi\phi \;,\; \chi\chi \;,\; VV)  
\end{equation}
for all operators in Table~\ref{tab:EFToperators}. As it is well known, higher dimensional operators such as the ones in our EFT can lead to perturbative partial wave unitarity violation at high energies, signaling a maximum value of the center of mass energy for its applicability. Therefore, in order to guarantee the validity of the EFT we impose in our simulation that the invariant mass of the DM pair $M_{\chi\chi,\phi\phi,VV}$ satisfies~\cite{Racco:2015dxa}:
\begin{equation}
\label{eq:inv-mass-cut}
 M_{\chi\chi,\phi\phi,VV} < \Lambda \;.
\end{equation}

In our statistical analysis we use the simplified likelihood approach given in Ref.~\cite{Collaboration:2242860}, taking into account the full correlation and covariance matrix provided in~\cite{CMS:2017tbk}. More specifically, we defined the likelihood function
\begin{equation}
\mathcal{L}(\Lambda, g_*, \theta) = \prod_i {\left(s_i(\Lambda,g_*) + b_i + \theta_i\right)^{n_i} e^{-(s_i(\Lambda,g_*) + b_i + \theta_i)} \over n_i!} e^{-\left({1\over 2} \theta^T V^{-1} \theta \right)} \;,
\label{eq:likelihood}
\end{equation}
where the $s_i(\Lambda, g_*)$ is the expected number of events of the DM signal in $i^{\rm th}$ bin, $b_i$ is the respective number of background events and $n_i$ is the number of observed events. In our case the signal cross section for each bin is a function of the coupling $g_*$ and the scale $\Lambda$.  For practical purposes we consider three different benchmark values for $g_*=4 \pi,\,\,6$ and $1$. The systematic uncertainties of the SM backgrounds and the DM signal are treated as nuisance parameters and they are approximated by zero-mean Gaussian variables $\theta_i$ and a covariance matrix $V$.

We define our test statistic function as
\begin{equation}
TS(\Lambda) \equiv \left\lbrace \begin{array}{ll} -2 \ln {\mathcal{L}(\Lambda,\hat{\theta}_\Lambda) \over \mathcal{L}(\hat{\Lambda}, \bar{\theta})} &\qquad \hat{\Lambda} \leqslant \Lambda \\
0 &\qquad \hat{\Lambda} > \Lambda 
\end{array}
\right. \;,
\end{equation}
where $\hat{\theta}_\Lambda$ is a $\theta$ vector that minimizes the logarithm of Eq.~\eqref{eq:likelihood} for a given value of $\Lambda$. On the other hand, $\hat{\Lambda}$ and $\bar{\theta}$ are the values of $\Lambda$ and of the $\theta$ vector that globally minimize the logarithm of Eq.~\eqref{eq:likelihood}. We find the upper limit on the scale of the mediator varying $\Lambda$ until $TS(\Lambda) =4$.\footnote{Here we neglect the small difference between $2\sigma$ exclusion and $95\%$ C.L. exclusion.}

In addition to the present limits we also  perform a projection of the CMS  reach  for an integrated luminosity of $300$ fb$^{-1}$. For this projection we assume a gaussian likelihood, that the number of background events scales with the luminosity and that the uncertainty on the background scales as the square root of the luminosity. We set a lower background limit to be $1\%$ of the background, based on post-fit numbers with respective background error provided by ATLAS and CMS~\cite{Aaboud:2016tnv,CMS:2017tbk}.

\section{Results}
\label{sec:results}

\begin{figure}[tb]
\vskip 0.8cm
(a)\hspace*{0.5\textwidth}\hspace{-0.2cm}(b)\\\vspace*{-2cm}\\ \\
        \includegraphics[width=0.5\textwidth]{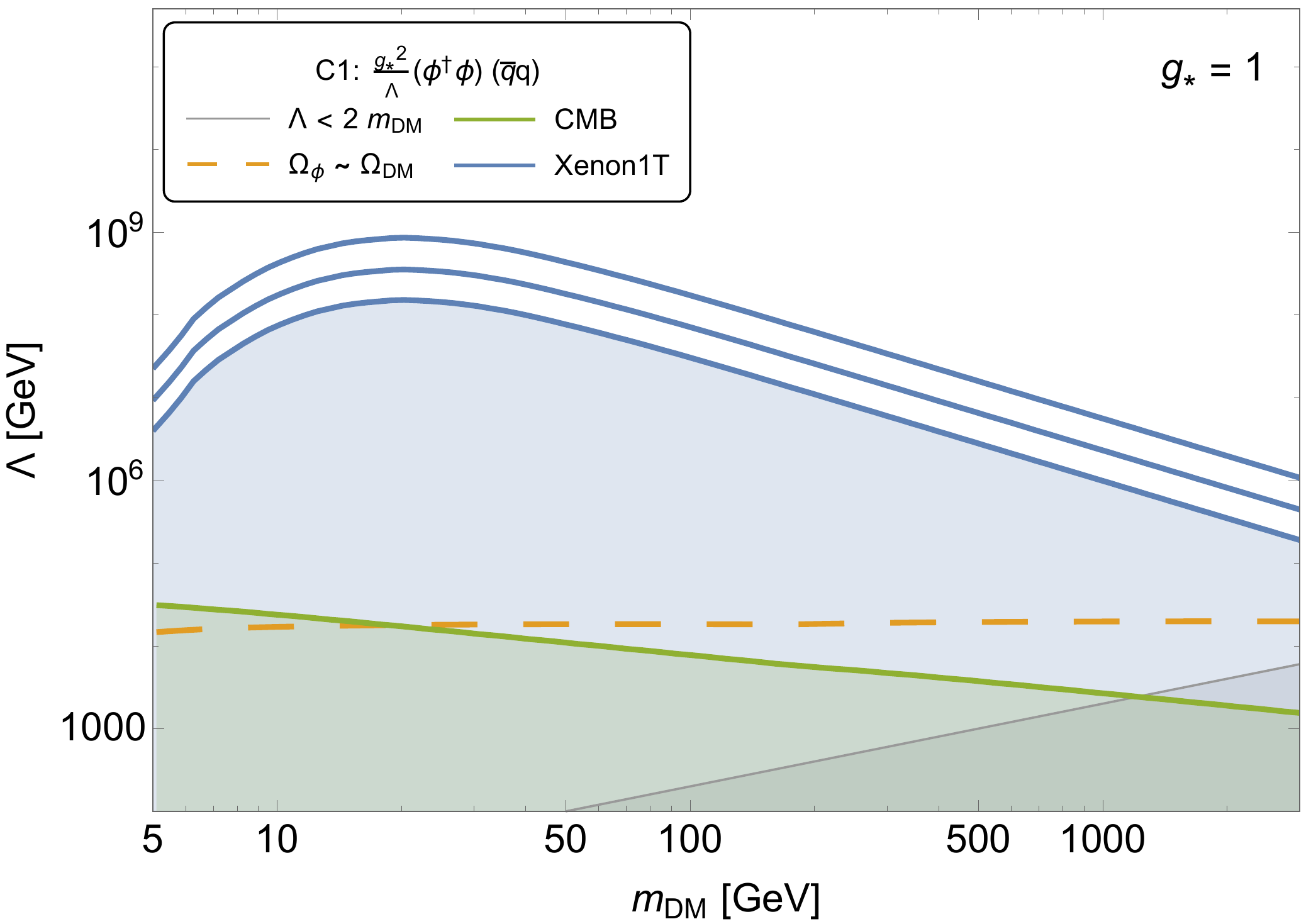}%
        \includegraphics[width=0.5\textwidth]{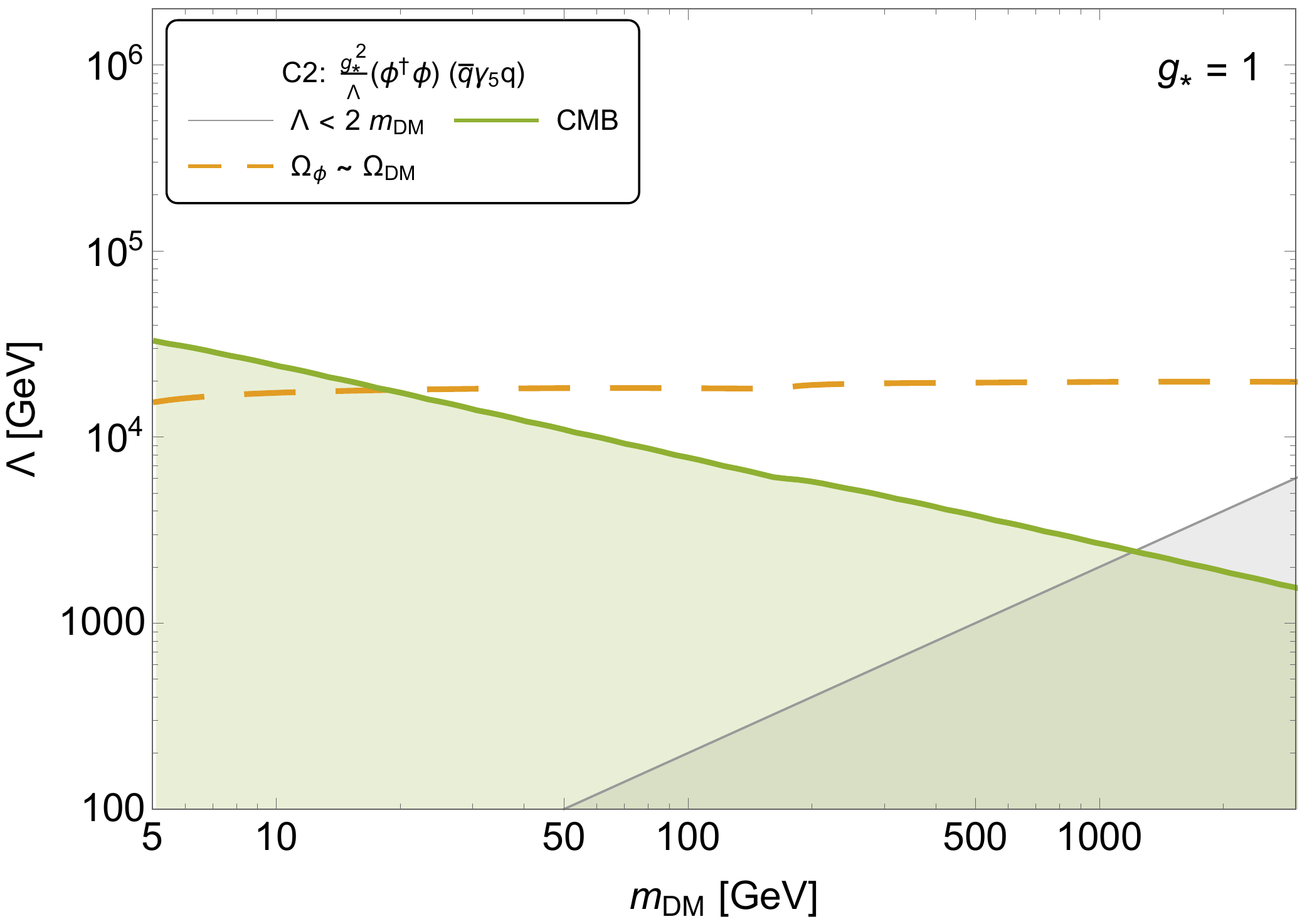}\\
\vskip 0.2cm
(c)\hspace*{0.5\textwidth}\hspace{-0.2cm}(d)\\\vspace*{-2cm}\\ \\
        \includegraphics[width=0.5\textwidth]{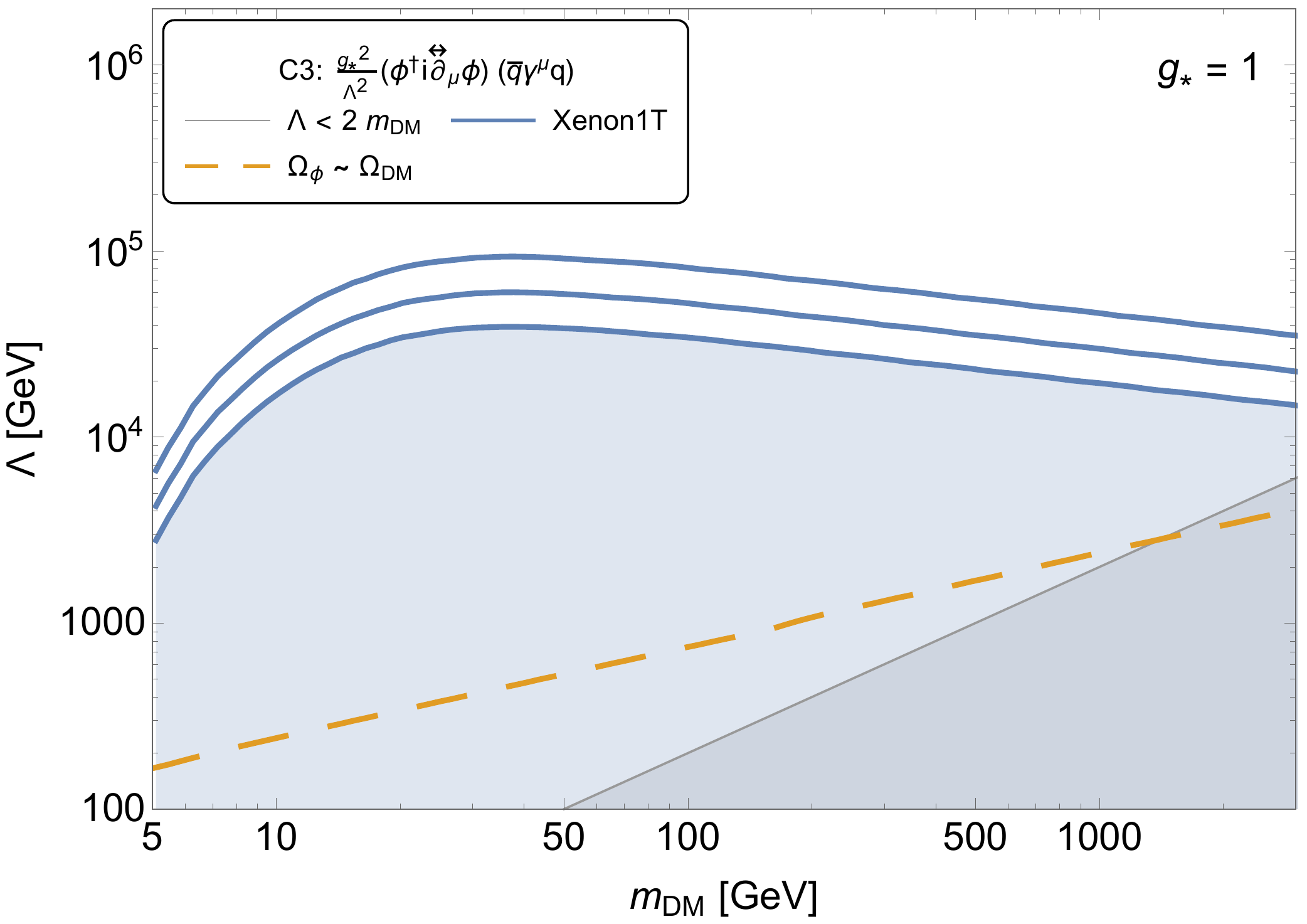}%
        \includegraphics[width=0.5\textwidth]{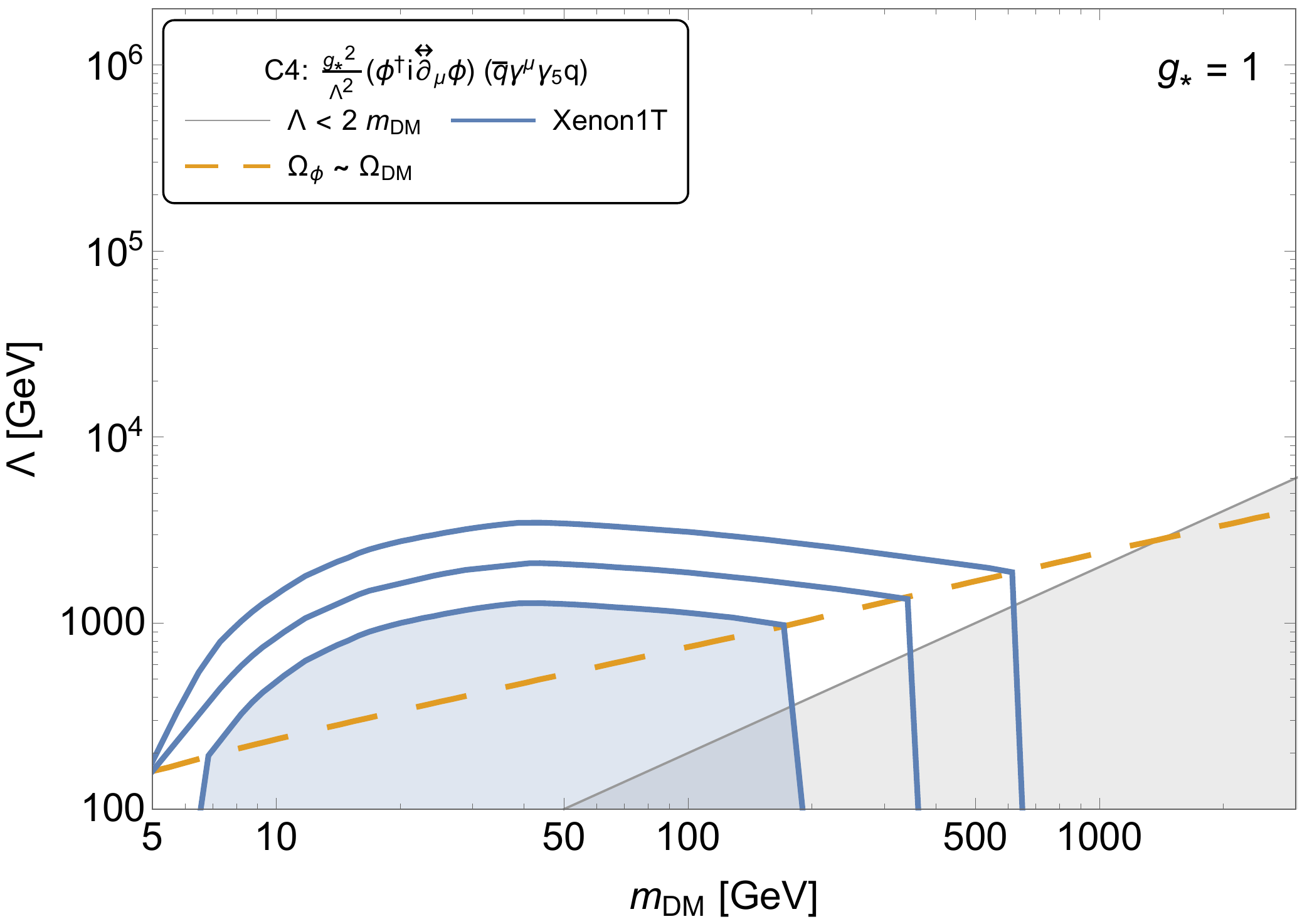}\\
\vskip 0.2cm
(e)\hspace*{0.5\textwidth}\hspace{-0.2cm}(f)\\\vspace*{-2cm}\\ \\
        \includegraphics[width=0.5\textwidth]{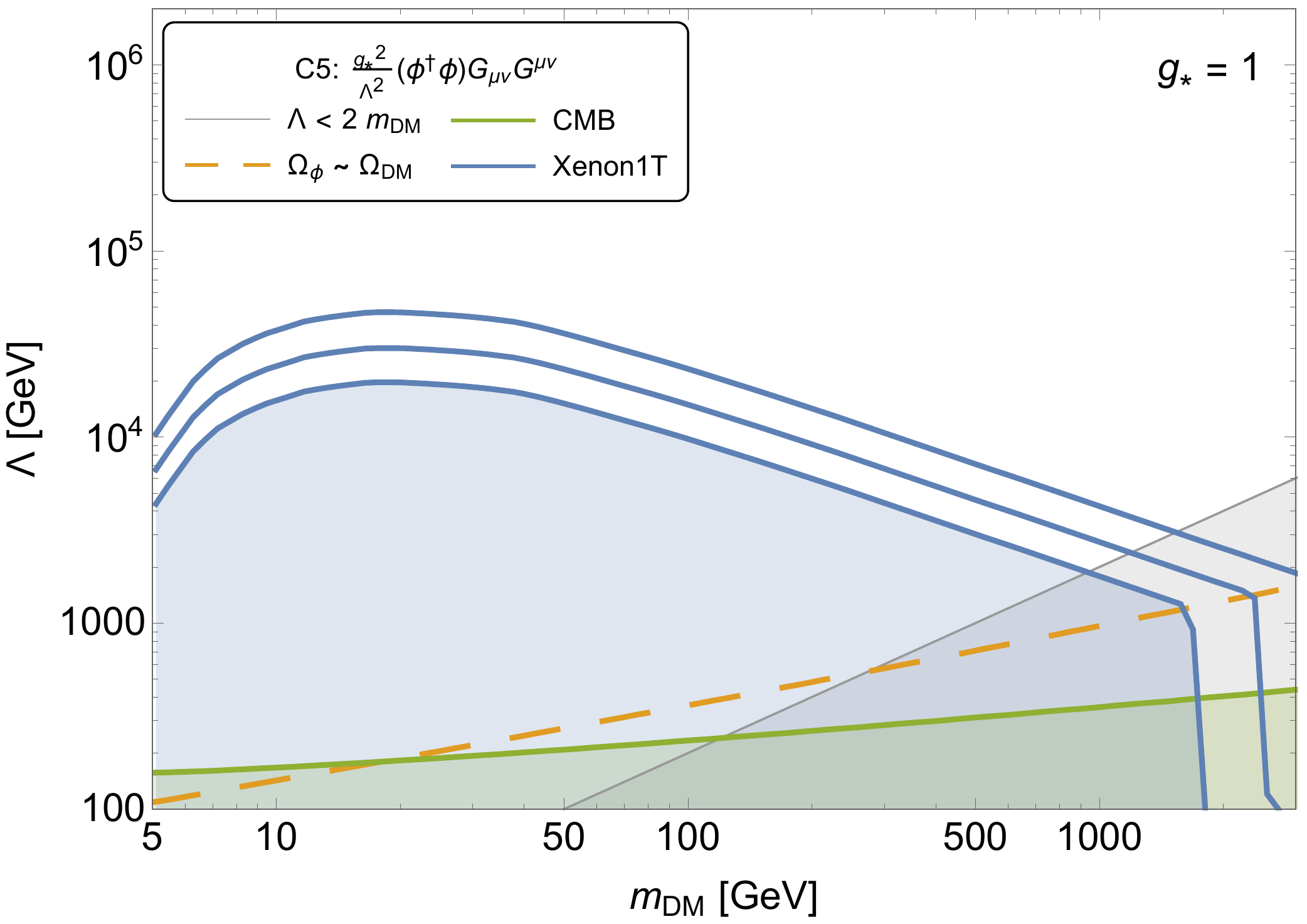}%
        \includegraphics[width=0.5\textwidth]{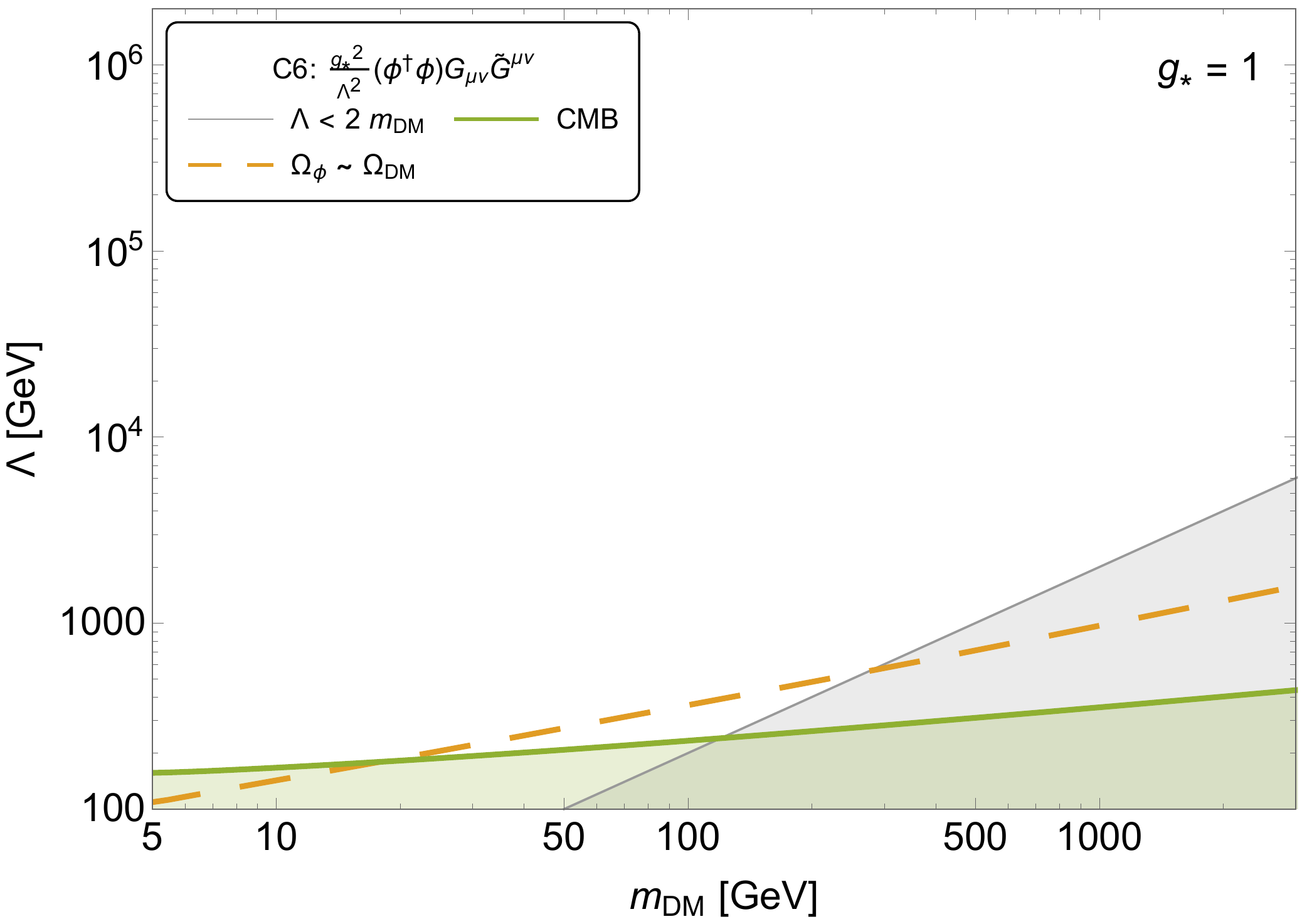} \\
        \caption{\label{fig:noncol-scalar} Non-collider constraints on
          the scalar C1--C6 operators, as indicated in the panel box:
          (i) constraints from SI DM DD searches (excluding the shaded
          blue region below the lowest blue contour), (ii) constraints
          from relic density (excluding the region above the yellow
          dashed line), (iii) constraints from the CMB (excluding the
          shaded green area below the solid green line) and (iv)
          constraints from the validity of the EFT (excluding the
          shaded grey area, where $\Lambda<2 m_{DM}$). The central and
          upper blue contour represent the central and lower values of
          the DM local density (see Section~\ref{sec:setup-dm-dd}).}
\end{figure}
\subsection{Non-collider constraints \label{non-collider-constraints}}

In this section we present combined non-collider constraints for all
operators under study. In particular, we obtain the bounds originating
from DM DD searches, indirect DM searches from the CMB and relic
density assuming the freeze-out mechanism and standard cosmology. The
results in this section are obtained with three different tools:
micrOMEGAs~\cite{Belanger:2013oya}, a modified version of the code
released with~\cite{DelNobile:2013sia} and
runDM~\cite{Crivellin:2014qxa,DEramo:2014nmf,DEramo:2016gos,runDM}.

\subsubsection*{Complex scalar DM}

\begin{figure}[tb]
\vskip 0.8cm
(a)\hspace*{0.5\textwidth}\hspace{-0.2cm}(b)\\\vspace*{-2cm}\\\\
        \includegraphics[width=0.5\textwidth]{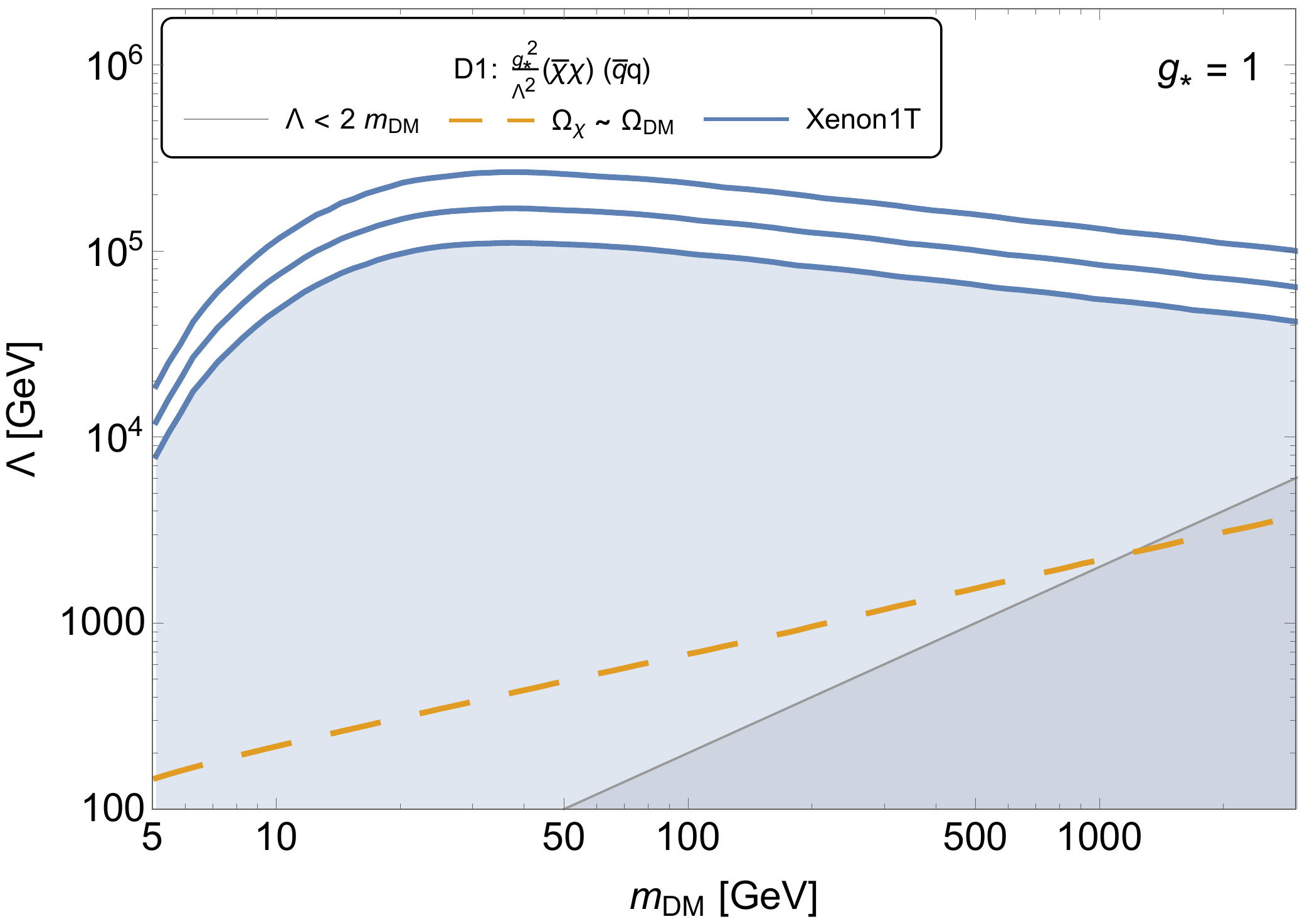}%
        \includegraphics[width=0.5\textwidth]{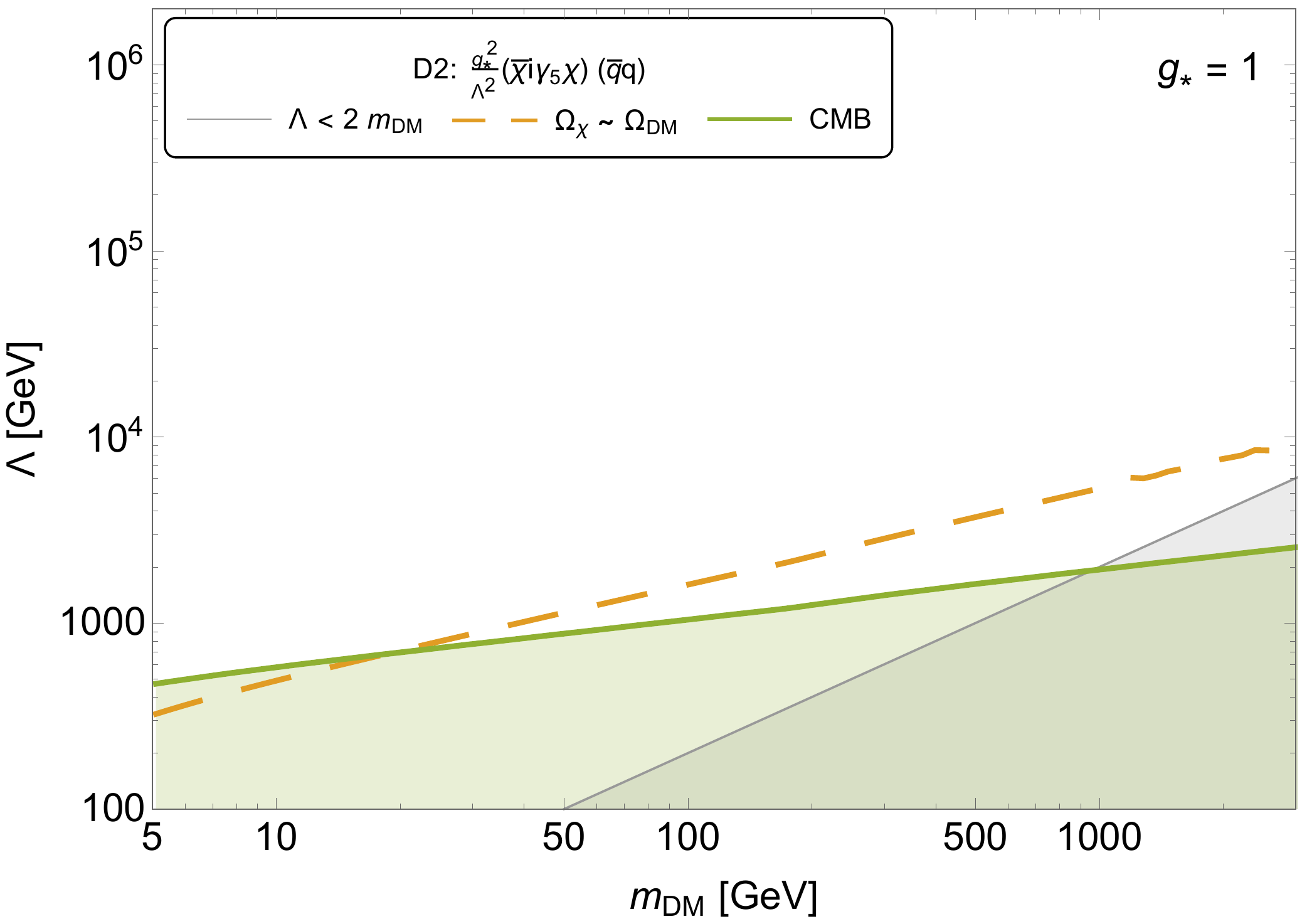} \\
\vskip 0.2cm
(c)\hspace*{0.5\textwidth}\hspace{-0.2cm}(d)\\\vspace*{-2cm}\\\\
        \includegraphics[width=0.5\textwidth]{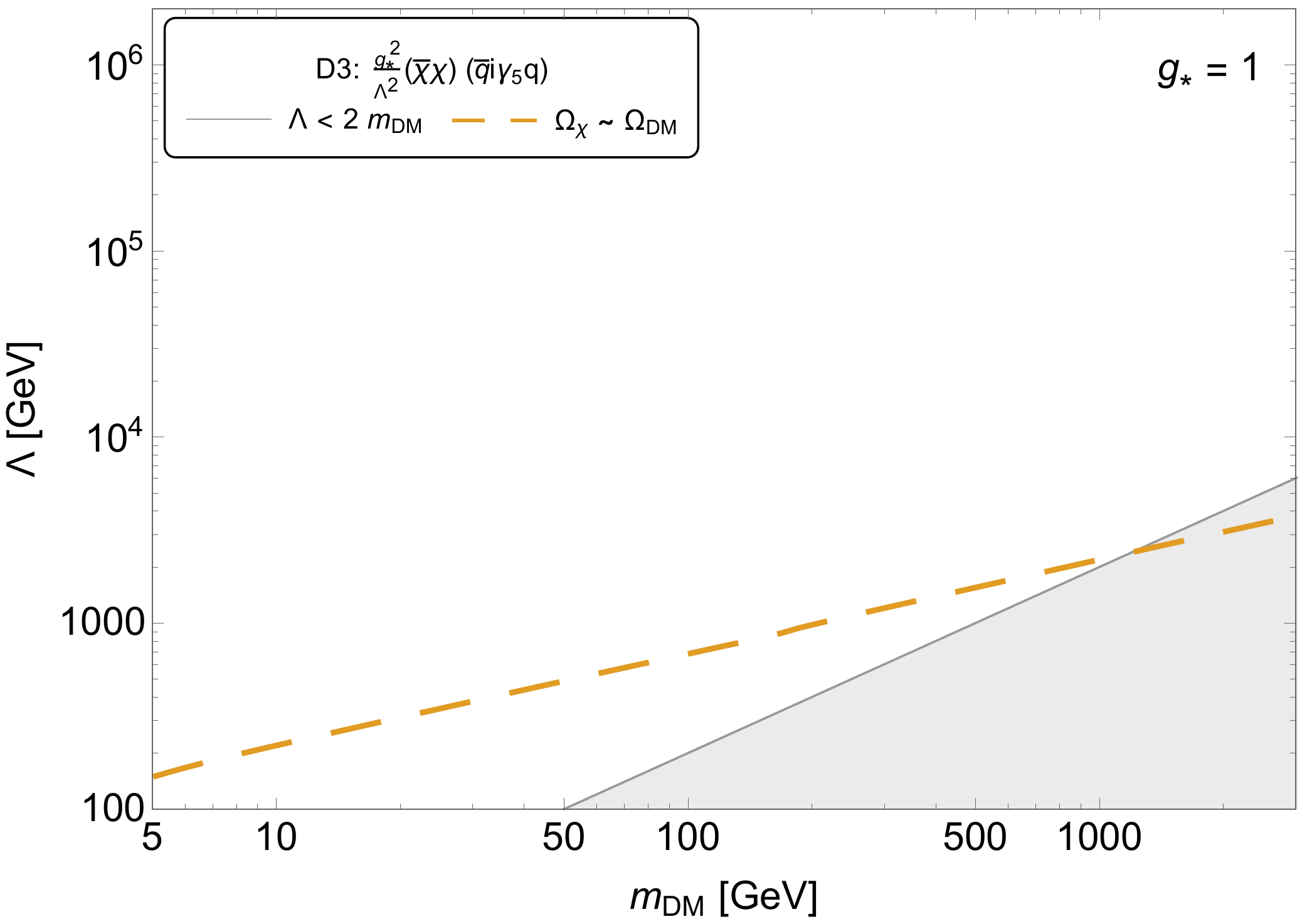}%
        \includegraphics[width=0.5\textwidth]{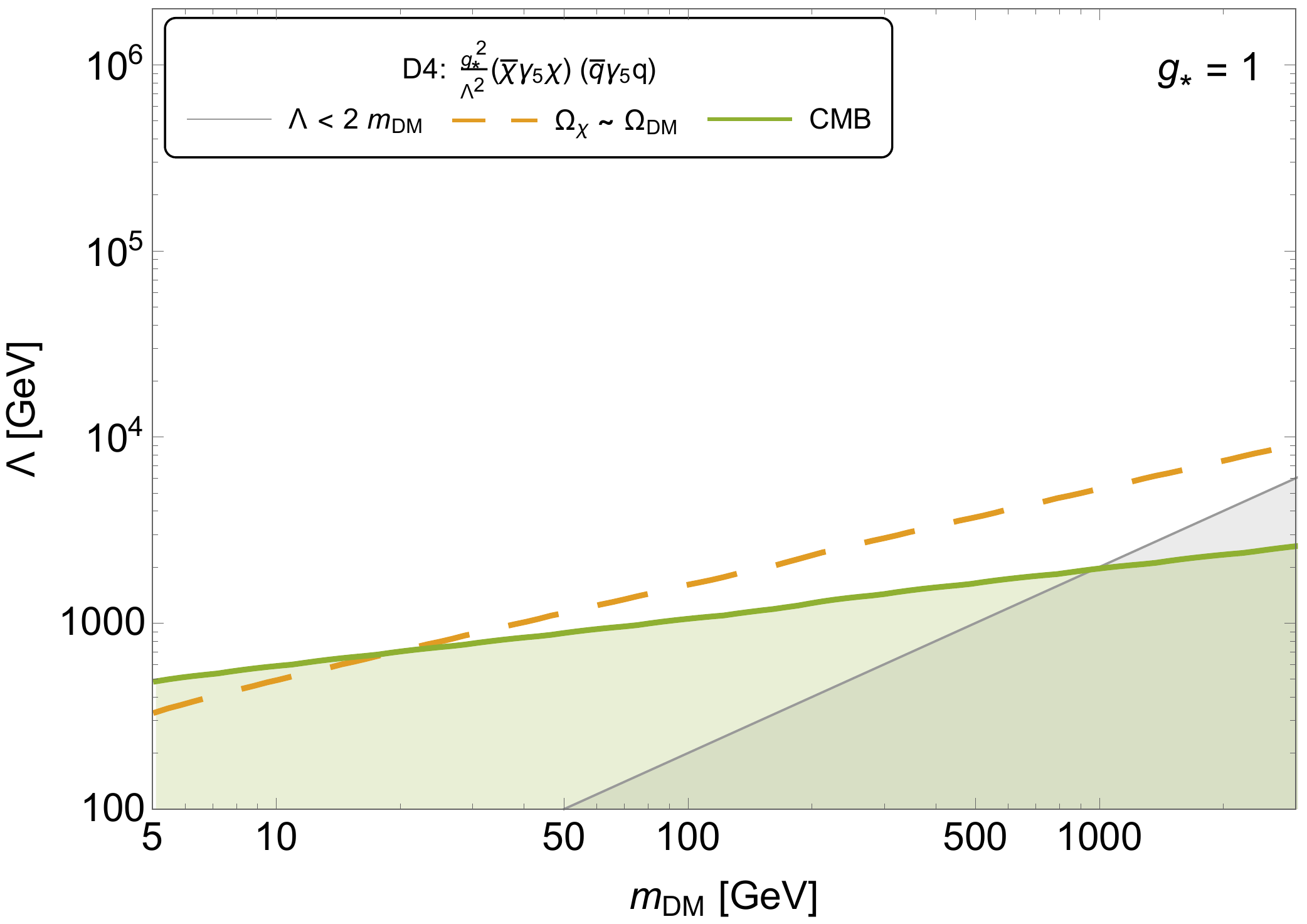} \\
        \caption{\label{fig:noncol-dirac1} As in
          Figure~\ref{fig:noncol-scalar}, for the fermion operators
          involving scalar and pseudoscalar currents D1--D4.}
\end{figure}

We start presenting in Fig.~\ref{fig:noncol-scalar} the results for
complex scalar DM. In all the panels, the area below the three blue
curves represents the region excluded by SI DM DD experiments, taking
into account the uncertainty on the local DM density as discussed in
Section~\ref{sec:setup-dm-dd}. The shaded blue region is the
conservative exclusion, while the middle and upper contours correspond
to the center (namely the one used by the experimental collaboration)
and the most optimistic exclusions, respectively. The dashed yellow
line corresponds instead to the region of parameter space in which the
predicted relic density matches the DM density observed by the Planck
collaboration. Above the dashed yellow curve, the predicted relic density is 
larger than the experimental one. The computation is done assuming that the observed DM
relic density is associated with only one particle, interacting with
the SM via only one of the effective operators listed in
Table~\ref{tab:EFToperators}. The region excluded by the CMB
measurement is indicated by a shaded green area below the solid green
line. Finally, the grey-shaded region represents the region
$\Lambda < 2 m_{DM}$, excluded by the requirement of the validity of
the EFT.

From Fig.~\ref{fig:noncol-scalar}(a) we can learn that the C1 operator
is strongly constrained by SI DM DD searches since this operator leads
to a cross section that is neither velocity nor transferred-momentum
suppressed. In fact, we can see from this panel that any scale
$\Lambda$ {\it below} 1000 TeV is excluded even for
$m_{DM}=1$~TeV. Moreover, the other two blue contours associated with
the astrophysical uncertainties show that the bounds can be tightened
by almost an order of magnitude. Therefore, it is important to be
conservative to make a robust exclusion of the parameter space, as we
do in the present work.  We can also see from
Fig.~\ref{fig:noncol-scalar}(a) that the relic density data exclude
$\Lambda$ above 20 TeV for all DM mass range under consideration,
since in this region the $\phi$ particle would over-close the
universe. Notwithstanding, one should note that this bound is quite
model dependent, {\it e.g.}  adding an additional particle that could
co-annihilate with the DM could change the relic density dramatically.
Finally, the CMB measurement leads to quite strong bounds that exclude
$\Lambda$ {\it below} 1-10 TeV for $g_\star=1$, depending on the DM
mass. This is due to the fact that the C1 operator gives rise to
s-wave annihilation cross section, which is strongly constrained by
the CMB bound.

We show in Fig.~\ref{fig:noncol-scalar}(b) the non-collider
constraints on the operator C2. As expected, the CMB and relic density
bounds on C2 are identical to those on C1. However, the SI DM DD
constraints are absent, since this operator violates
parity. Therefore, the limit from CMB plays a crucial role for this
operator, excluding the parameter space below the solid green line for
$g_\star = 1$. The unconstrained window can be tightened for DM masses above
$\simeq 20$ GeV using the relic density information if we assume that
there is no process that leads to DM co-annihilation.

We display in Fig.~\ref{fig:noncol-scalar}(c) the limits for the
operator C3, that contains a vector quark current. The bounds coming
from SI DM DD are weaker than those on the operator C1 since C3 is a
dimension six operator, while C1 is dimension five. The uncertainty on
this constraint is roughly a factor of 3 due to the higher dimensionality 
of the operator. Moreover, the annihilation cross section for the C3 
operator is velocity suppressed (p--wave), therefore the effect on the 
CMB spectrum is negligible.  The complementarity of the relic density 
and DD DM bounds completely rule out WIMP models that give rise at 
low energy to this operators only.

The bounds on the C4 operator, that contains a pseudo-vector quark
current, are shown in Fig.~\ref{fig:noncol-scalar}(d).  At variance
with the operator C2, the SI DM DD limits are non-vanishing for
operator C4 since the running and mixing effects play an important
role; for further details see Section~ \ref{sec:eft-run}. Notice that
the sudden drop of the SI DM DD constraint around $m_{DM} \simeq 200$
GeV is due to the rescaling of the DM density distribution given in
Eq.~\eqref{eq:rescale-rho}. Similarly to the operator C3, C4 is also
velocity suppressed, and consequently it is not constrained by the
CMB.

Finally, Figs.~\ref{fig:noncol-scalar}(e) and~\ref{fig:noncol-scalar}(f) 
present the non-collider bounds for the C5 and C6 operators, involving the gluon field strength. As we 
can see, there is no DD bound on C6 due to its parity violating nature. The bound on C5 is instead important, 
given the large gluon content of the proton. For DM masses below $10$ GeV, the bound is comparable to the one 
on the C3 operator. For larger masses, the bound weakens because the SI cross section is suppressed by an 
additional factor of $m_{DM}^2$ with respect to the C3 case (see Eq.~\eqref{eq:lambda_GG}). Still, the SI bound 
is always the dominant one, in the region in which the EFT is valid. The constraints of the operator C6 are instead much weaker,
since it does not contribute to SI DM DD, being dominated by the CMB
data and relic density.

From Fig.~\ref{fig:noncol-scalar} we can notice that the slope of the
CMB constraint is negative for the C1 and C2 operators, in contrast
with it being positive for C5 and C6. This happens because the leading
term for the annihilation cross section is independent of the DM mass
for C1 and C2, while it is proportional to the DM mass squared for C5
and C6.
This picture is generic -- 
if the amplitude for DM pair annihilation  is not velocity suppressed, 
then using dimensional analysis one finds that:
a) for dimension five  operators, the annihilation cross section is constant with respect to the DM mass, leading to the slight negative slope because the ``effective" photon spectrum for the CMB constraint drops with the invariant mass; 
b) for dimension six  operators the annihilation cross section is proportional to $M_{DM}^2$ and one observes positive slope in $(M_{DM}, \Lambda)$ plane for the CMB constraint.


\subsubsection*{Dirac fermion DM}
\begin{figure}[tb]
\vskip 0.8cm
(a)\hspace*{0.5\textwidth}\hspace{-0.2cm}(b)\\\vspace*{-2cm}\\\\
        \includegraphics[width=0.5\textwidth]{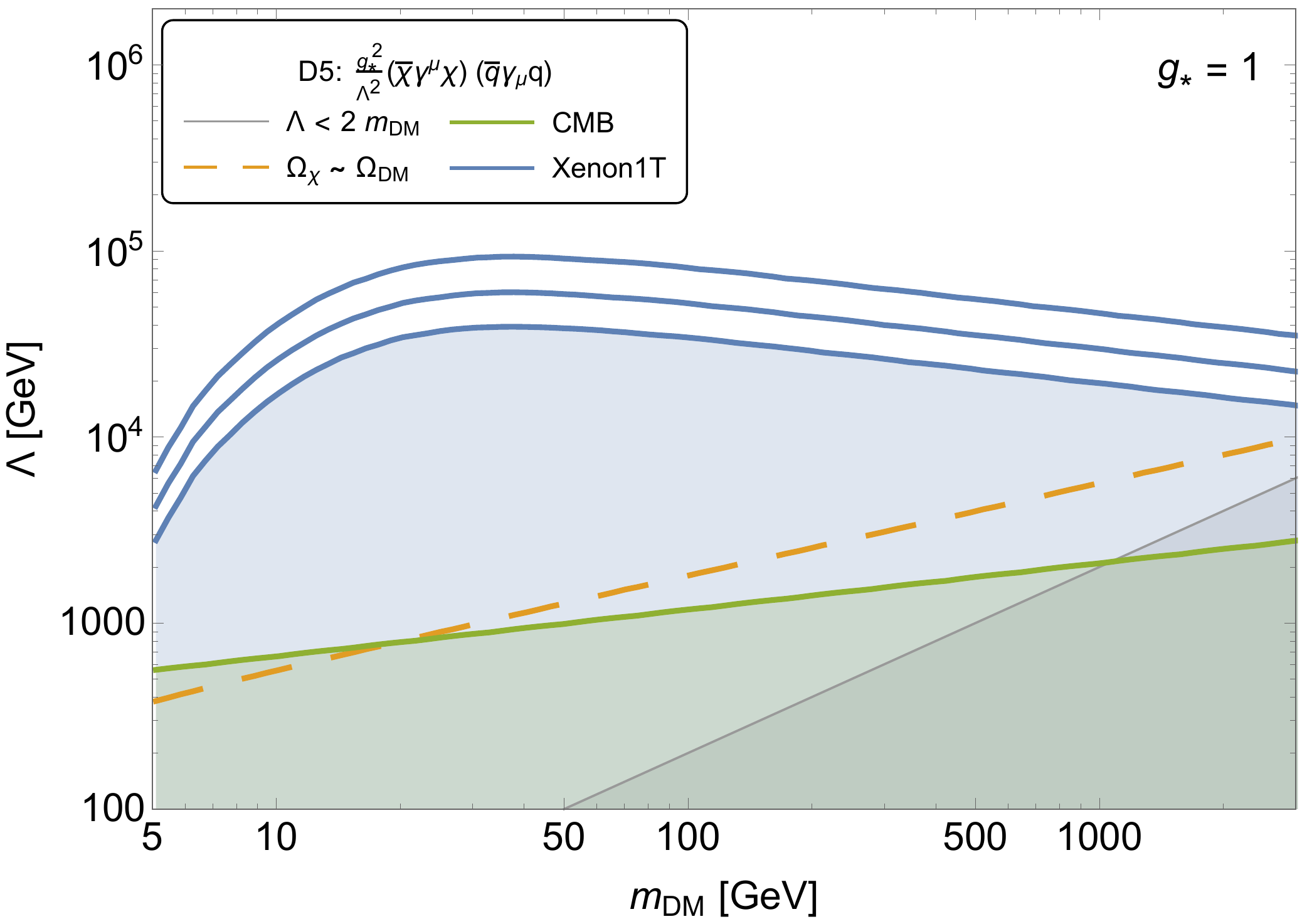}%
        \includegraphics[width=0.5\textwidth]{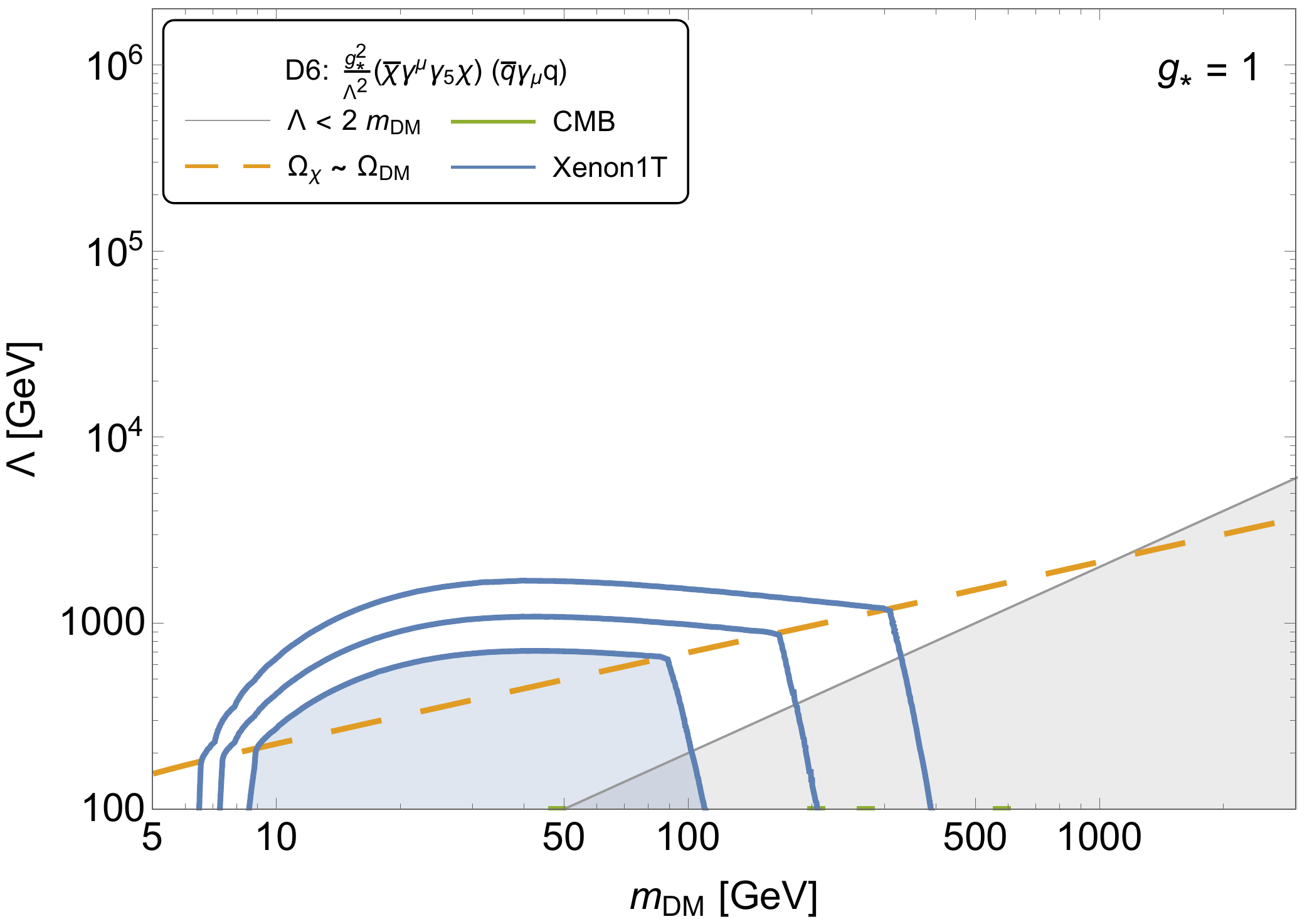}\\
\vskip 0.2cm
(c)\hspace*{0.5\textwidth}\hspace{-0.2cm}(d)\\\vspace*{-2cm}\\\\
        \includegraphics[width=0.5\textwidth]{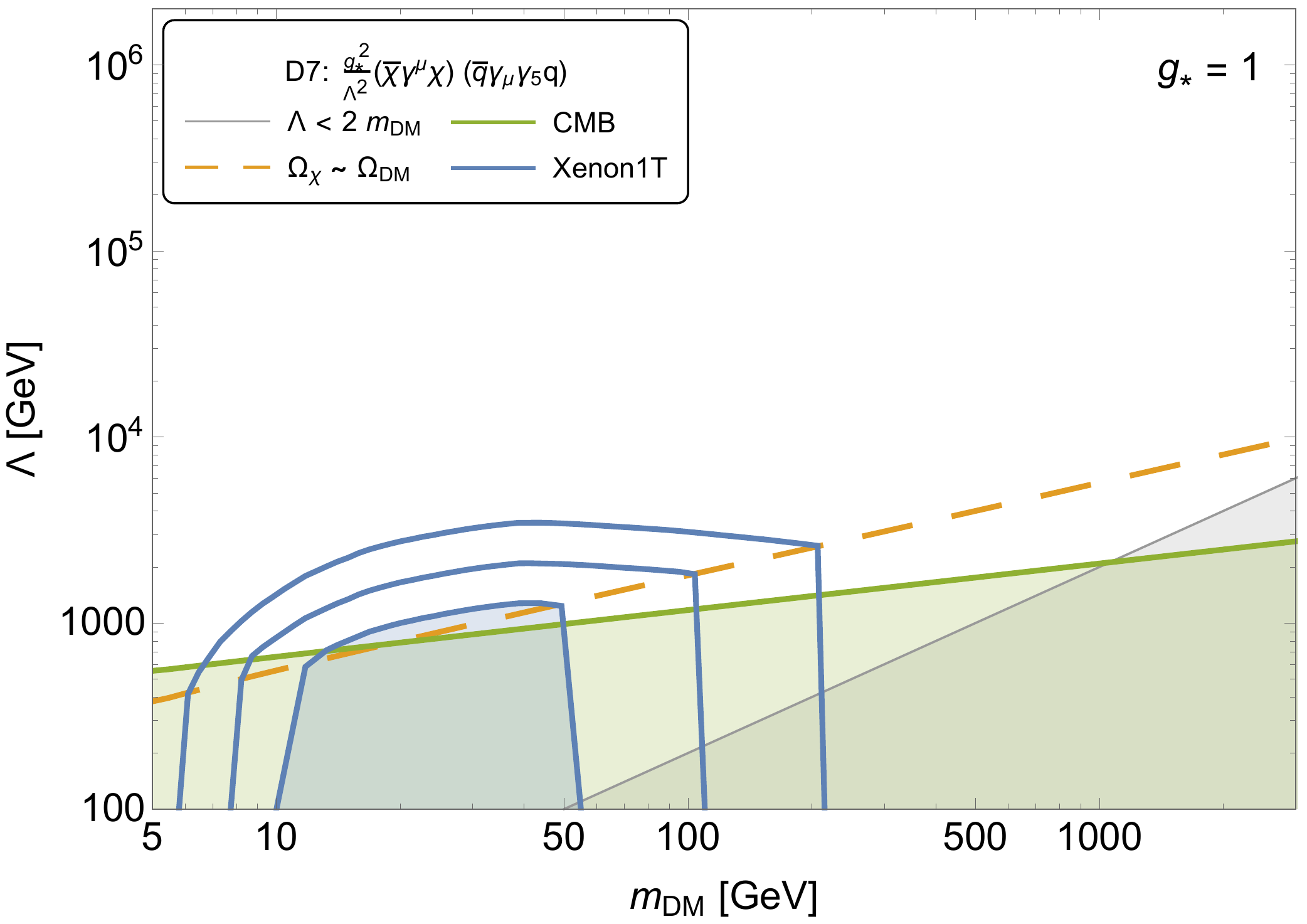}%
        \includegraphics[width=0.5\textwidth]{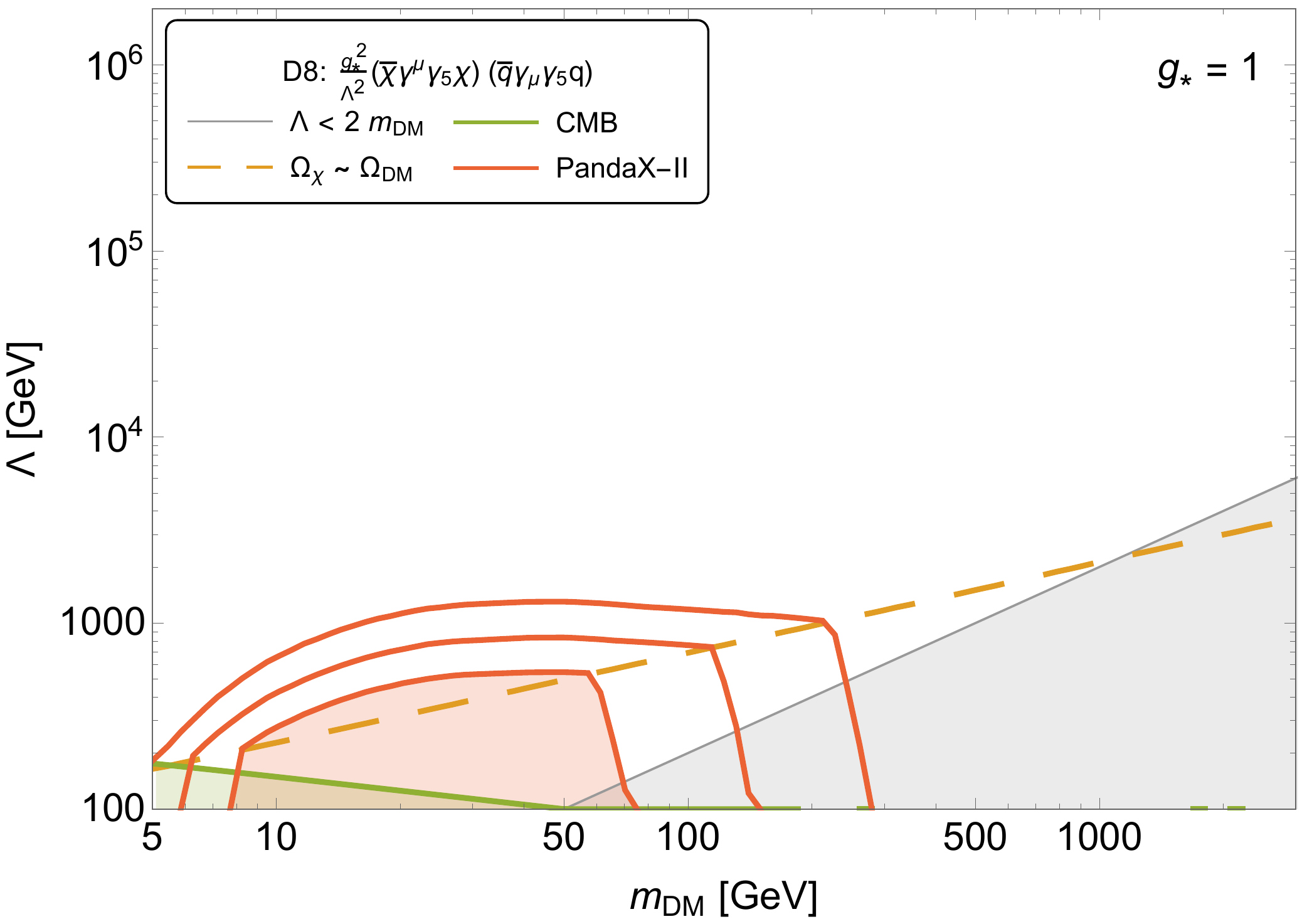}\\
    \caption{\label{fig:noncol-dirac2} As in
      Figure~\ref{fig:noncol-scalar} for the D5--D8 fermion operators, 
      involving vector and vector-axial currents. The solid red lines
      and the respective shaded area indicate the SD DM DD bounds from
      PandaX-II. The uncertainty due to the different values of the local 
	  DM density is presented in an analogous way as for SI DM DD.}
\end{figure}

We present in Figure~\ref{fig:noncol-dirac1} the non-collider constraints for the operators D1--D4, that exhibit scalar or pseudo-scalar interactions of fermion Dirac DM with quarks.
From this figure we can see the operator D1 is strongly constrained by the SI DM DD searches while the operators D2, D3 and D4 are not bound by these searches since they give rise either to spin-dependent interactions (D3 and D4) or the spin-independent cross section is momentum suppressed (D2). Notice that the uncertainty of the SI DM DD limits on D1 is of the order of 3 since this operator is of dimension six; this is true for all operators D1--D10.
On the other hand, the CMB data lead to constraints on the operators D2 and D4 since they exhibit an s--wave annihilation cross section, while D1 and D3 are not bound by CMB due to their p-wave cross sections.
The relic density constraint set a strong upper bound on the operators D2, D3 and D4 and rule out the operator D1 unless there is a mechanism to avoid the overproduction of DM in the early universe.
As before, the theoretical consistency of our framework, {\it i.e.} the EFT validity region, becomes important just for large DM masses.

\begin{figure}[tb]
\vskip 1cm
(a)\hspace*{0.5\textwidth}\hspace{-0.2cm}(b)\\\vspace*{-2cm}\\\\
        \includegraphics[width=0.5\textwidth]{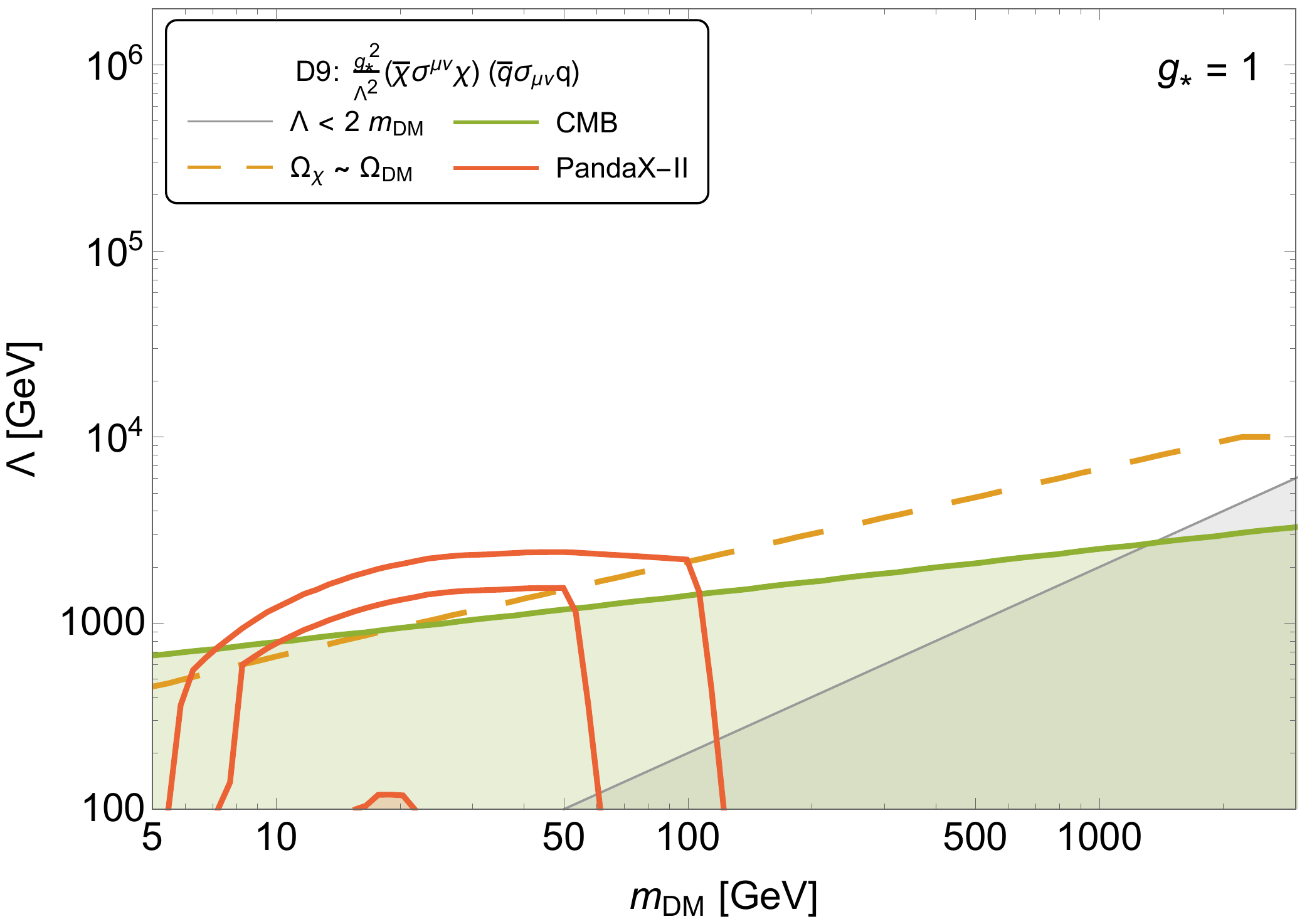}%
        \includegraphics[width=0.5\textwidth]{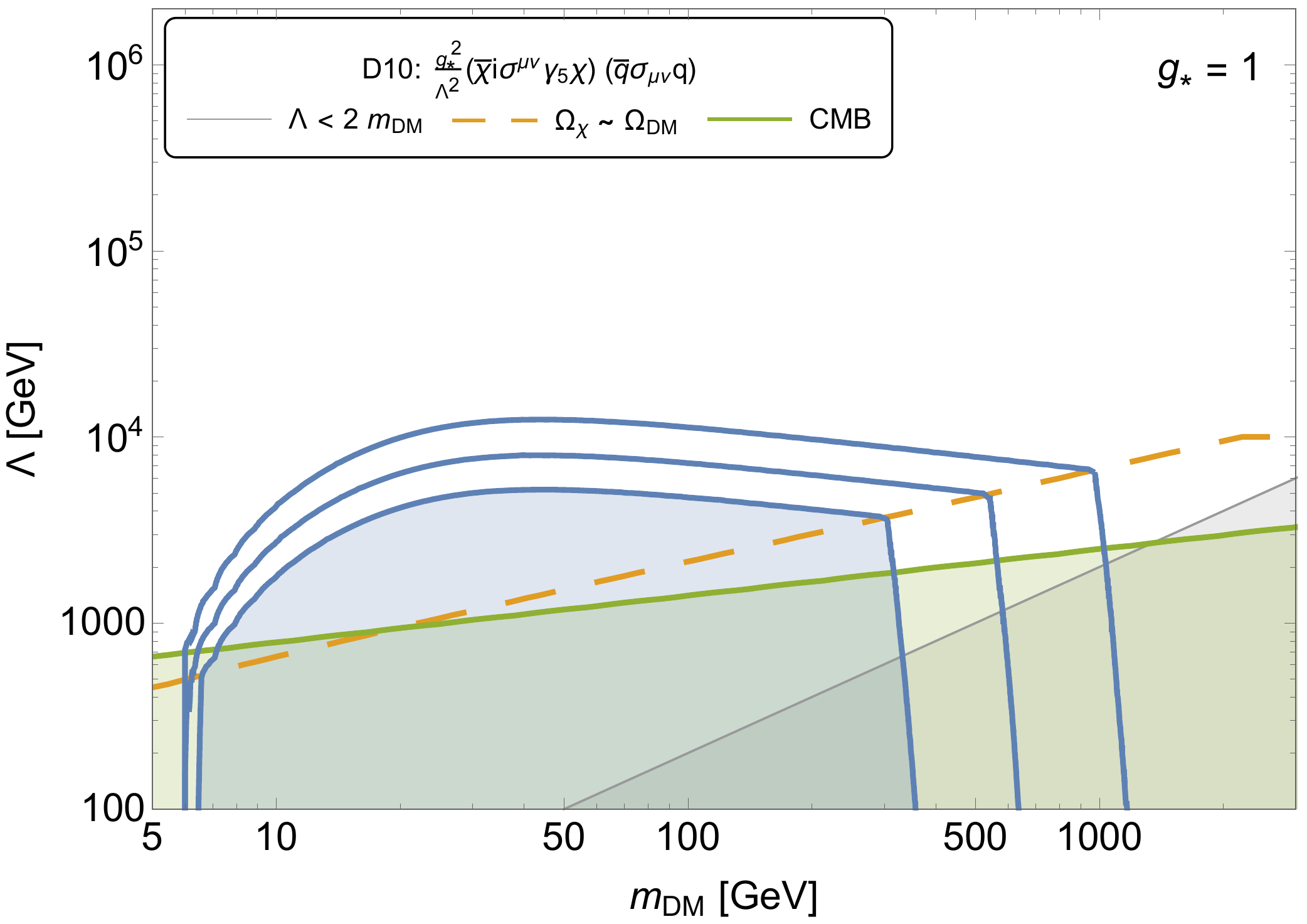}\\
        \caption{\label{fig:noncol-dirac3} As in
          Figure~\ref{fig:noncol-scalar} for the D9--D10 fermion
          operators involving tensor currents.}
\end{figure}

We show in Fig.~\ref{fig:noncol-dirac2} the non-collider limits on the
fermion operators D5--D8, which contain vector and axial vector
currents. The SI DM DD bound on the D5 operator is very strong, since
its SI cross section is unsuppressed. On the other hand, the
constraints on the operator D6 are milder, since its SI cross section
is both velocity and nuclear recoil momentum suppressed. The operator
D7 has in principle only SD cross sections. However, as discussed in
Section~\ref{sec:eft-run}, the running from the higher scale mixes the
operators D7 and D5, leading to SI limits on D7. The same is true for
the D8 operator, which in the running to low energy mixes into
D6. Notwithstanding, the strongest limits on D8 come from the
PandaX-II SD searches~\cite{Fu:2016ega}, since the D6 SI cross section
is momentum suppressed in contrast with the D8 SD one.
We can see from the Fig.~\ref{fig:noncol-dirac2} that the CMB data put
stronger constraints on the D5 and D7 operators since they lead to
unsuppressed s-wave annihilation cross sections. The limits on the
operators D8 are much weaker since its s-wave cross section is dumped
by a factor $m_q^2/m_{DM}^2$, while the operator D6 is not limited due
to its p-wave annihilation cross section.

Finally, we show in Figure~\ref{fig:noncol-dirac3} the non-collider
limits on the operators D9 and D10 which exhibit tensor currents.
The DD cross section due to operator D9 is spin dependent, however, it
is not suppressed.  Notice that the conservative limit on D9 is
restricted to a very narrow DM mass range. To understand this result,
it is interesting to compare the operators D8 and D9, since they have
the same low energy limit. Despite the numerical coefficient of D9
being twice the one of D8~\cite{DelNobile:2013sia}, the constraints on
D8 are stronger than the ones on D9 because the bounds on the latter
are weakened by the rescaling in the relic density given in
Eq.~(\ref{eq:rescale-rho}).
On the other hand, the SI bounds on D10 are the most stringent ones up
to DM masses of the order of 400 GeV despite momentum suppressed cross
section.
The CMB bound on both operators D9 and D10 is important in part of the
parameter space, since these operators exhibit s-wave annihilation
cross sections. Moreover, for both operators, the relic density
leaves only a small window below $\Lambda\sim 10$ TeV unconstrained.

Before concluding, let us observe that also in the fermion DM case the CMB 
bound has positive slope, as was happening in the case of the C5 and C6 
operators. Again, this is due to the fact that when the annihilation 
cross section is not velocity suppressed, it grows as $m_{DM}^2$.

\subsubsection*{Vector DM}

Before we present the non-collider bounds on vector DM we would
like to stress that the operators V1--V12 exhibit high energy
asymptotic behaviors that correspond to dimension-seven and -eight
operators~\cite{Belyaev:2016pxe}, and therefore we included additional
powers of $m_{DM}/\Lambda$ in their parametrisation; see
Table~\ref{tab:EFToperators}.  With this parametrisation the limits on
$\Lambda$ from the LHC and non-collider experiments are of the same
order as the one for scalar and fermion DM operators; for further
details see Ref.~\cite{Belyaev:2016pxe}.

\begin{figure}[tb]
\vskip 0.8cm
(a)\hspace*{0.5\textwidth}\hspace{-0.2cm}(b)\\\vspace*{-2cm}\\\\
        \includegraphics[width=0.5\textwidth]{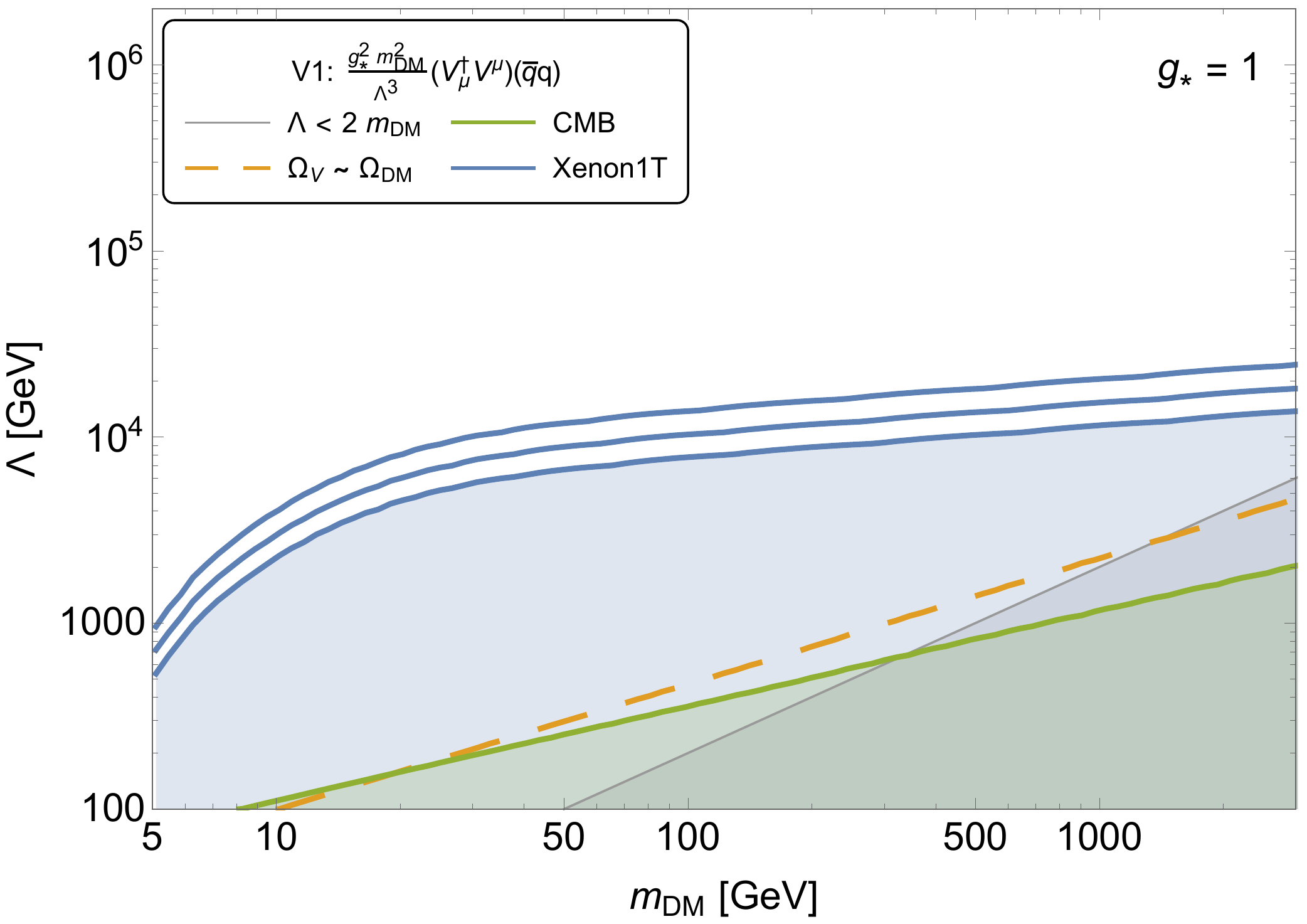}%
        \includegraphics[width=0.5\textwidth]{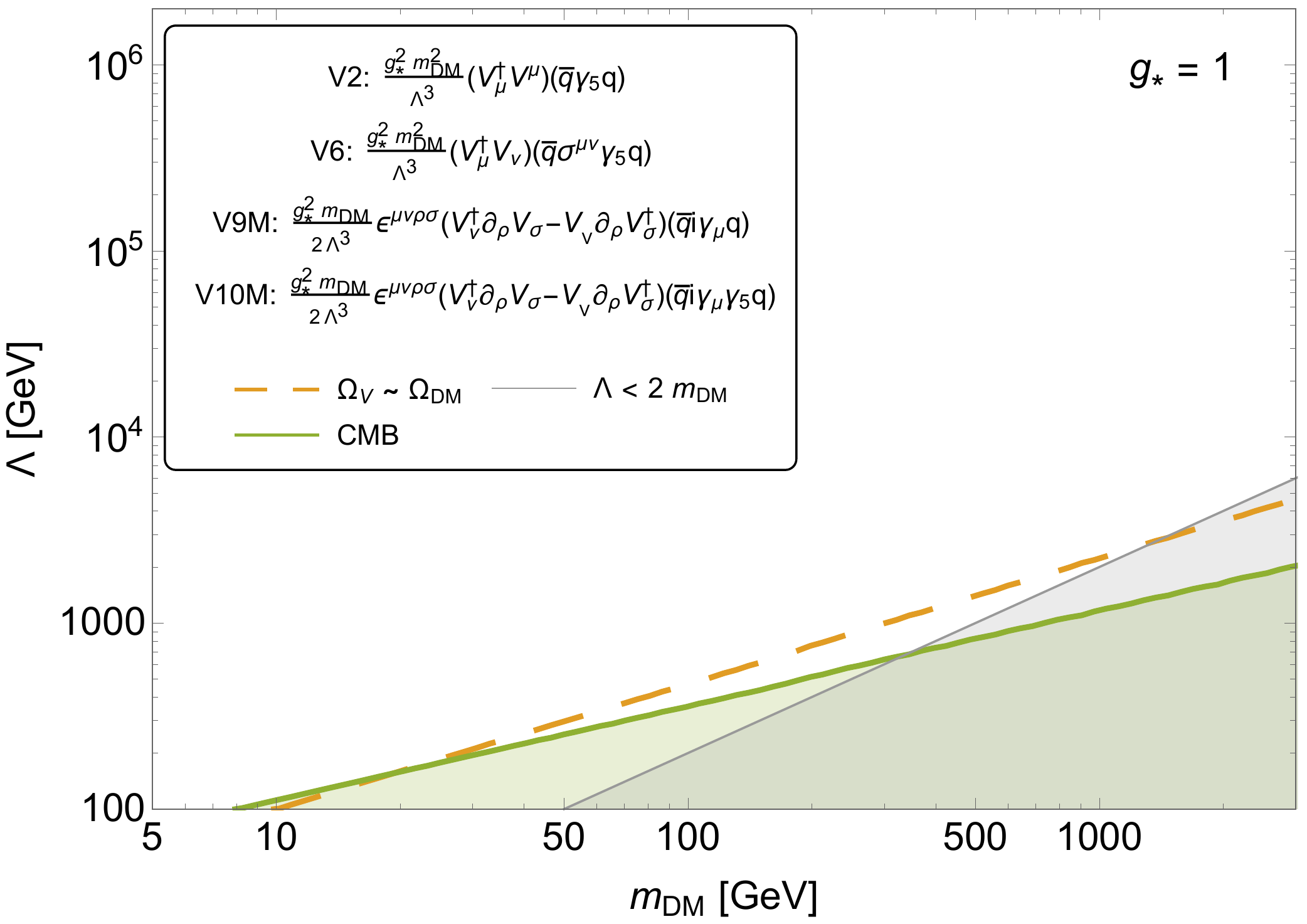}\\
\vskip 0.2cm
(c)\hspace*{0.5\textwidth}\hspace{-0.2cm}(d)\\\vspace*{-2cm}\\\\
        \includegraphics[width=0.5\textwidth]{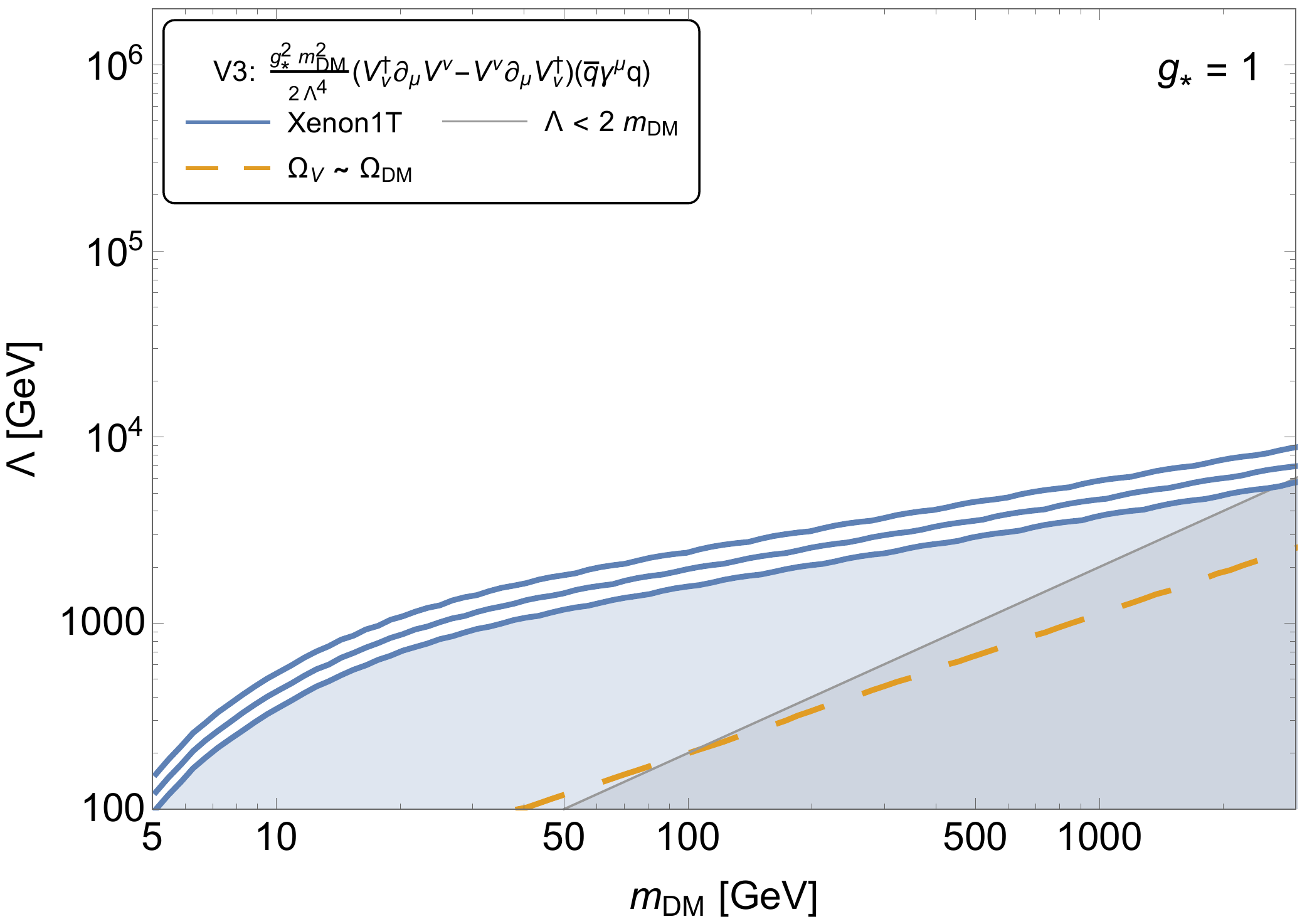}%
        \includegraphics[width=0.5\textwidth]{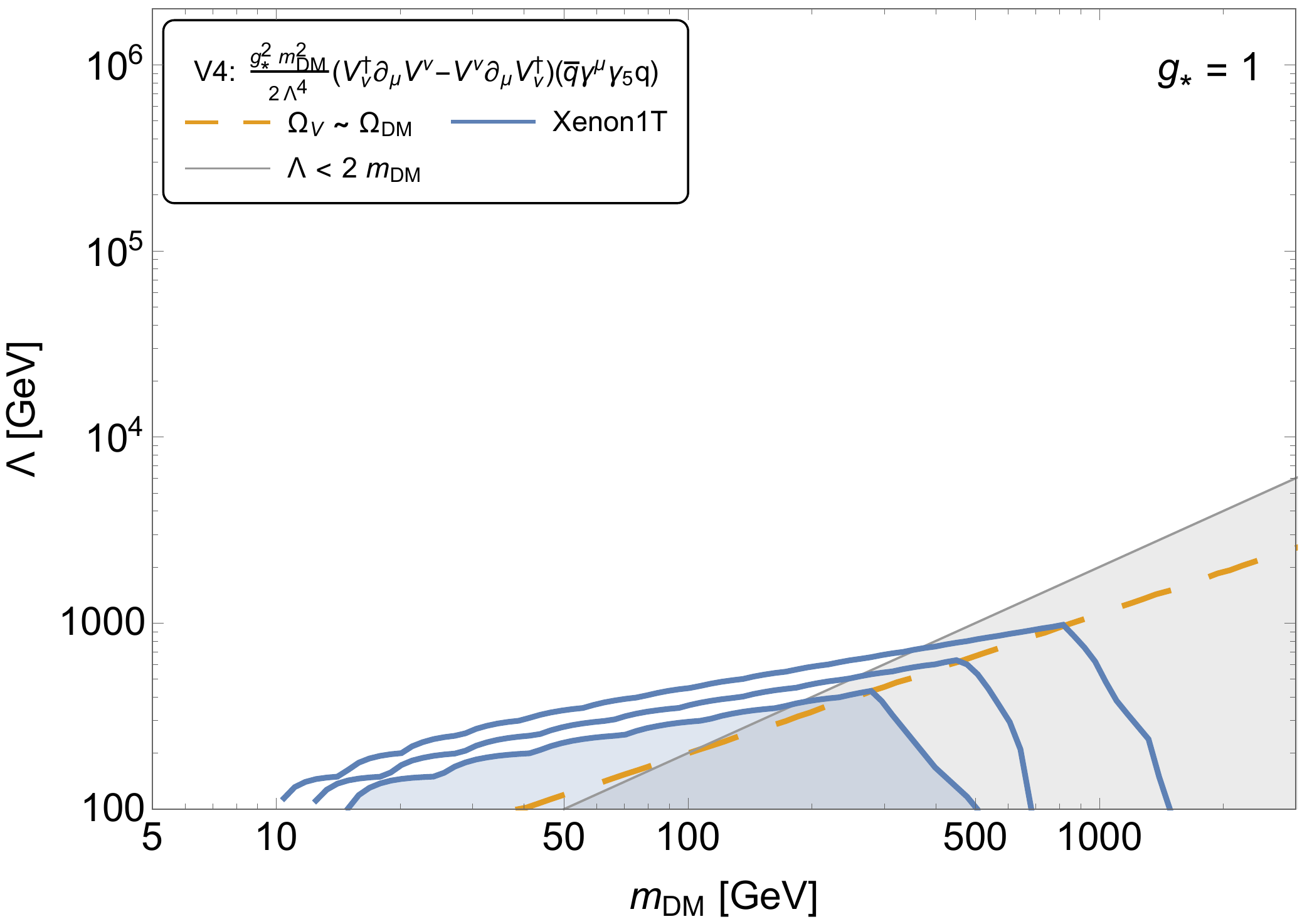}\\
\vskip 0.2cm
(e)\hspace*{0.5\textwidth}\hspace{-0.2cm}(f)\\\vspace*{-2cm}\\\\
           \includegraphics[width=0.5\textwidth]{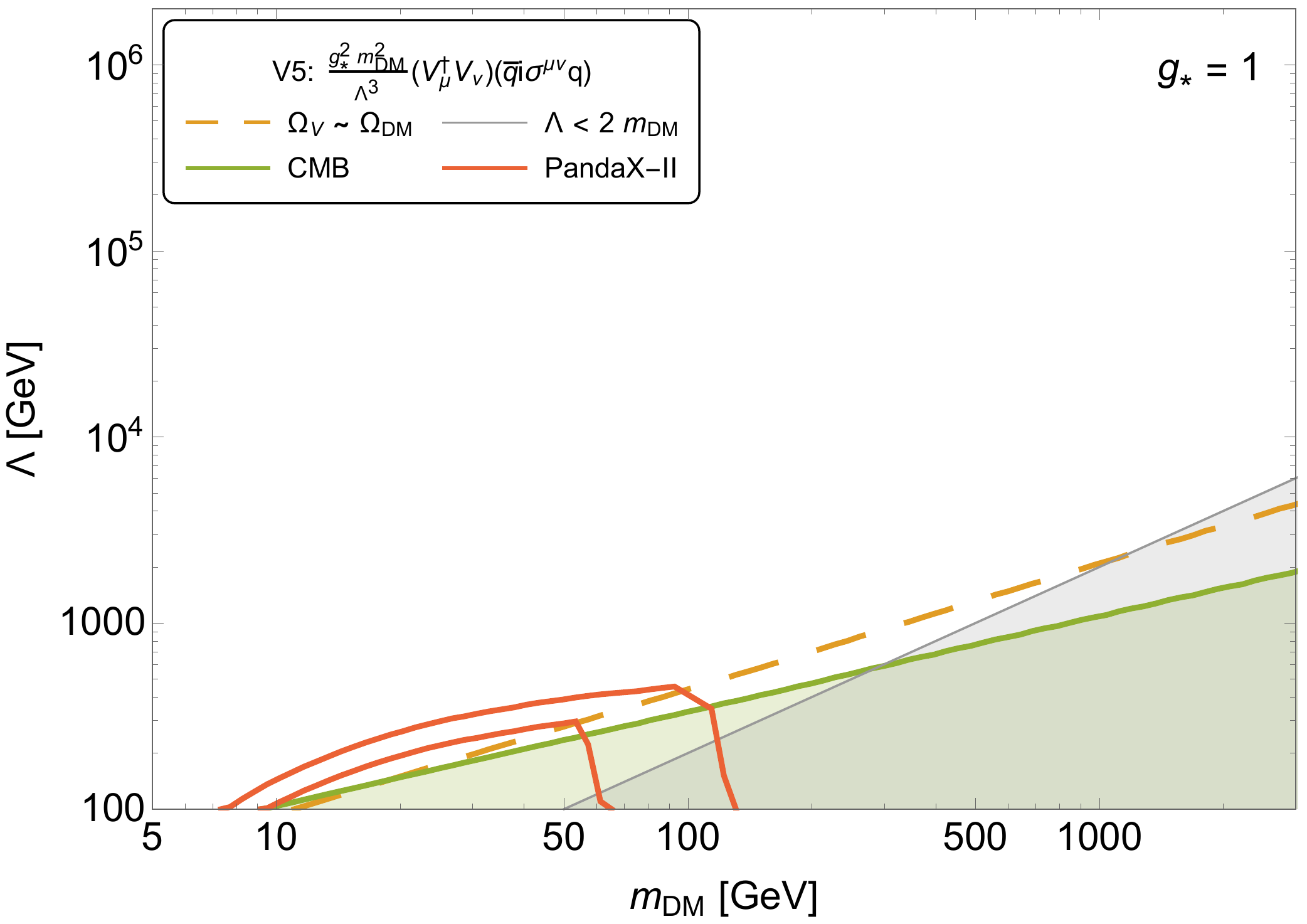}%
           \includegraphics[width=0.5\textwidth]{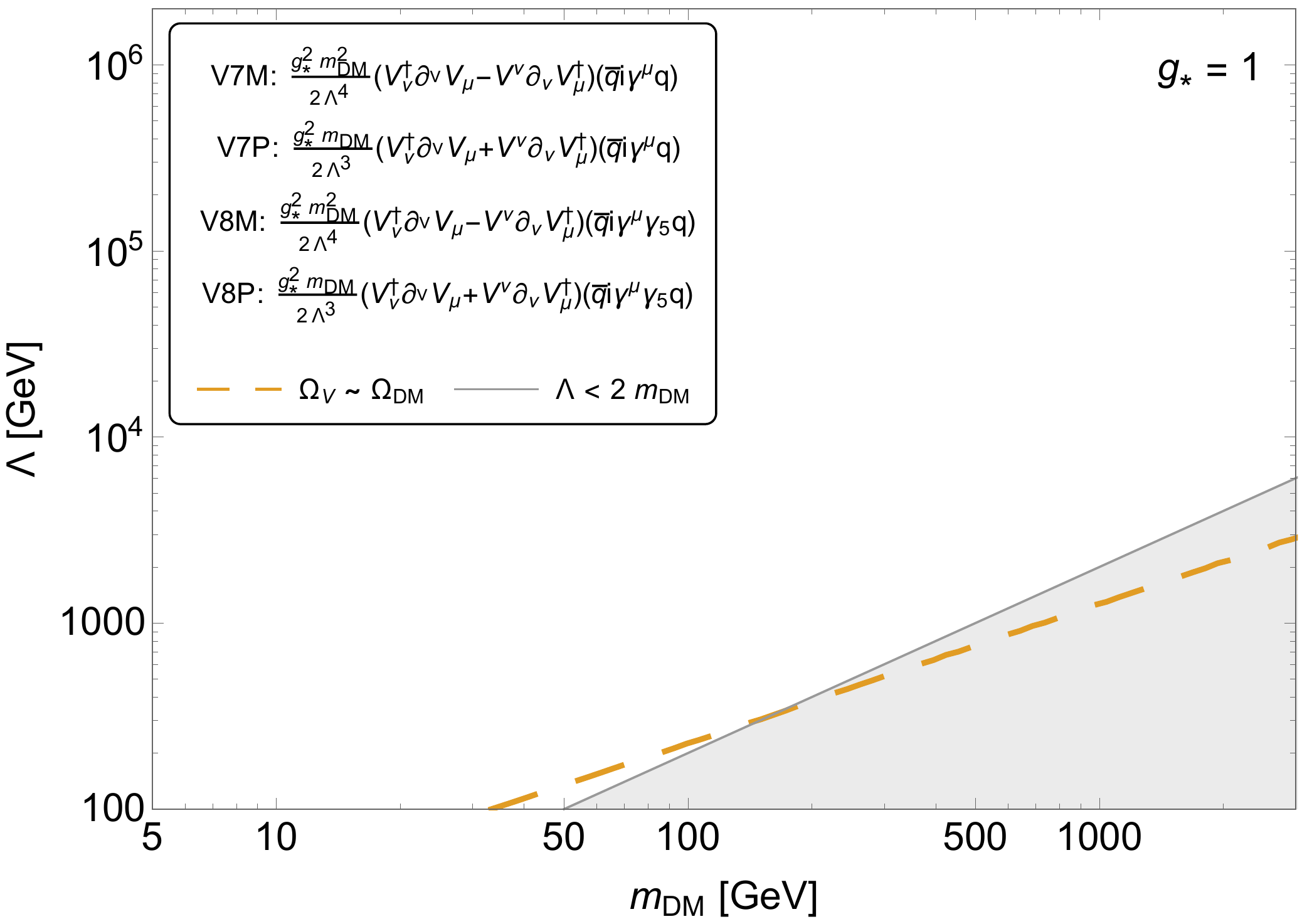}\\
\vskip 0.2cm
(g)\hspace*{0.5\textwidth}\hspace{-0.2cm}(h)\\\vspace*{-2cm}\\\\
              \includegraphics[width=0.5\textwidth]{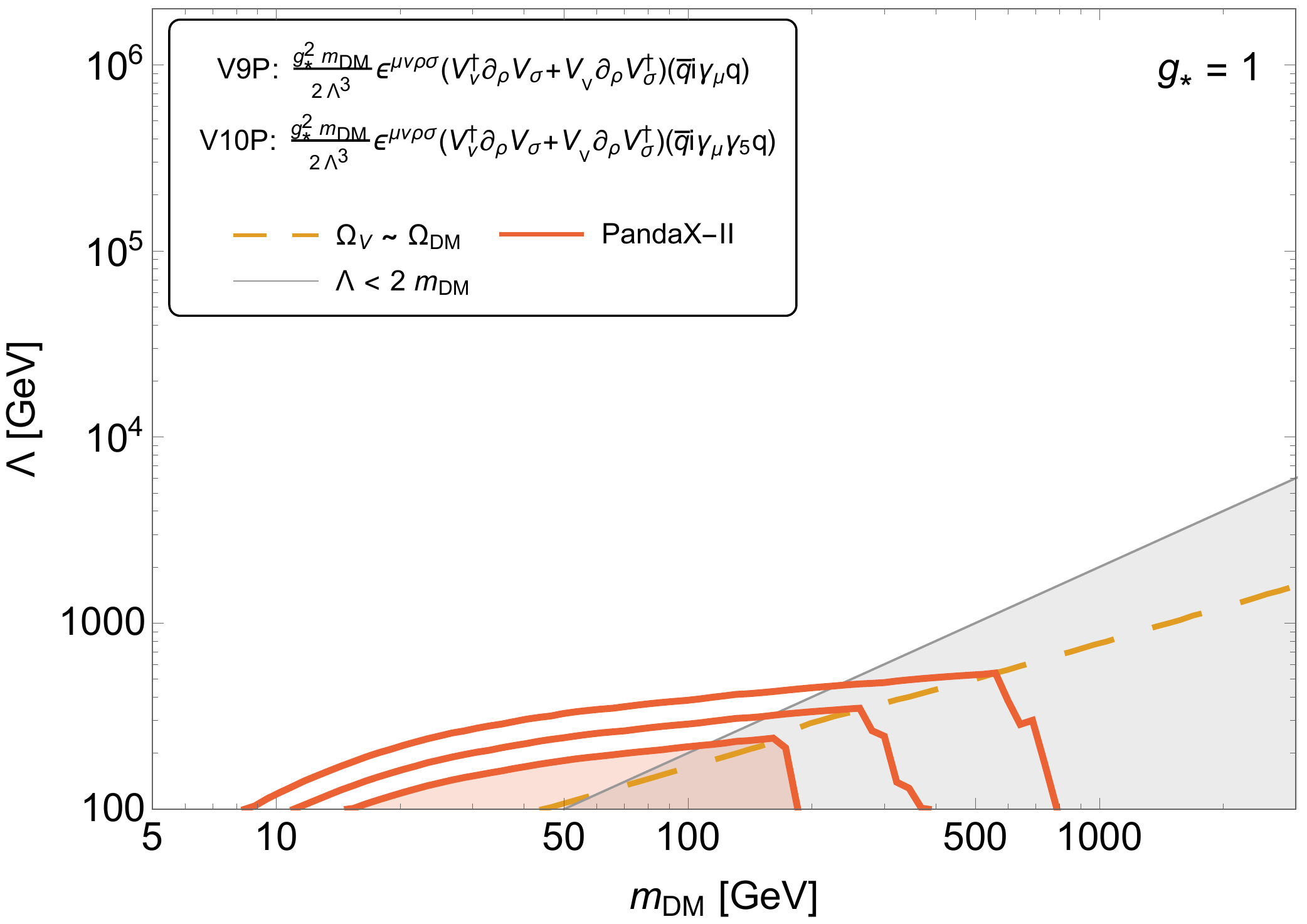}%
              \includegraphics[width=0.5\textwidth]{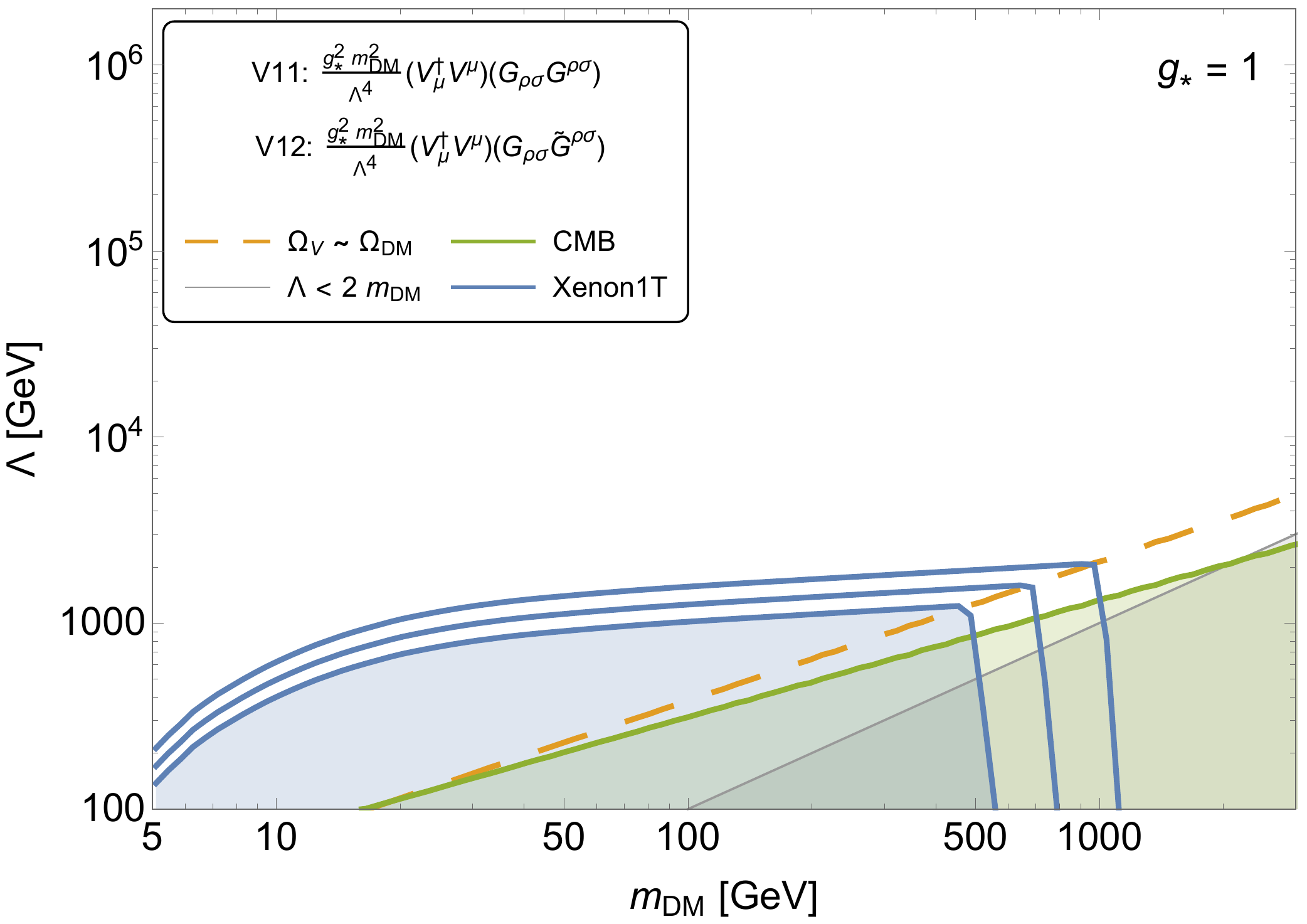}\\
\vskip -0.5cm
\caption{\label{fig:noncol-vector} Non-collider
                constraints (DD, CMB, relic density and validity of
                the EFT) for the operators involving vector DM. We
                group the operators with sensibly similar bounds. The
                conventions are same of
                Fig.~\ref{fig:noncol-dirac2}. Notice that in panel (h) the SI bound applies only to the V11 operator.}
\end{figure}

We present the non-collider limits on the effective operators
containing vector DM in Fig.~\ref{fig:noncol-vector}. When the
exclusion plots are very similar, {\it e.g.} as is the case for the
V2, V6, V9M and V10M operators, for instance, we show only one
representative plot. Let us discuss these results starting with DD
bounds.  The most stringent SI constraints are on the V1 and V3
operators, since these operators lead to non-suppressed SI cross
sections. We can see from panels~\ref{fig:noncol-vector}(a)
and~\ref{fig:noncol-vector}(c) that for $g_\star =1$, the constraint on
$\Lambda$ can reach 10 TeV for large DM masses around $\simeq$1
TeV. Another operator for which the SI constraints are relevant is
V4, although it exhibits a pseudo-vector quark
current. Notwithstanding, it develops a small mixing with the V3
operator due to the running from the scale $\Lambda$ to the 1 GeV
scale. As can be seen from Fig.~\ref{fig:noncol-vector}(d), the SI DD
limits on V4 are weaker than those on V1 or V3; still, they dominate
the constraints for DM masses in the range $\simeq$ 20--300 GeV. Also the V11 
operator is quite constrained by SI searches, which are the dominant bound 
for $m_{DM} \lesssim 500$ GeV (notice that, although the bounds on V11 and V12 are presented in the same panel, the SI bound applies only to the V11 operator, since V12 is parity-violating and do not contribute to the SI cross section). The
other twelve operators involving vector DM either do not
contribute to SI processes or their contributions are
velocity/momentum suppressed, and consequently, are not bound by SI DM
DD searches. It is interesting to observe that the astrophysical
uncertainties amount to a factor of 2.

The SD DM DD data can be used to constrain a few operators that lead
to SD cross sections that are unsuppressed. We can see from
Fig.~\ref{fig:noncol-vector}(g) that for the V9P and V10P operators,
the most constraining limits in the DM mass range $\simeq$ 20--200 GeV
stem from SD DM DD searches.  On the other hand, the SD bounds on the
operator V5, see Fig.~\ref{fig:noncol-vector}(e), are rather
weak. Nevertheless, it can be the most stringent one for DM masses
between 10 and 50 GeV depending on the local DM density.

The CMB data constrain operators exhibiting s-wave DM annihilation
channels that, in the case of vector DM, are V1, V2, V5, V6, V9M,
V10M, V11, and V12.  In the case of the operator V1, the DM bounds are
looser than the ones coming from SI DM DD, as we can see from
Fig.~\ref{fig:noncol-vector}(a). On the other hand, for the operator
V5, the CMB limits are tighter than the conservative SD DM DD ones;
see Fig.~\ref{fig:noncol-vector}(e). For the remaining operators
constrained by CMB data, {\it i.e.} V2, V6, V9M, V10M and V12
the CMB bounds are the more robust ones on these operators as we can
learn from Figs.~\ref{fig:noncol-vector}(b) and
~\ref{fig:noncol-vector}(h).  Nevertheless, the analyses of the relic
density predicted by the operators in the last class indicate that vector 
operators are either ruled out (V1, V3, V4, V9P, and V10P) or very 
strongly constrained (V2, V5, V6, V7M, V7P, V8M, V8P, V9M, V10M, V11, 
and V12).
As remarked before, these last limits can be evaded with simple
modifications of the model.

It is interesting to notice that the operators V7P, V7M, V8P, and V8M
are only bound by the relic density data and EFT validity region, as
we can see from Fig.~\ref{fig:noncol-vector}(f).\footnote{ The relic 
density bounds shown correspond to the ones on V7M and V8M. The 
limits on V7P and V8P are very similar differing by the fact that 
the line describing this limit starts at $m_{DM}\sim20$ 
GeV.}

\clearpage

\subsection{LHC sensitivity and its complementarity to non-collider
  constraints}

In this section we present the LHC bounds on all the operators listed in Table~\ref{tab:EFToperators} in the $(m_{DM},\Lambda)$ plane, analogously to what has been done for the non-collider constraints in the previous section. Applying the
analysis described in Section~\ref{sec:collider-setup} we obtain
limits on the DM EFT operators using the CMS monojet and mono $W/Z$
(hadronically decaying) searches~\cite{CMS:2017tbk} at LHC Run 2 with
an integrated luminosity of 35.9 fb$^{-1}$. Moreover, we also assess
the LHC potential to probe the DM EFT for a projected integrated
luminosity of 300 fb$^{-1}$.

Our results are shown in Figs.~\ref{fig:coll-scal}--\ref{fig:coll-vector2}. In these figures, the red shaded area represents the LHC exclusion region at 95\% CL for $g_\star=1$. Furthermore, the area inside the solid orange (blue) curve is excluded at 95\% CL by the presently available CMS mono-jet data for $g_\star=6$ ($g_\star = 4\pi$). For the sake of comparison, we also display in these figures the excluded region for $g_\star=1$ due to non-collider searches, represented by the light purple shaded area.  We do not include the relic density constraints into the ``non-collider" excluded area since these bounds are model-dependent and can easily be evaded, {\it e.g.}  by the addition of co-annihilating DM partners. The region where the EFT approach is not valid ($\Lambda > 2 m_{DM}$) is represented by the grey triangle in the right bottom side of the figures. Finally, the regions inside the dashed red, orange and blue curves represent our estimates for the future 95\%CL exclusion by the LHC with 300 fb$^{-1}$, taking $g_\star=1$, $6$ or $4\pi$, respectively.

A common feature of the monojet excluded regions is that it exhibits
upper and lower limits in $\Lambda$ for a fixed DM mass. The lower limit is caused by the cut given in Eq.~\eqref{eq:inv-mass-cut}, that
guarantees that the EFT is applied only within its validity limit.
Furthermore, this cut has an important effect when the bound on
$\Lambda$ is low. On the other hand, for higher values of $\Lambda$
(above few TeV) the impact of this cut on the cross sections is
negligible for scalar operators. We will see below that there are also
important differences in the effects of this cut for different DM spin
cases since the DM pair invariant mass distributions depend strongly
on the DM spin, as discussed in Ref.~\cite{Belyaev:2016pxe}.

\subsubsection*{Complex scalar DM}

\begin{figure}[tbp]
\vskip 0.8cm
(a)\hspace*{0.5\textwidth}\hspace{-0.2cm}(b)\\\vspace*{-2cm}\\\\
\includegraphics[width=0.5\textwidth]{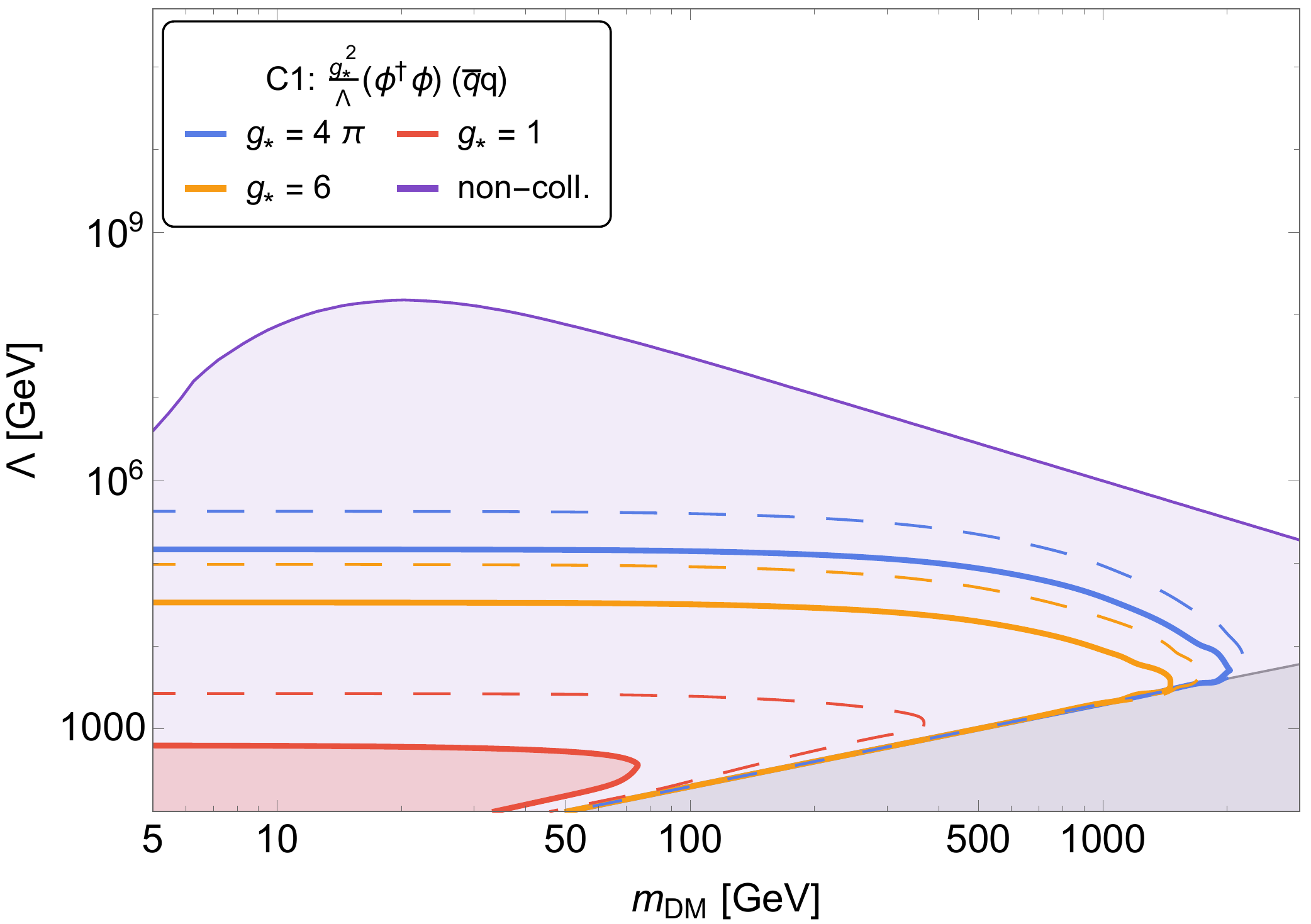}%
\includegraphics[width=0.5\textwidth]{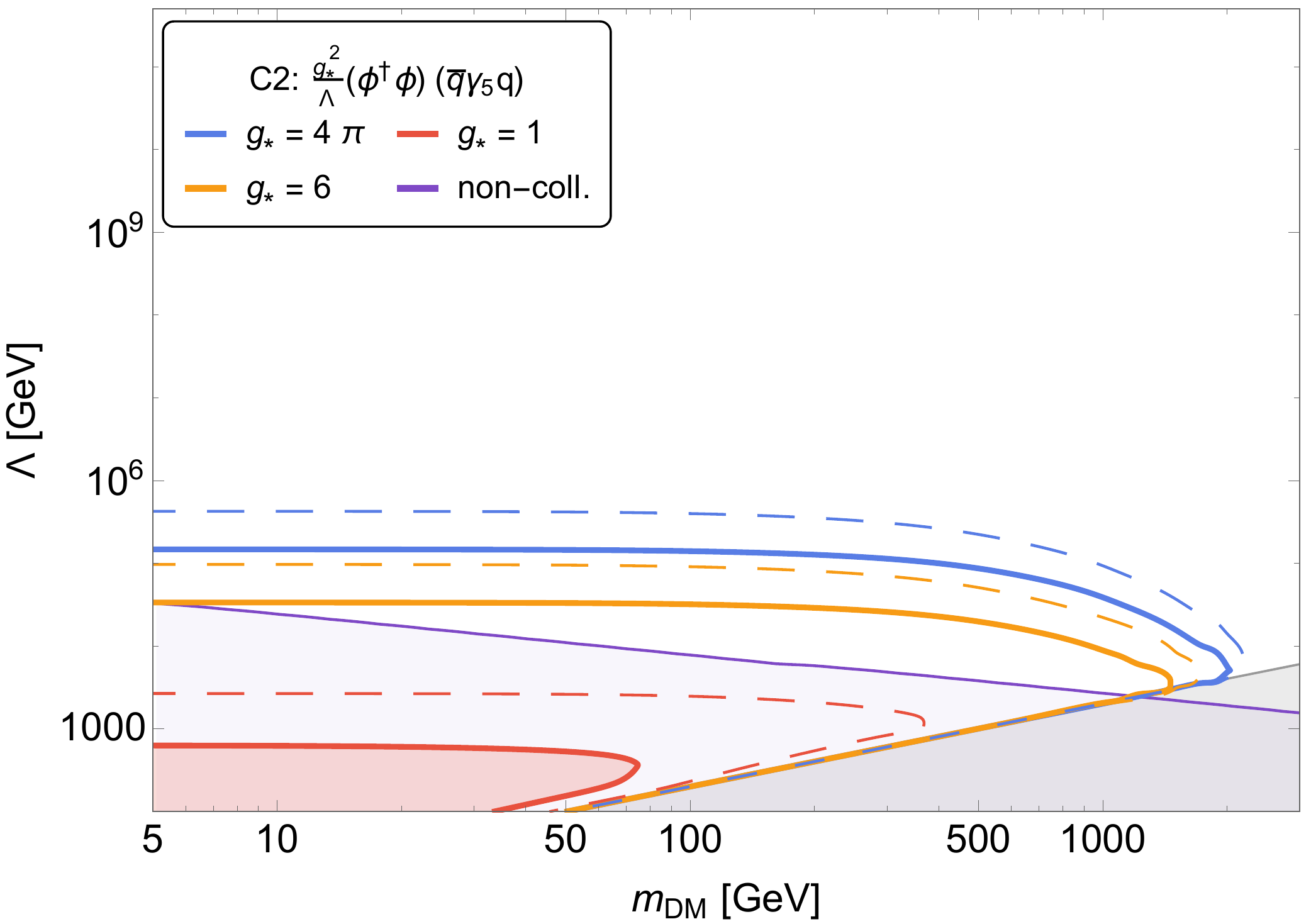}\\
\vskip 0.2cm
(c)\hspace*{0.5\textwidth}\hspace{-0.2cm}(d)\\\vspace*{-2cm}\\\\
\includegraphics[width=0.5\textwidth]{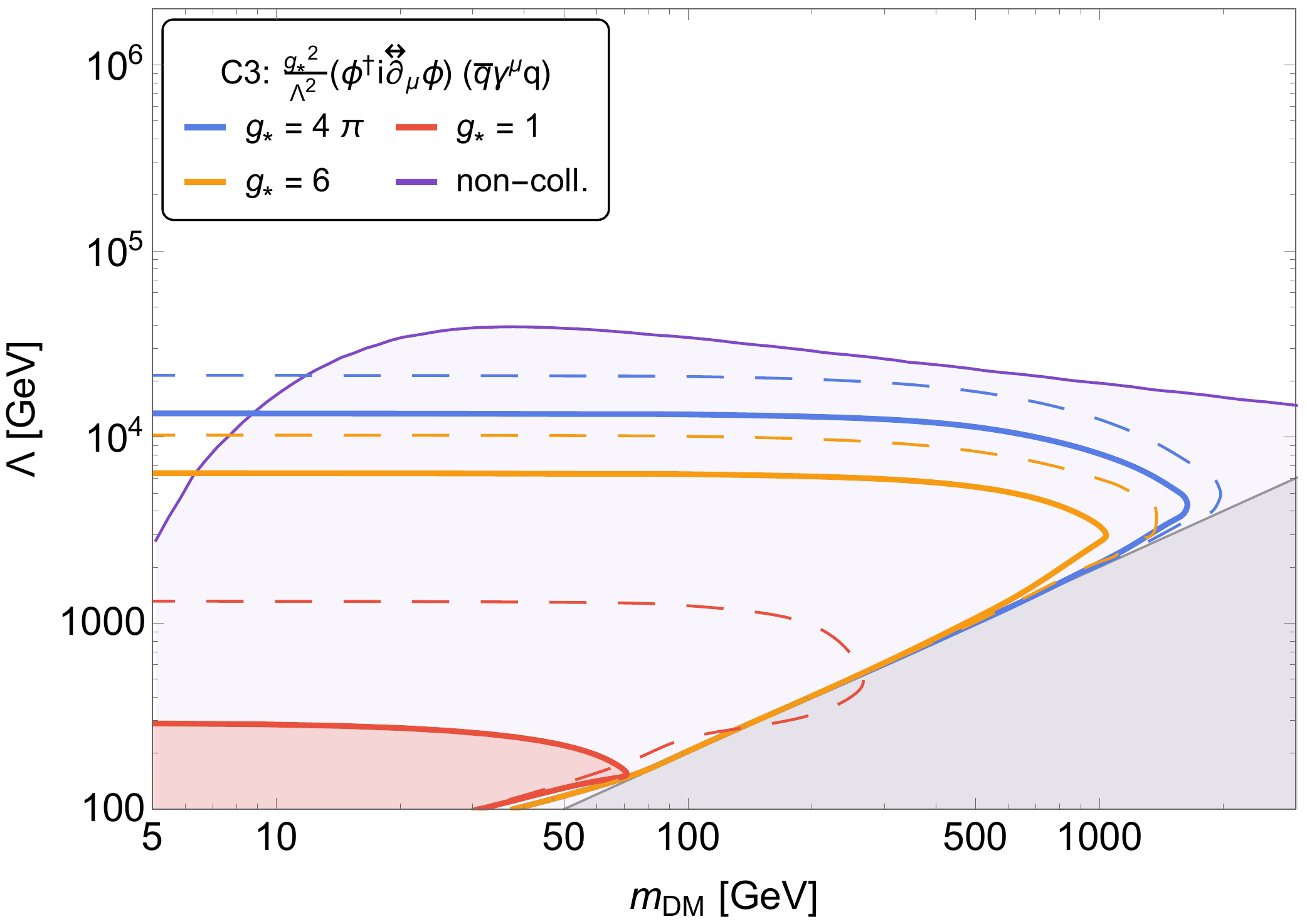}%
\includegraphics[width=0.5\textwidth]{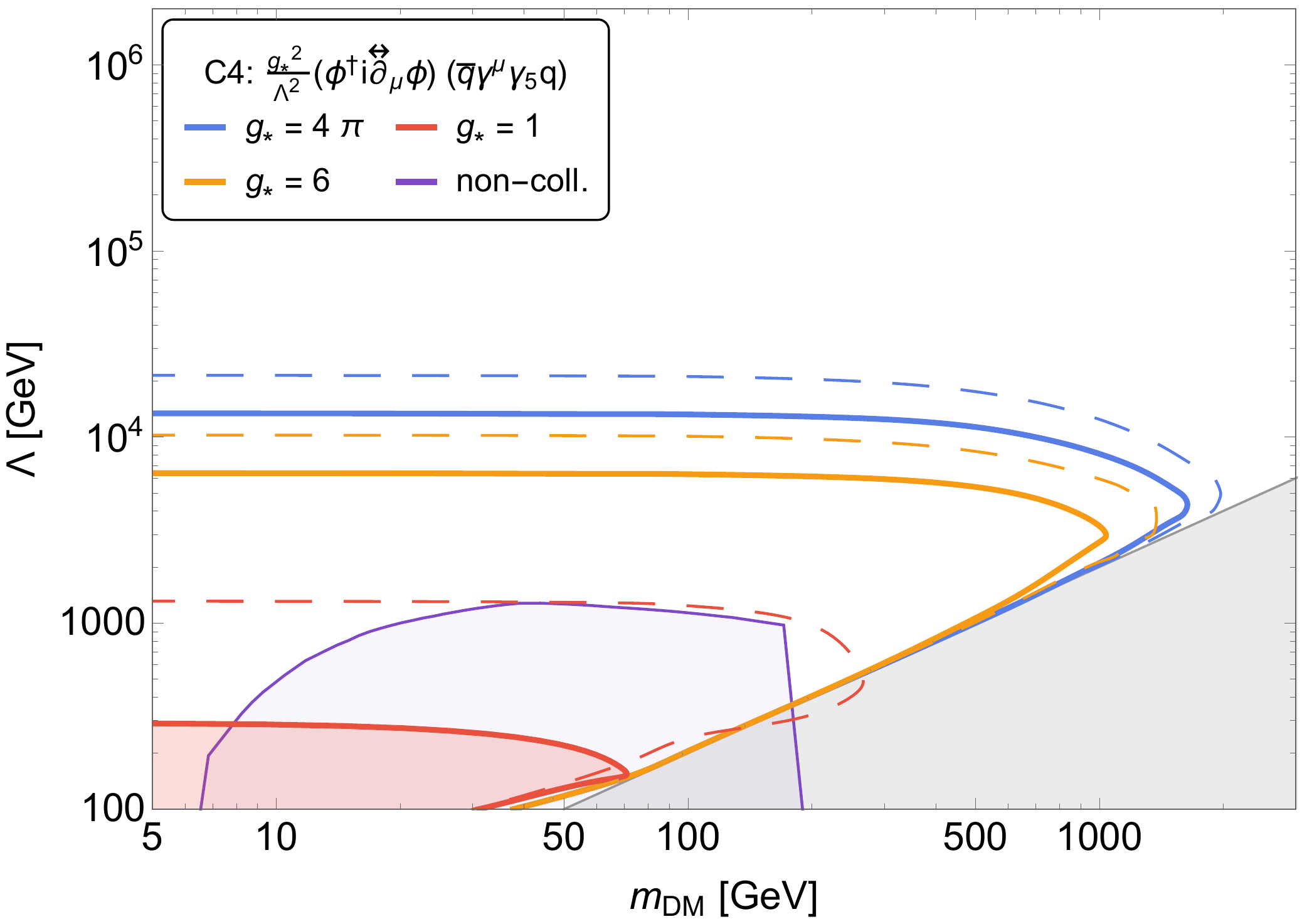}\\ 
\vskip 0.2cm
(e)\hspace*{0.5\textwidth}\hspace{-0.2cm}(f)\\\vspace*{-2cm}\\\\
\includegraphics[width=0.5\textwidth]{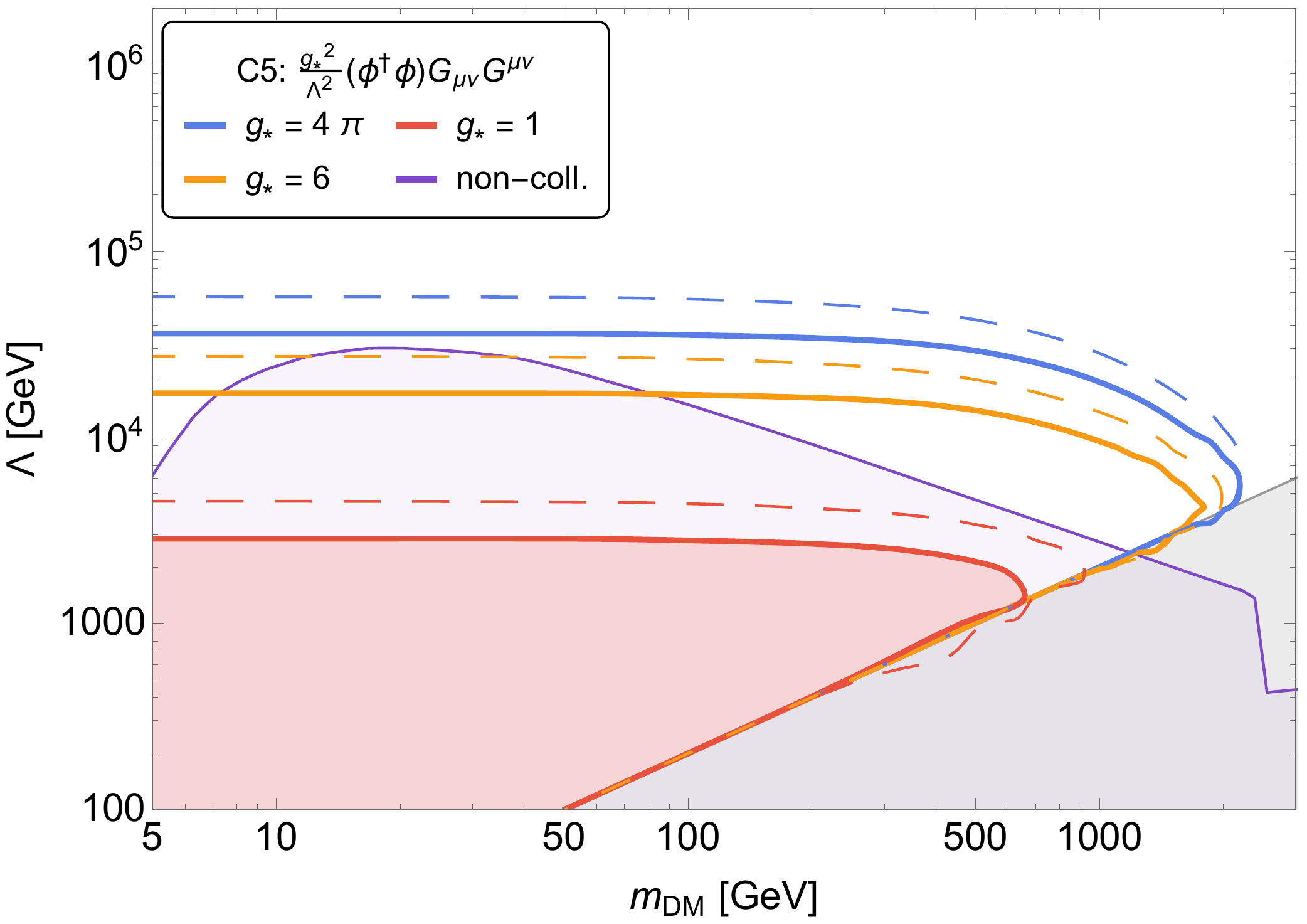}%
\includegraphics[width=0.5\textwidth]{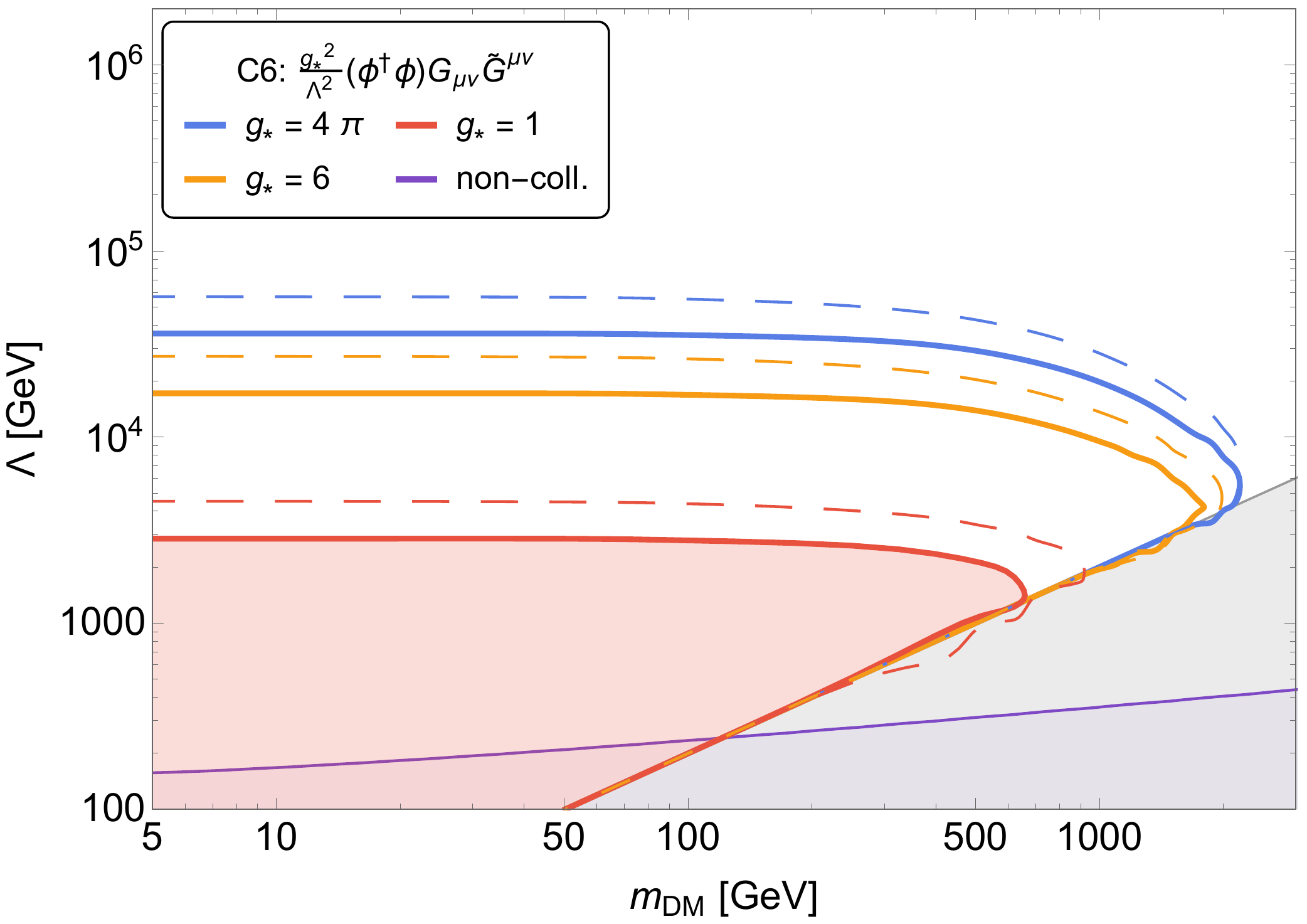}    
\caption{\label{fig:coll-scal} LHC monojet constraints on EFT
  operators with scalar DM, as indicated in each panel. The area inside 
  the red, orange and blue solid curves is excluded by current LHC 
  data at 95\% CL for $ g_\star =1$, $6$ and $4\pi$, respectively. The
  projected LHC limits for 300 fb$^{-1}$ are indicated by dashed thin
  lines. The combined exclusion regions from CMB and DM DD searches
  for $g_\star=1$ are given by the light-purple area below the purple
  curve. The grey triangular region marks the region where the EFT approach is
  not valid.  }
\end{figure}

We present in Fig.~\ref{fig:coll-scal} the CMS monojet bounds on the
EFT operators C1--C6 that contain scalar DM states. From the top left
panel of this figure, we can learn that the LHC data is able to
exclude up to DM masses of $\simeq 70$ GeV for $\Lambda \lesssim 600$
GeV and $g_\star=1$. We can also see how these bounds evolve with the
coupling $g_\star$: as $g_\star$ varies, the upper limit on $\Lambda$
does not scale as $g^2_\star$ as naively expected. The reason for this
behavior is the cut in Eq.~(\ref{eq:inv-mass-cut}), that reduces more
significantly the available phase space for smaller $g_\star$ since this is
associated to smaller $\Lambda$ upper limits. This behavior is true
for all operators containing vector and pseudo-vector quark currents.
Moreover, this panel shows that the conservative non-collider bounds
are the strongest ones for $g_\star=1$ This situation persists even
when a larger integrated luminosity is accumulated.
\smallskip

The top right panel of Fig.~\ref{fig:coll-scal} shows the collider
bounds on the operator C2.  As we can see, the collider limits on C1
and C2 are the same. Moreover, the most stringent limits on this
operators originates from non-collider data, irrespective of the
integrated luminosity.  In addition, from the middle left panel
of this figure we see that results for the operator C3 are similar to
the ones for C1 and C2, except that the collider limit on $\Lambda$ is
reduced.

The scale of $\Lambda$ which the LHC can probe for scalar DM EFT
operators strongly depends on the effective operator: it is about 0.6
TeV for the C1 and C2 operators, it is only about 0.3 TeV for the C3
and C4 operators, and it is about 3 TeV for the C5 and C6 operators.
The LHC searches are the least sensitive for the operators C3 and C4
for two reasons: first of all, these operators contain explicit
momentum dependence through the derivative, and LHC cuts on the
monojet are not hard enough to enhance this operator.  Secondly the
invariant DM pair mass distribution is shifted to higher values than
for other scalar DM operators (see detailed discussion in
Ref.~\cite{Belyaev:2016pxe}), therefore the cut given by the
Eq.~(\ref{eq:inv-mass-cut}) reduces their signal more then for other
EFT operators with scalar DM.

From Fig.~\ref{fig:coll-scal} we can also see that LHC plays an
important complementary role in probing the DM parameter space for the
C4 and C6 operators. Notice that the LHC searches are especially
important to test the operator C6 that involve gluons.  In
fact, the sensitivity of non-collider experiments to the EFT parameter
space for this operator is very poor, as one can see from
Figure~\ref{fig:coll-scal}(f). Moreover, even for the
operators C1, C2 and C3, the LHC could help elucidate the nature of DM
by independently probing the DM parameter space, especially if some DM
signals would take place at collider and non-collider experiments.

In the case that the DM interactions are stronger, {\it e.g.}
$g_\star = 4 \pi$, we can scale the CMB and DD limits shown in
Fig.~\ref{fig:coll-scal} by $g_\star^2$ ($g_\star$) for dimension 5 (6) operators to compare with the LHC
results. The conclusions for $g_\star=4 \pi$ are the same as the ones
above for $g_\star=1$, except for the range of masses where the DM DD
is dominant that shrinks to the interval 20--140 GeV. Moreover, we
project that the LHC limits on C4 will be the strongest ones for an
integrated luminosity of 300 fb$^{-1}$.


\begin{figure}[tb]
\vskip 0.8cm
(a)\hspace*{0.5\textwidth}\hspace{-0.2cm}(b)\\\vspace*{-2cm}\\\\
\includegraphics[width=0.5\textwidth]{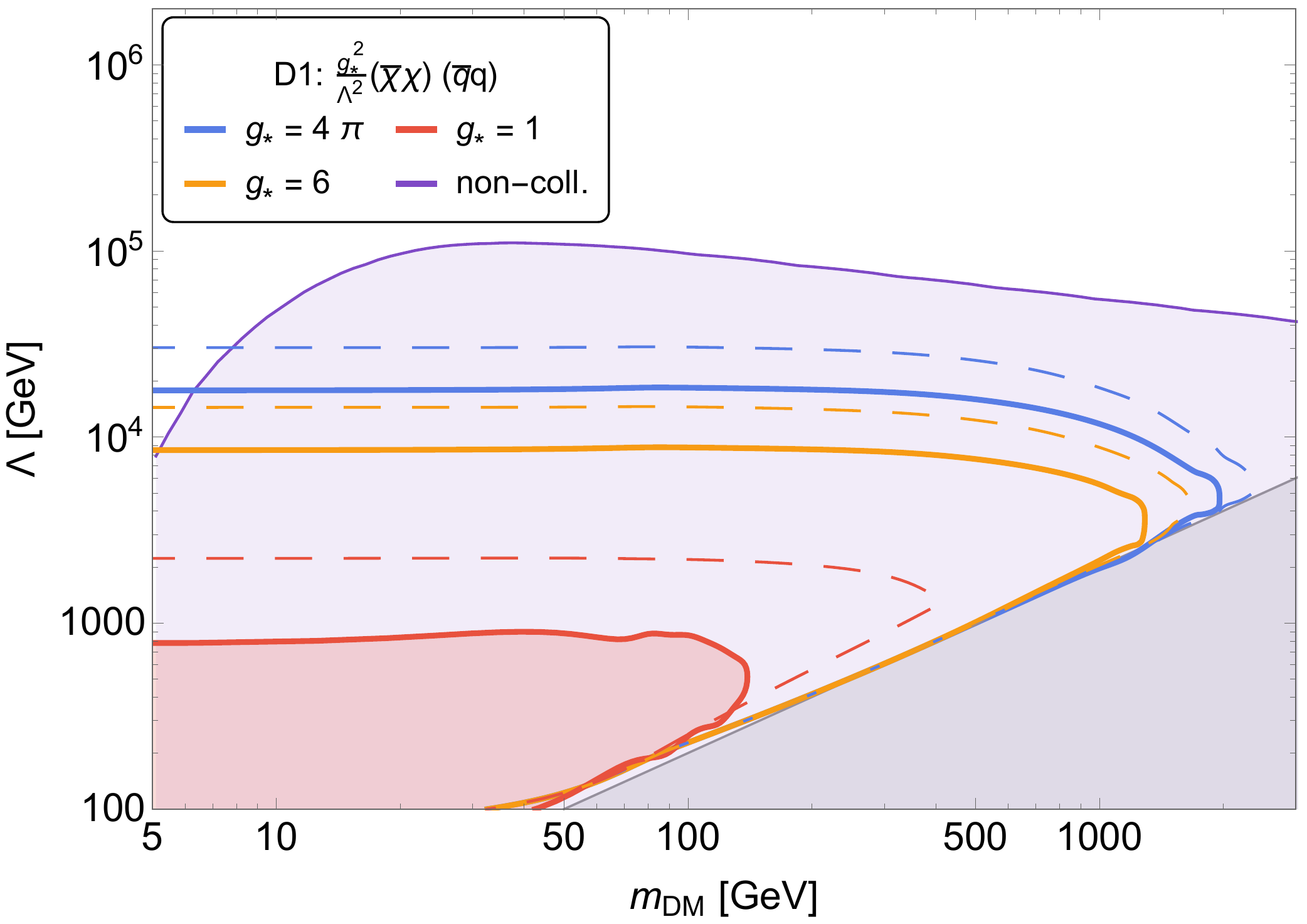}%
\includegraphics[width=0.5\textwidth]{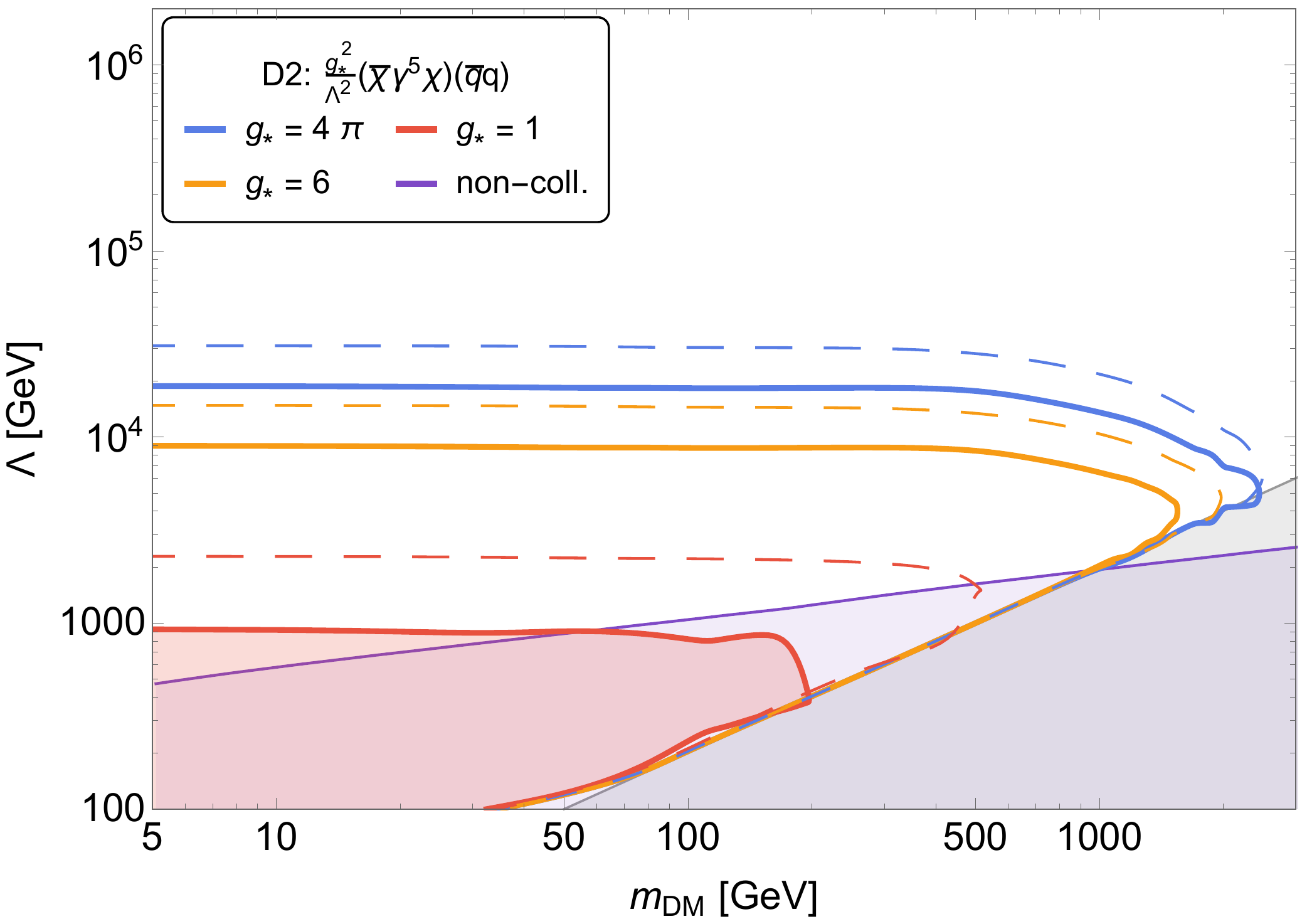}\\
\vskip 0.2cm
(c)\hspace*{0.5\textwidth}\hspace{-0.2cm}(d)\\\vspace*{-2cm}\\\\
\includegraphics[width=0.5\textwidth]{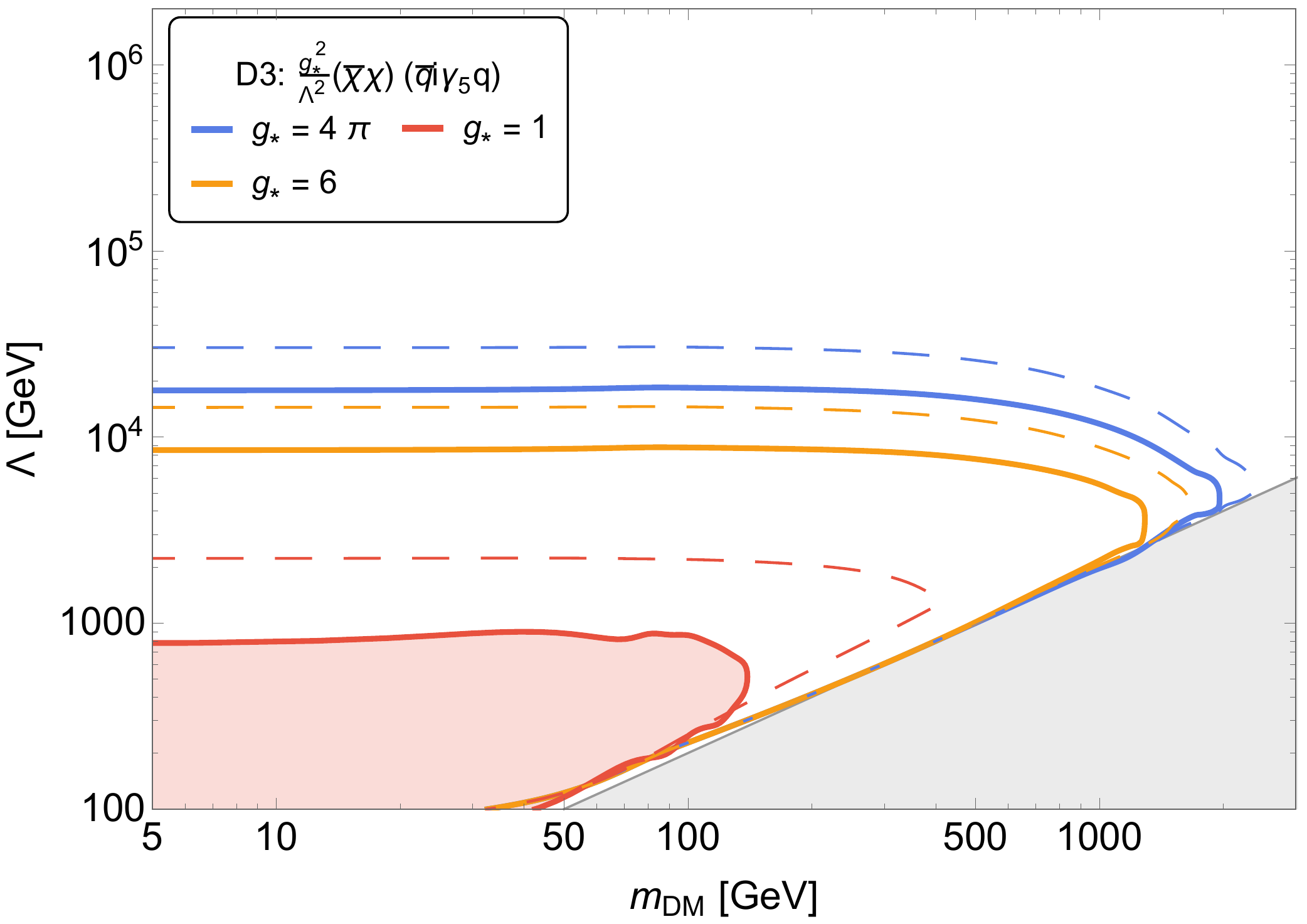}%
\includegraphics[width=0.5\textwidth]{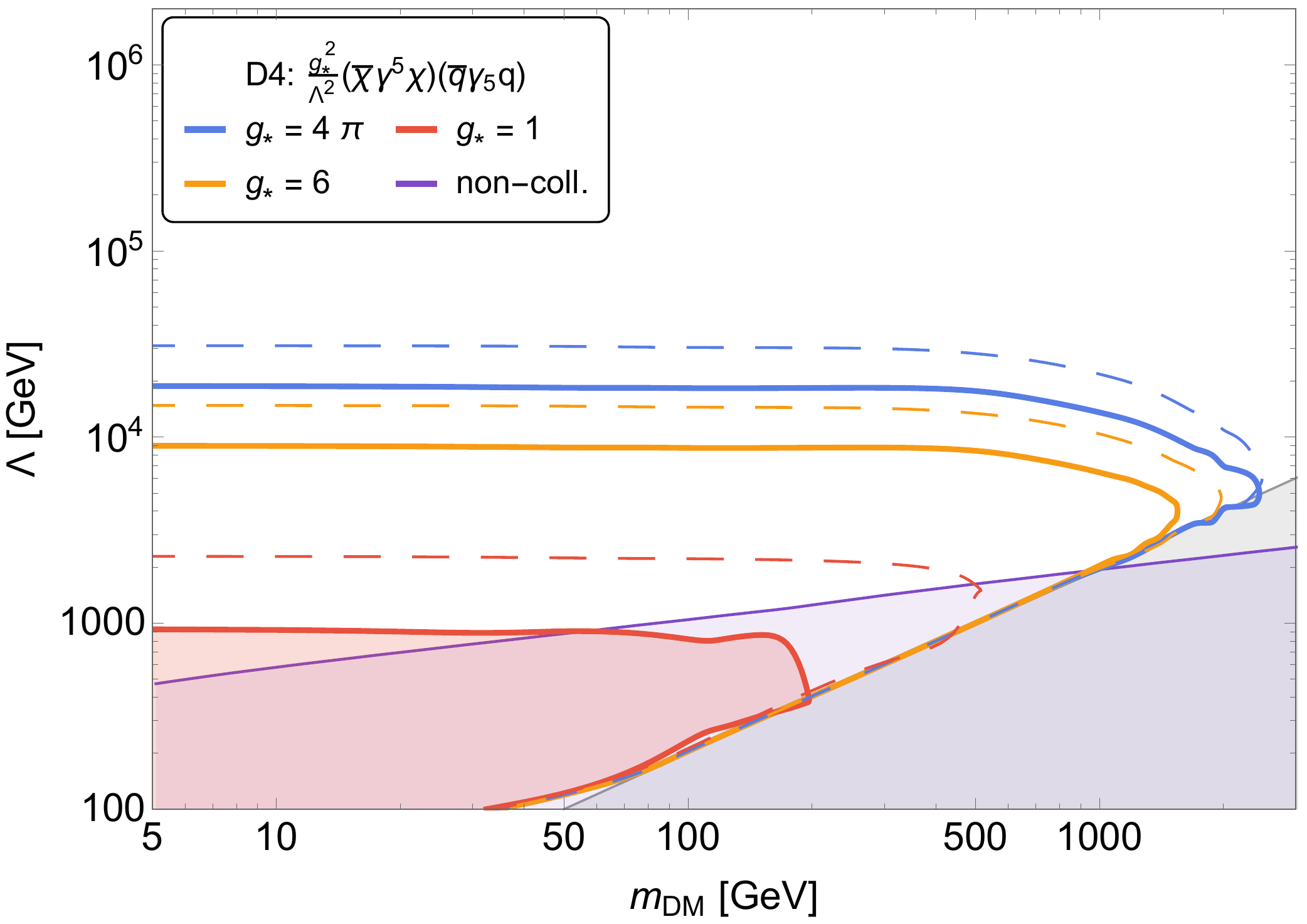} 
\\
\vskip 0.2cm
(e)\hspace*{0.5\textwidth}\hspace{-0.2cm}(f)\\\vspace*{-2cm}\\\\
\includegraphics[width=0.5\textwidth]{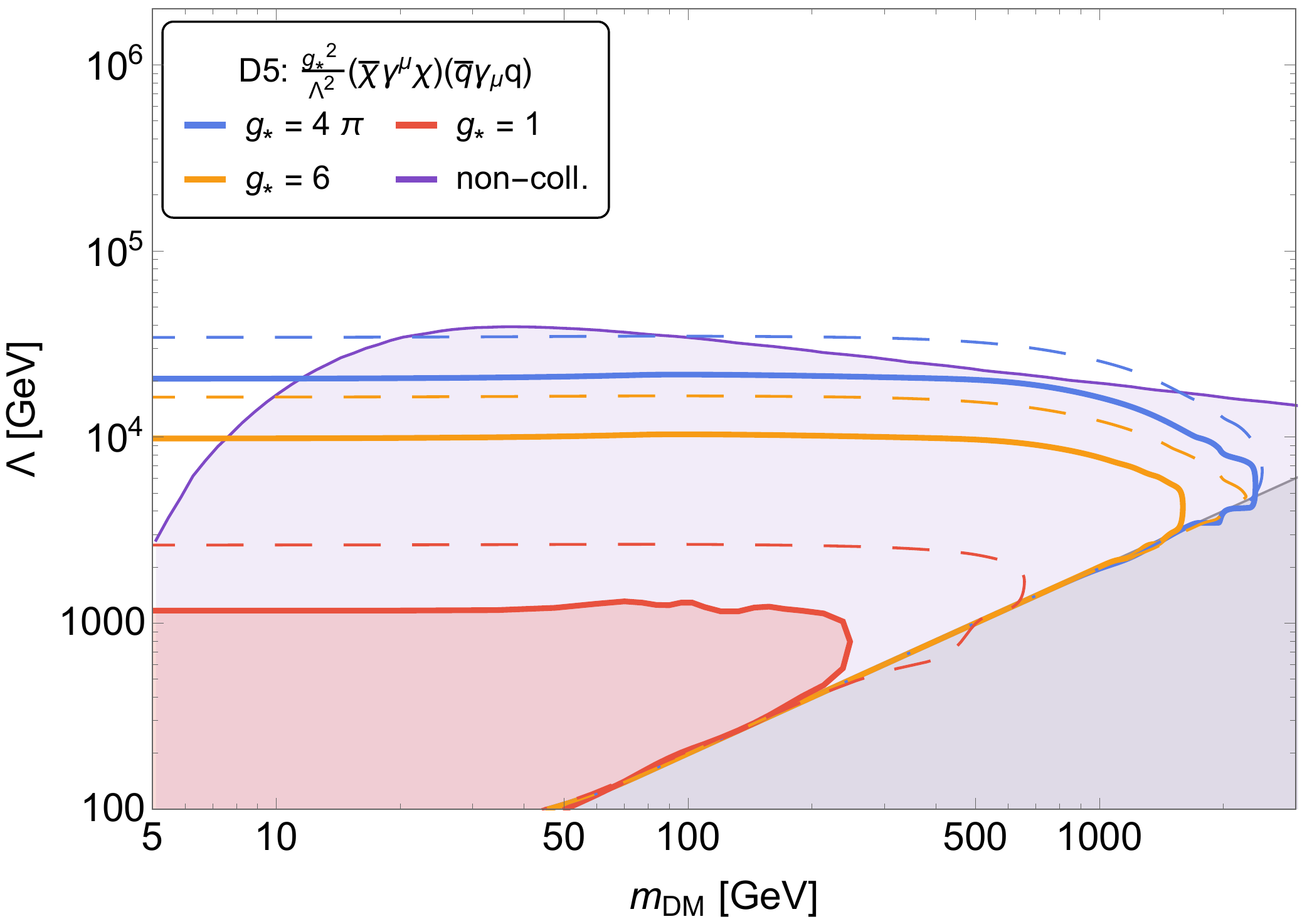}%
\includegraphics[width=0.5\textwidth]{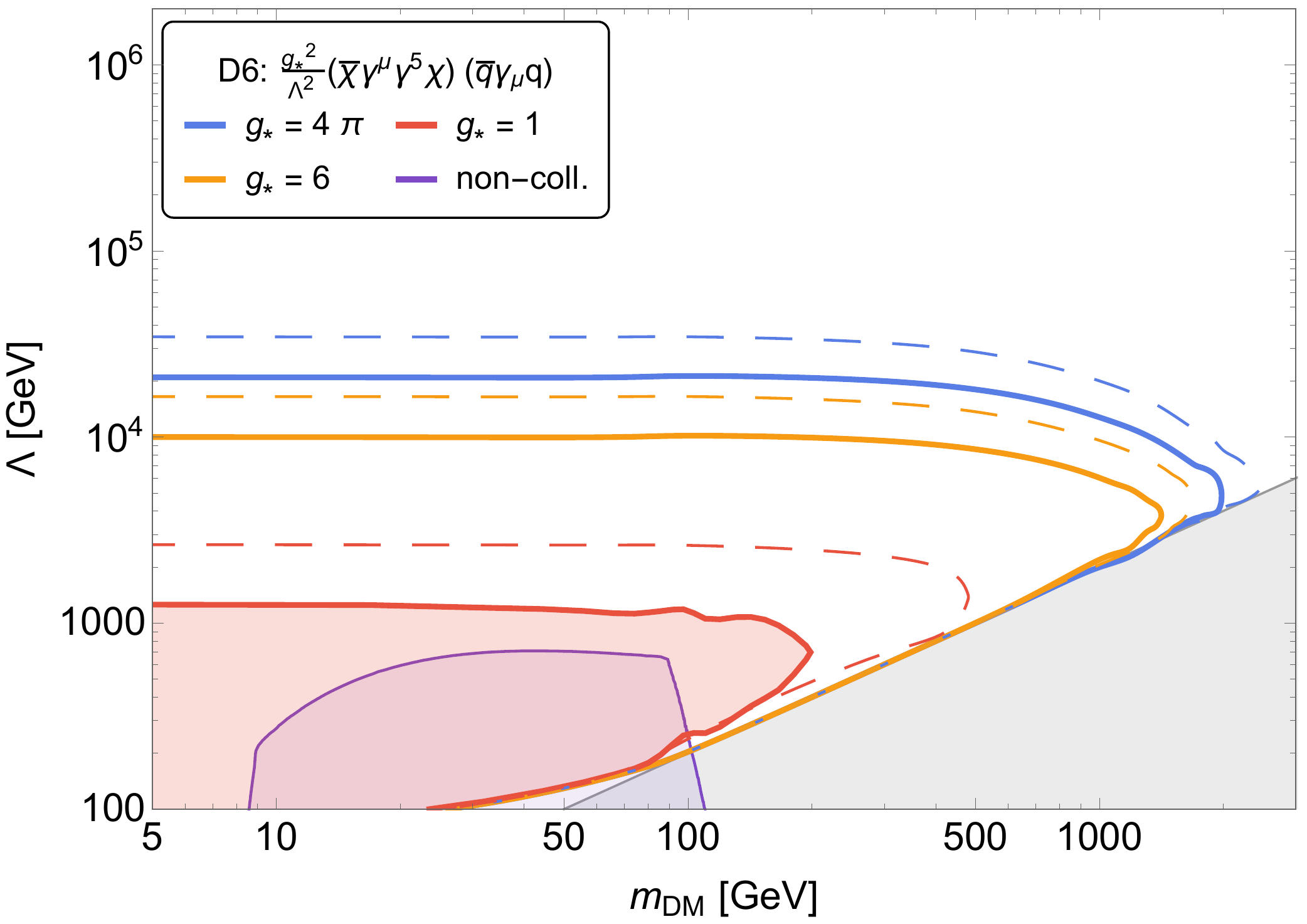}\\   
\vskip 0.2cm
(g)\hspace*{0.5\textwidth}\hspace{-0.2cm}(h)\\\vspace*{-2cm}\\\\
\includegraphics[width=0.5\textwidth]{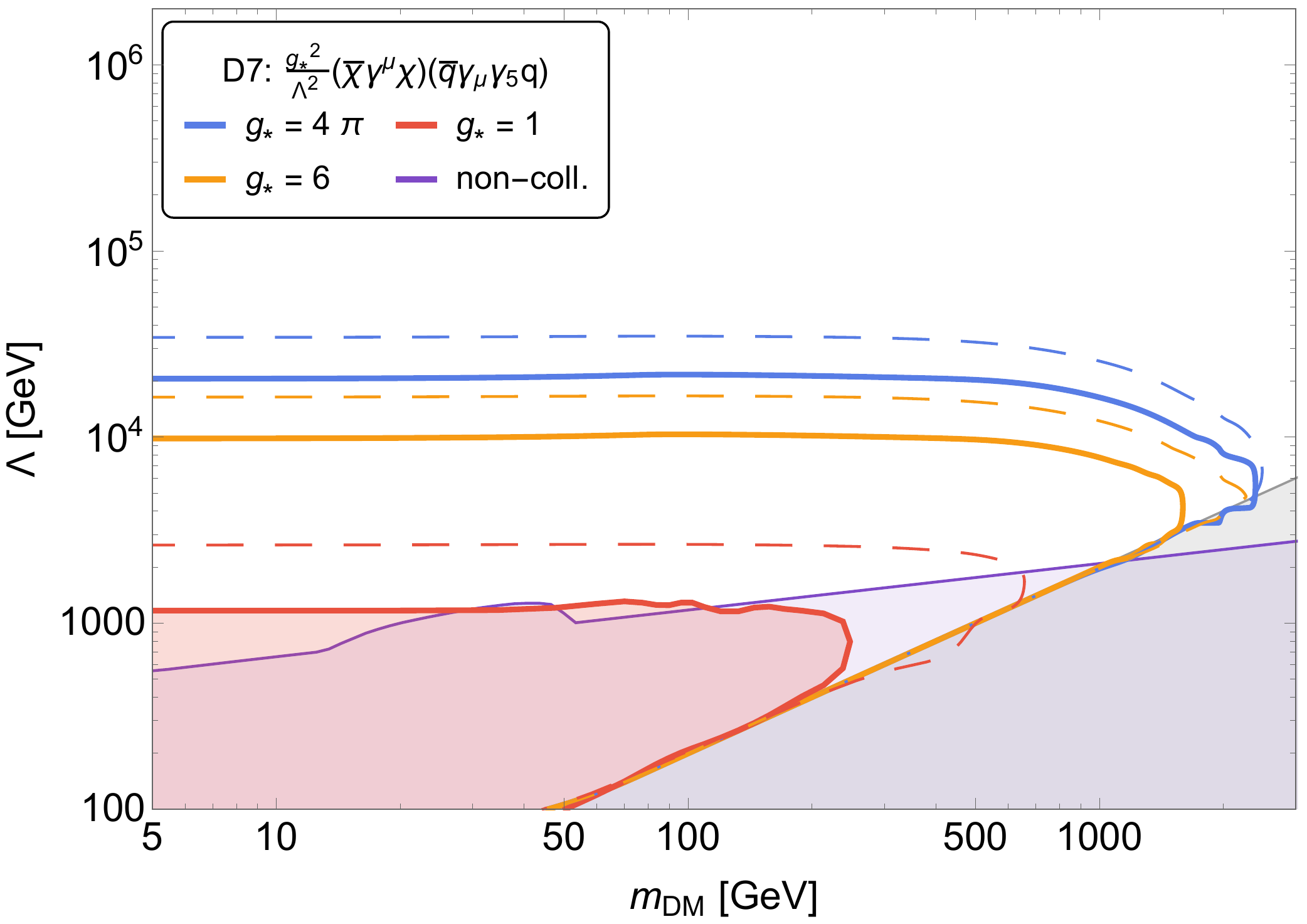}%
\includegraphics[width=0.5\textwidth]{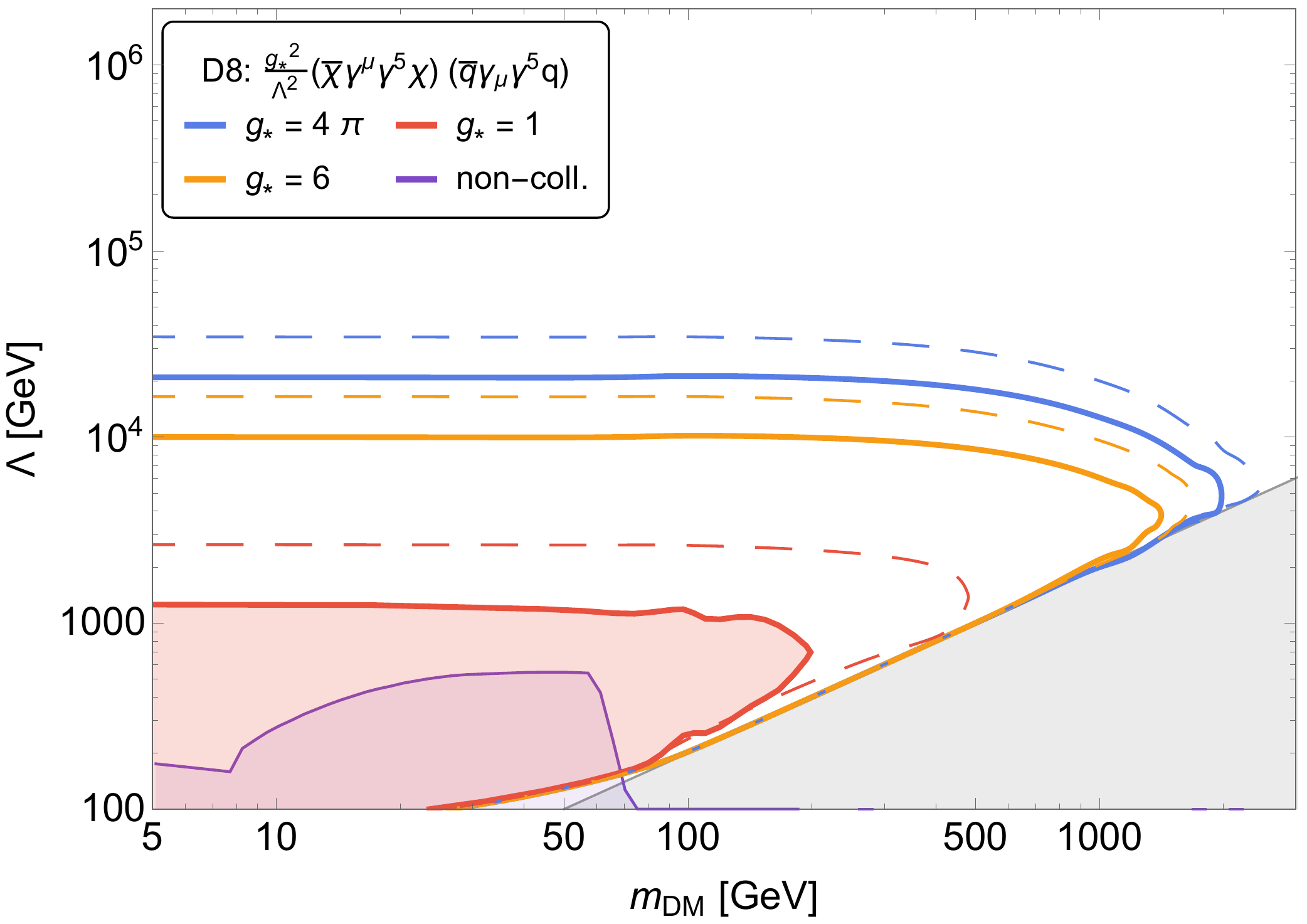}
\caption{\label{fig:coll-dirac2} Constraints from the LHC monojet
  searches on the operators D1-- D8 that contain Dirac
  fermion DM. The conventions are as in Fig.~\ref{fig:coll-scal}.  }
\end{figure}
\clearpage

\subsubsection*{Dirac fermion DM}

From Fig.~\ref{fig:coll-dirac2} we can see that the LHC monojet data
exclude DM masses up to $\simeq$ 150--200 GeV and
$\Lambda \lesssim 900$ GeV for the operators containing scalar and
pseudo-scalar quark currents and $g_\star=1$. 
Furthermore, in the case of effective operators containing vector and
pseudo-vector quark currents, the LHC excluded region is slightly
larger for $g_\star =1$: $m_{DM} \lesssim 200$ GeV and $\Lambda
\lesssim 1.1$ TeV.
In the case of couplings
approaching the strongly interacting regime ($g_\star = 4 \pi$) the
exclusion region is extended to $m_{DM} \lesssim 2$ TeV and
$\Lambda \lesssim 20$ TeV for all operators D1--D8.

Fig.~\ref{fig:coll-dirac2} also allows us to see the complementarity
between the collider and non-collider searches. First of all, the DM
SI DD bounds are more stringent than the LHC ones for operators that
possess unsuppressed contributions to DM SI DD, {\it i.e.} D1 and D5;
see panels (a) and (e) of this figure. 
On the other hand, the LHC bounds are stronger than the non-collider
ones for the operators D6 and D8 since these operators only exhibit
velocity suppressed contributions to DM SI DD searches. The same
conclusion applies to the operator D3 that is not limited by neither
the CMB data nor the DD searches.
Moreover, the CMB data and the LHC monojet searches complement nicely
each other for the operators D2, D4 and D7 because the LHC searches
dominates the bounds for DM masses smaller than $\simeq 50$ GeV while
the CMB limits are stronger above this mass.

\begin{figure}[tbp]
\vskip 1cm
(a)\hspace*{0.5\textwidth}\hspace{-0.2cm}(b)\\\vspace*{-2cm}\\\\
\includegraphics[width=0.5\textwidth]{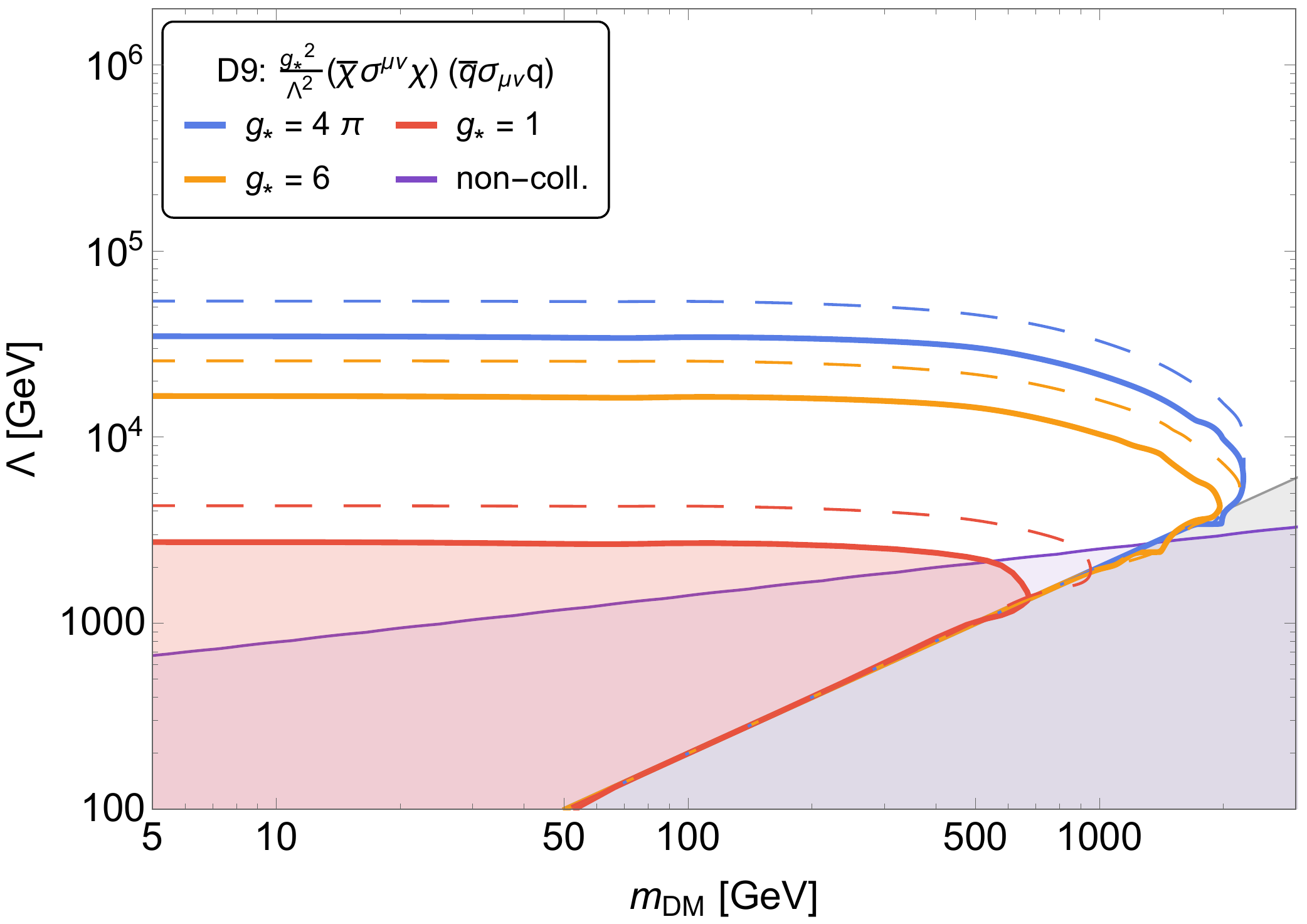}%
\includegraphics[width=0.5\textwidth]{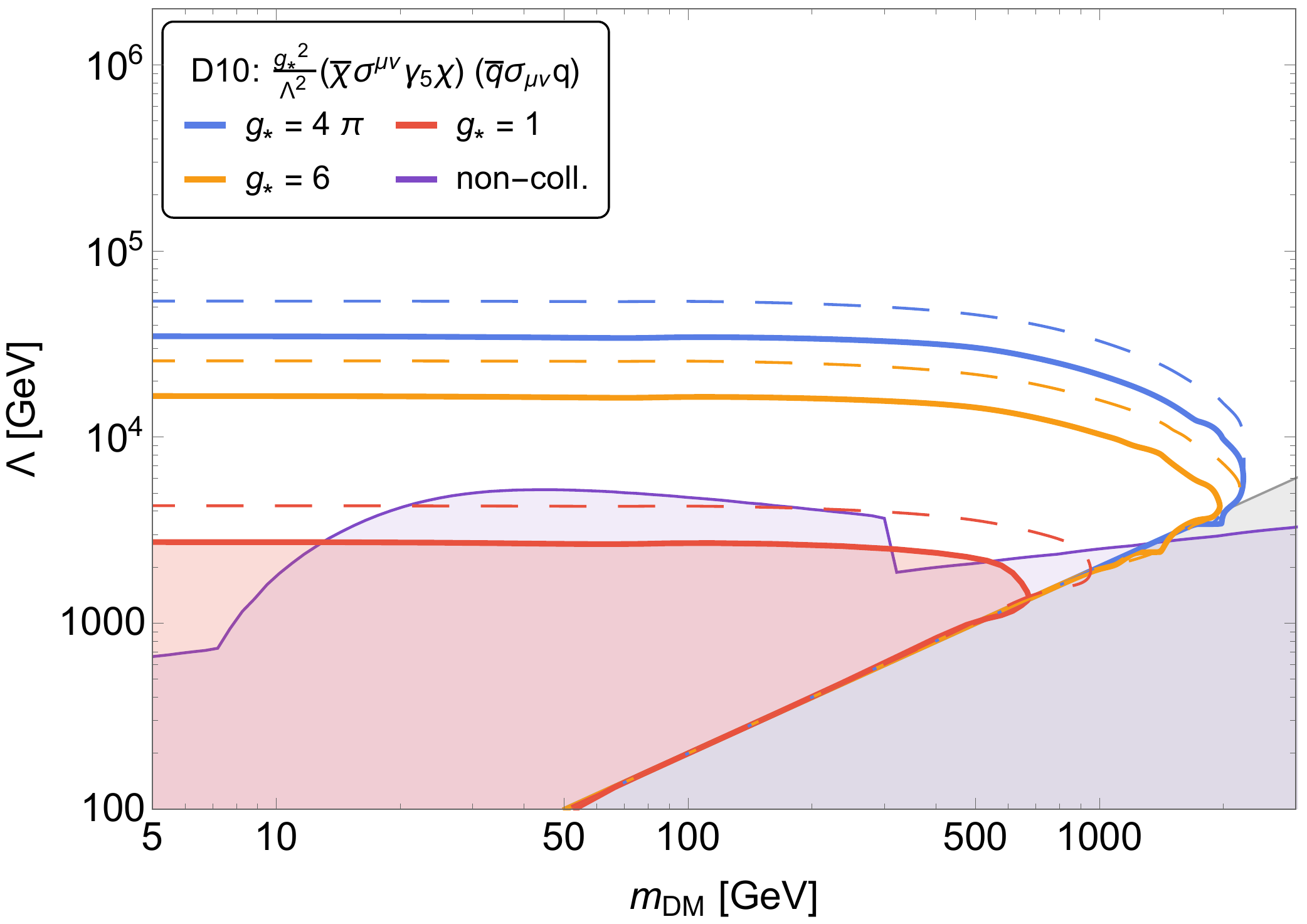}
\caption{\label{fig:coll-dirac3} Constraints from the LHC monojet
  searches on the operators D9 and D10 that contain Dirac fermion
  DM. The conventions are as in Fig.~\ref{fig:coll-scal}. }
\end{figure}

Fig.~\ref{fig:coll-dirac3} depicts the LHC limits on the operators D9
and D10 that contain a tensor quark current.
The LHC monojet searches excludes the region $m_{DM} \lesssim 600$ GeV
and $\Lambda \lesssim 3$ TeV for these operators. For larger couplings
$g_\star = 4 \pi$ the region is expanded to $m_{DM} \lesssim 2$ TeV
and $\Lambda \lesssim 35$ TeV. 
For the operator D9 and $g_\star=1$, the CMB bounds dominates in a
small region for $m_{DM} \gtrsim $ 600 GeV while the LHC limits are
stronger below this mass. 
The most stringent bounds on the operator D10 have three different
sources depending on the DM mass for $g_\star=1$. At low DM masses
$\simeq 15$ GeV the bounds are dominated by the LHC searches while for
heavier masses $\gtrsim 300$ GeV the most important constraints come
from the CMB data. On the other hand, the SI limits are more stringent
in the mass window $15 \lesssim m_{DM} \lesssim 300$ GeV.

As for the case of scalar DM, DD and CMB limits on the Dirac fermion
DM rescale as $g_\star^2$ ($g_\star$) for dimension 5 (6) operators.  
In the case of $g_\star = 4 \pi$ the
region where the CMB constraints are more stringent than the collider
ones on D2, D4, D7, and D10 changes to DM masses in excess of 350,
350, 400, and 620 GeV respectively. In the case of the other effective
operators for Dirac fermion DM the conclusions remain the
approximately the same for the value of $g_\star$.

\subsubsection*{Vector DM}
\label{sec:coll-vector-dm}

\begin{figure}[tb]
\vskip 0.8cm
(a)\hspace*{0.5\textwidth}\hspace{-0.2cm}(b)\\\vspace*{-2cm}\\\\
\includegraphics[width=0.5\textwidth]{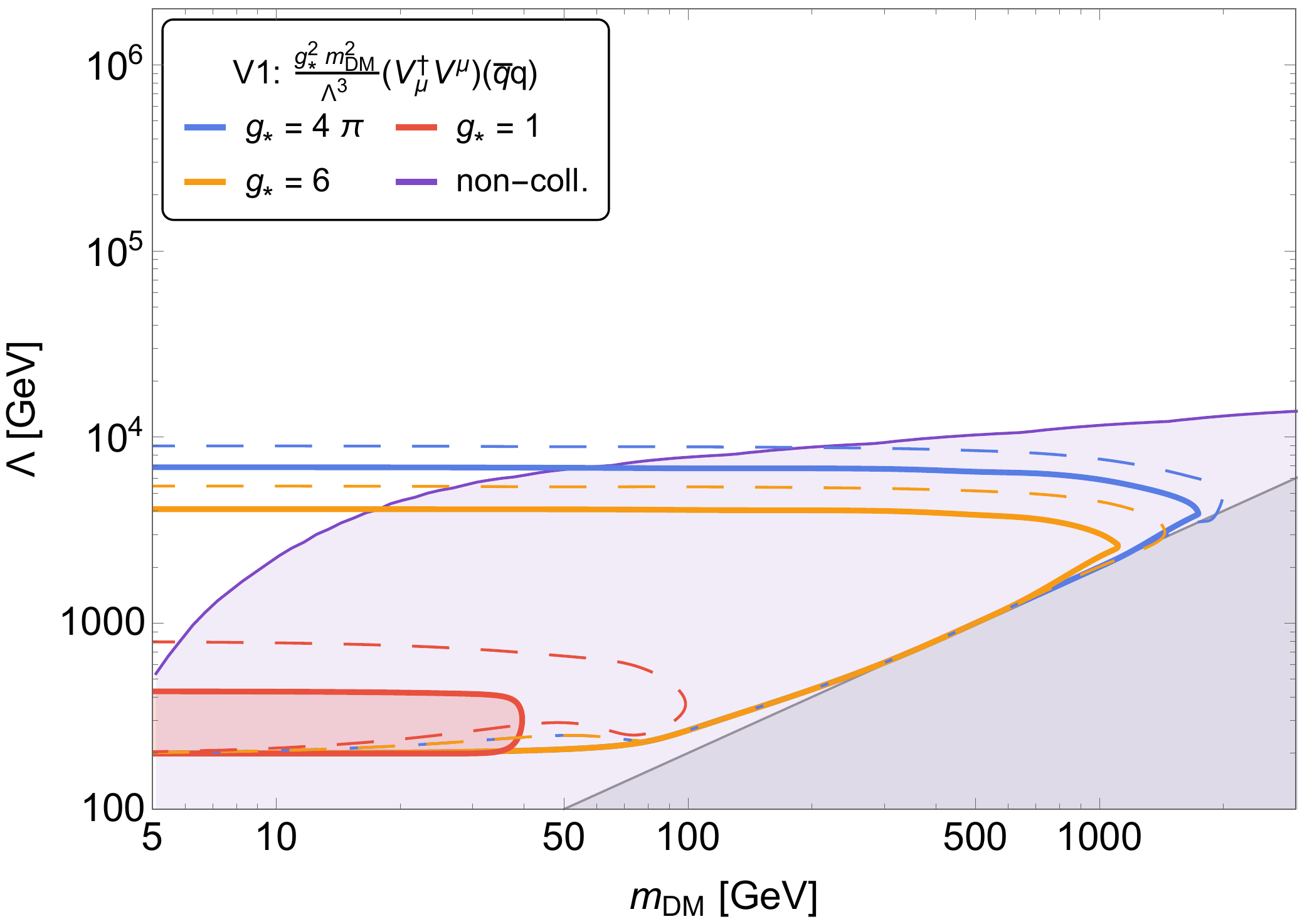}%
\includegraphics[width=0.5\textwidth]{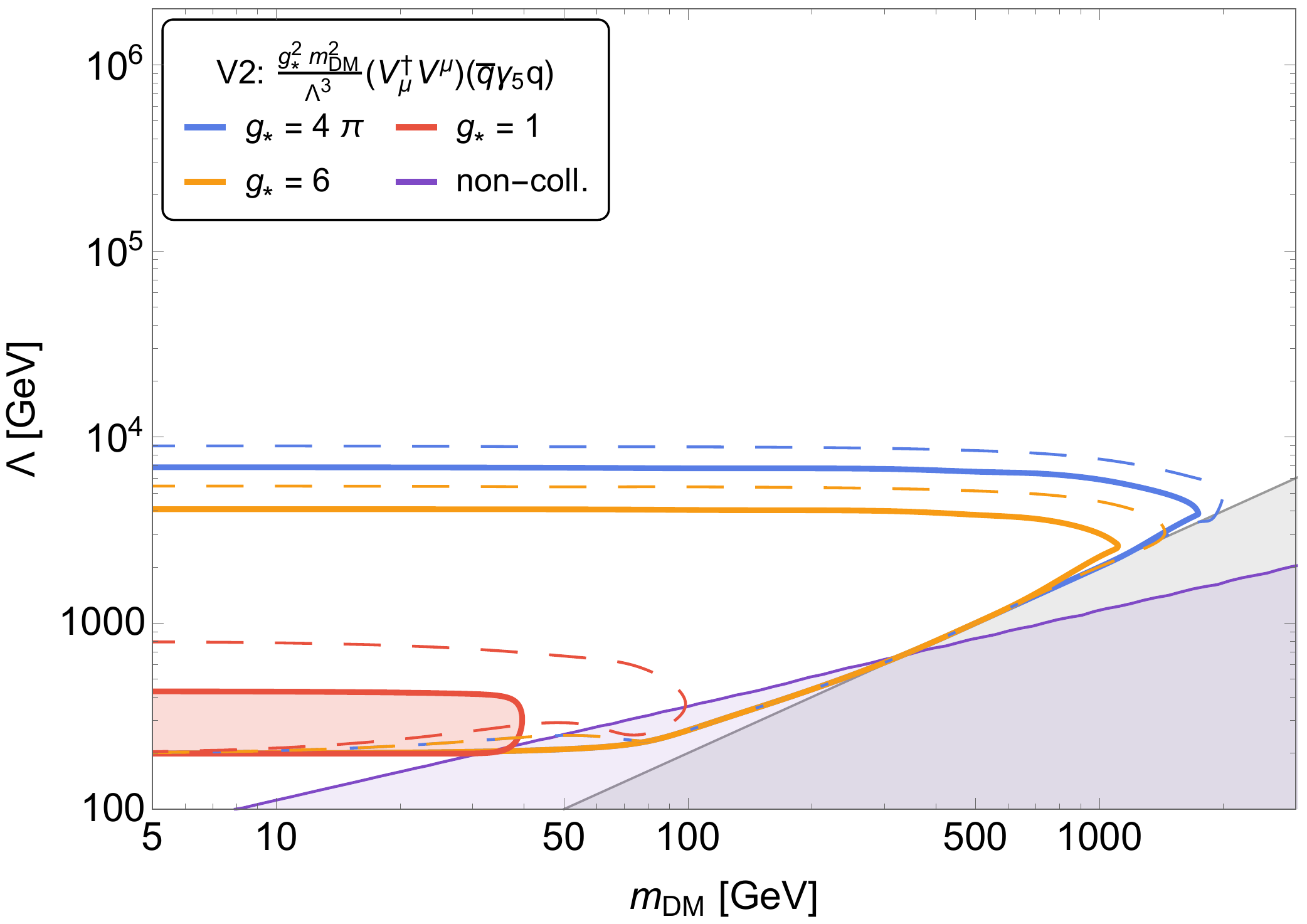}\\
\vskip 0.2cm
(c)\hspace*{0.5\textwidth}\hspace{-0.2cm}(d)\\\vspace*{-2cm}\\\\
\includegraphics[width=0.5\textwidth]{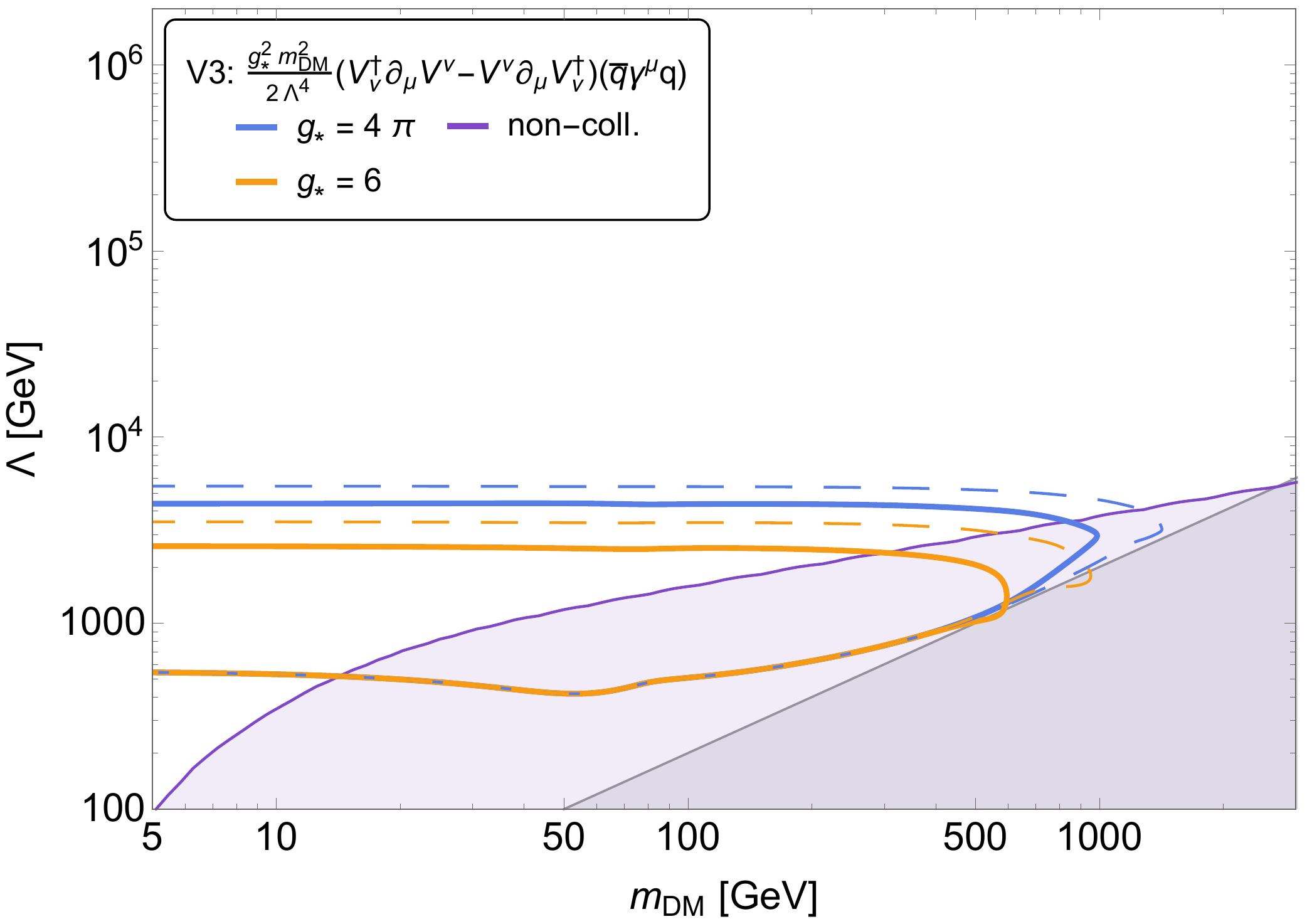}%
\includegraphics[width=0.5\textwidth]{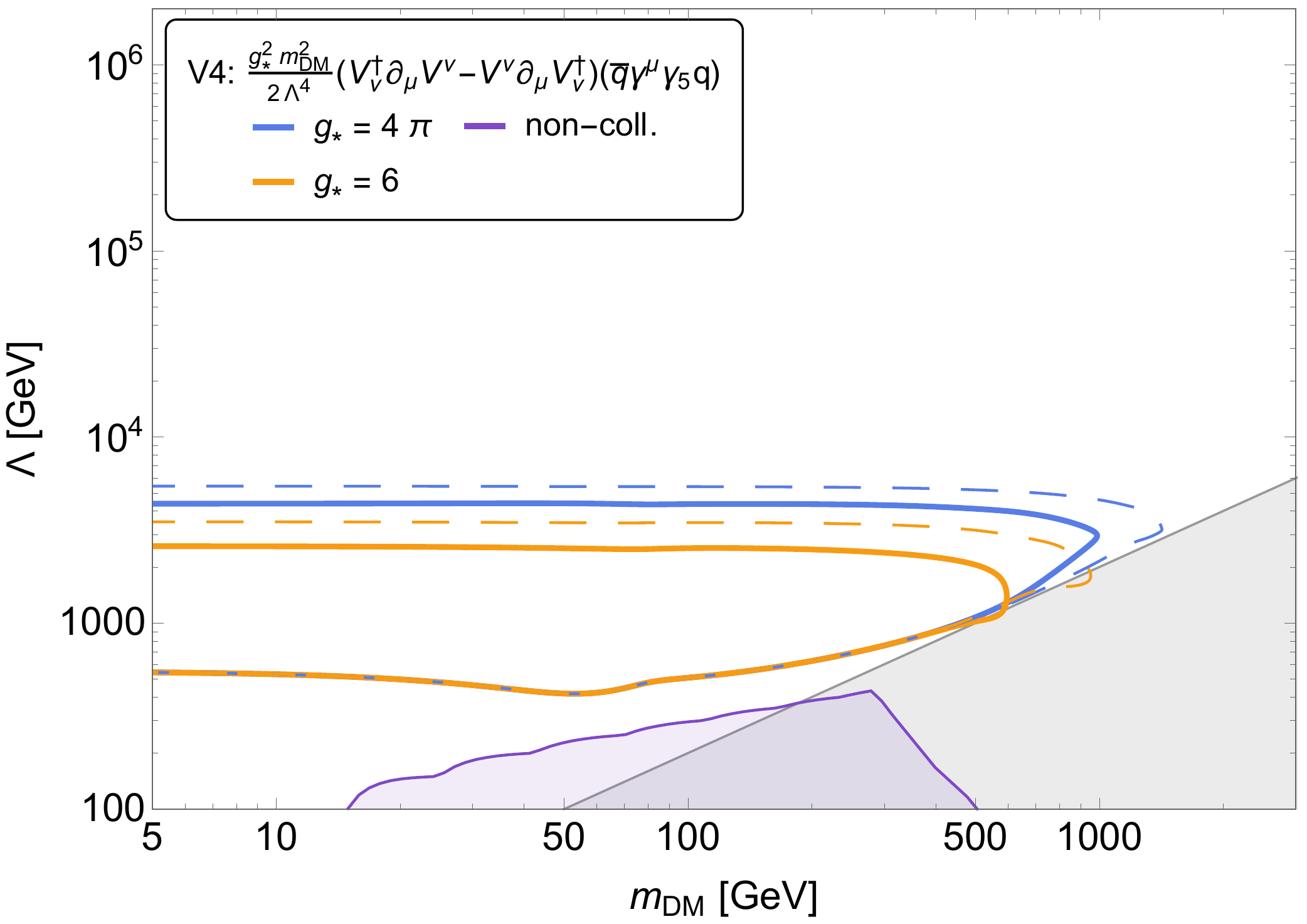}\\
\caption{\label{fig:coll-vector1} Constraints from LHC monojet
  searches on the EFT operators V1, V2, V3, and V4 that exhibit
  vector DM. The conventions are as in Fig.~\ref{fig:coll-scal}.}
\end{figure}

\begin{figure}[tbp]
\vskip 0.8cm
(a)\hspace*{0.5\textwidth}\hspace{-0.2cm}(b)\\\vspace*{-2cm}\\\\
\includegraphics[width=0.5\textwidth]{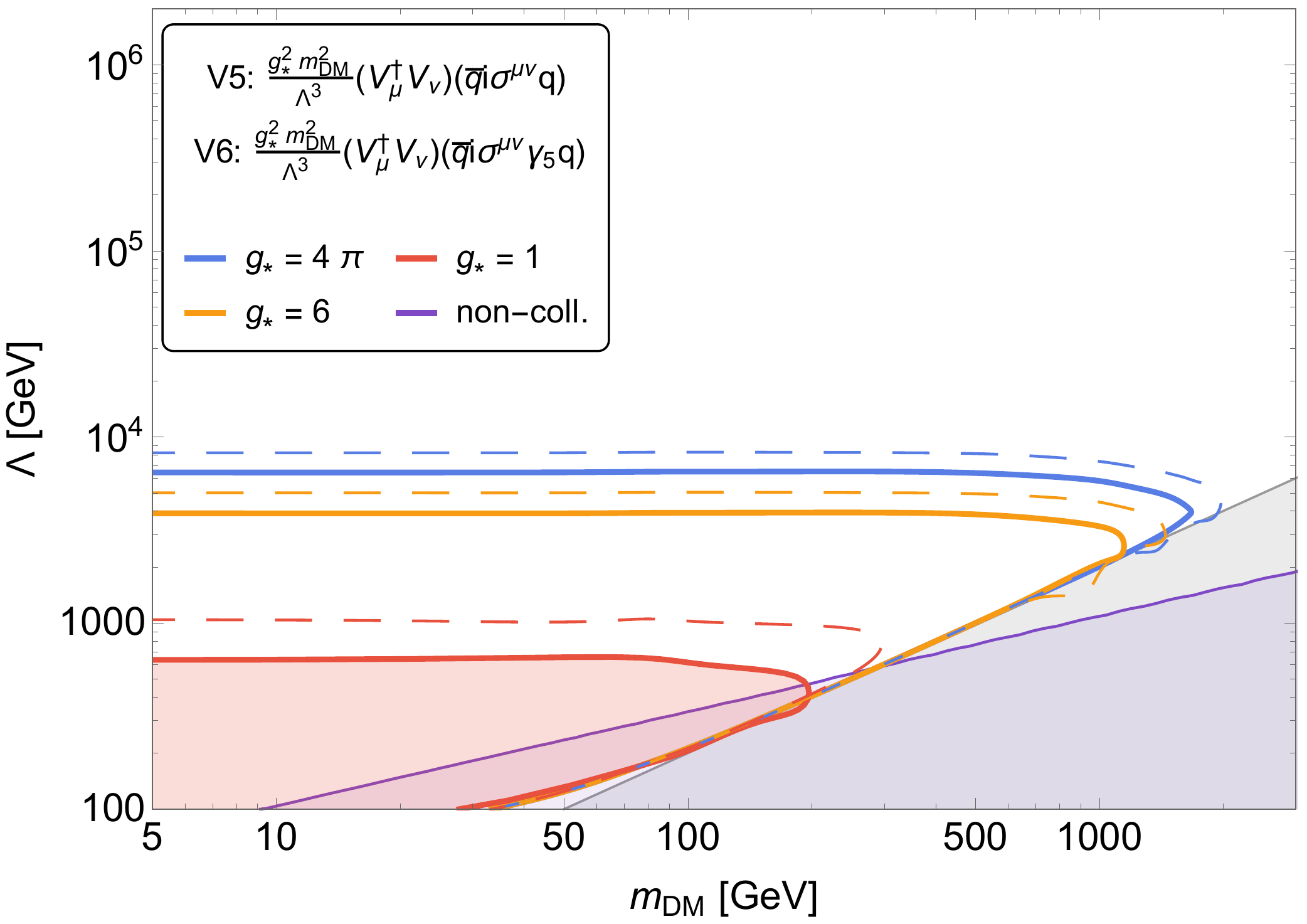}
\includegraphics[width=0.5\textwidth]{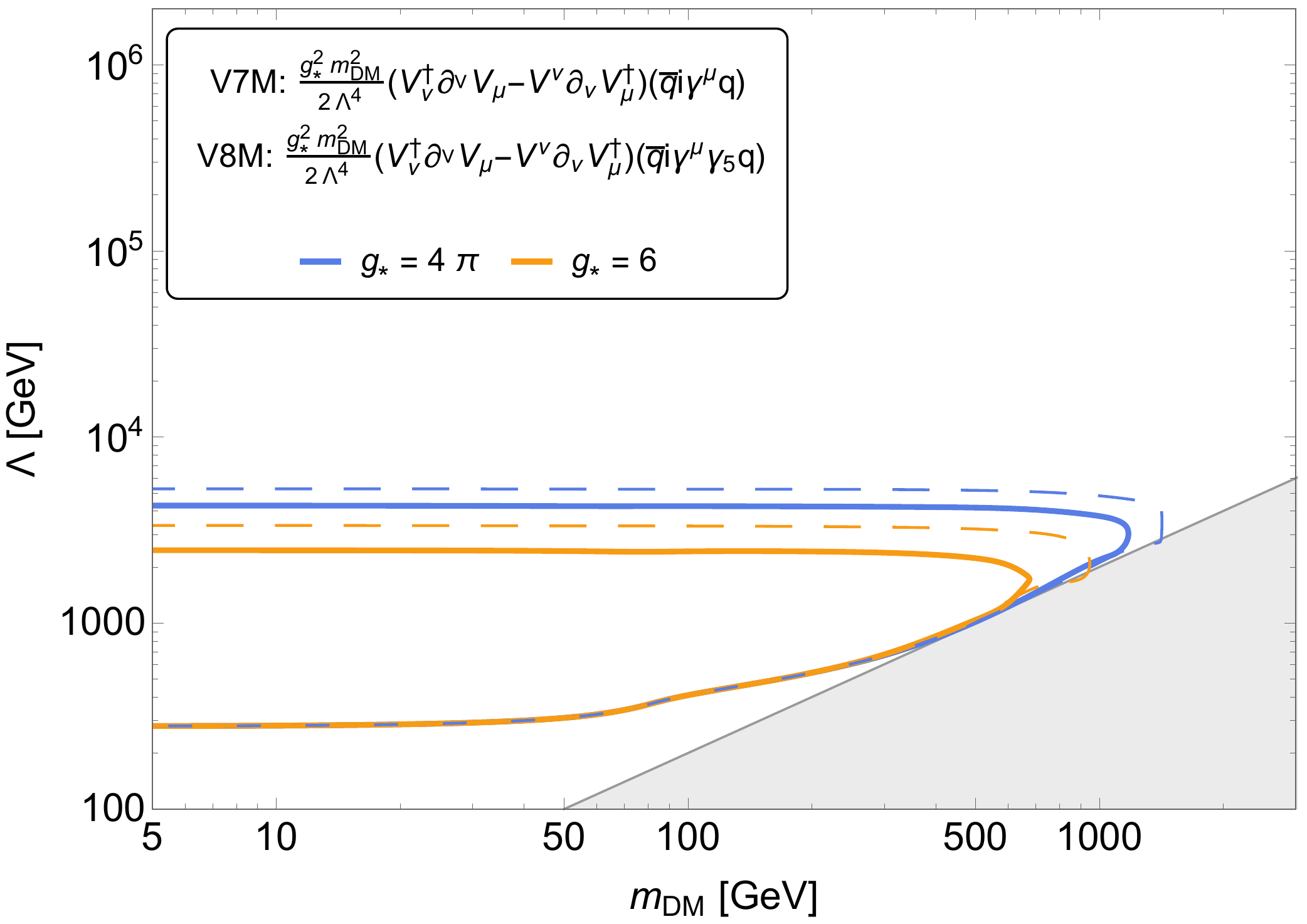} \\
\vskip 0.2cm
(c)\hspace*{0.5\textwidth}\hspace{-0.2cm}(d)\\\vspace*{-2cm}\\\\
\includegraphics[width=0.5\textwidth]{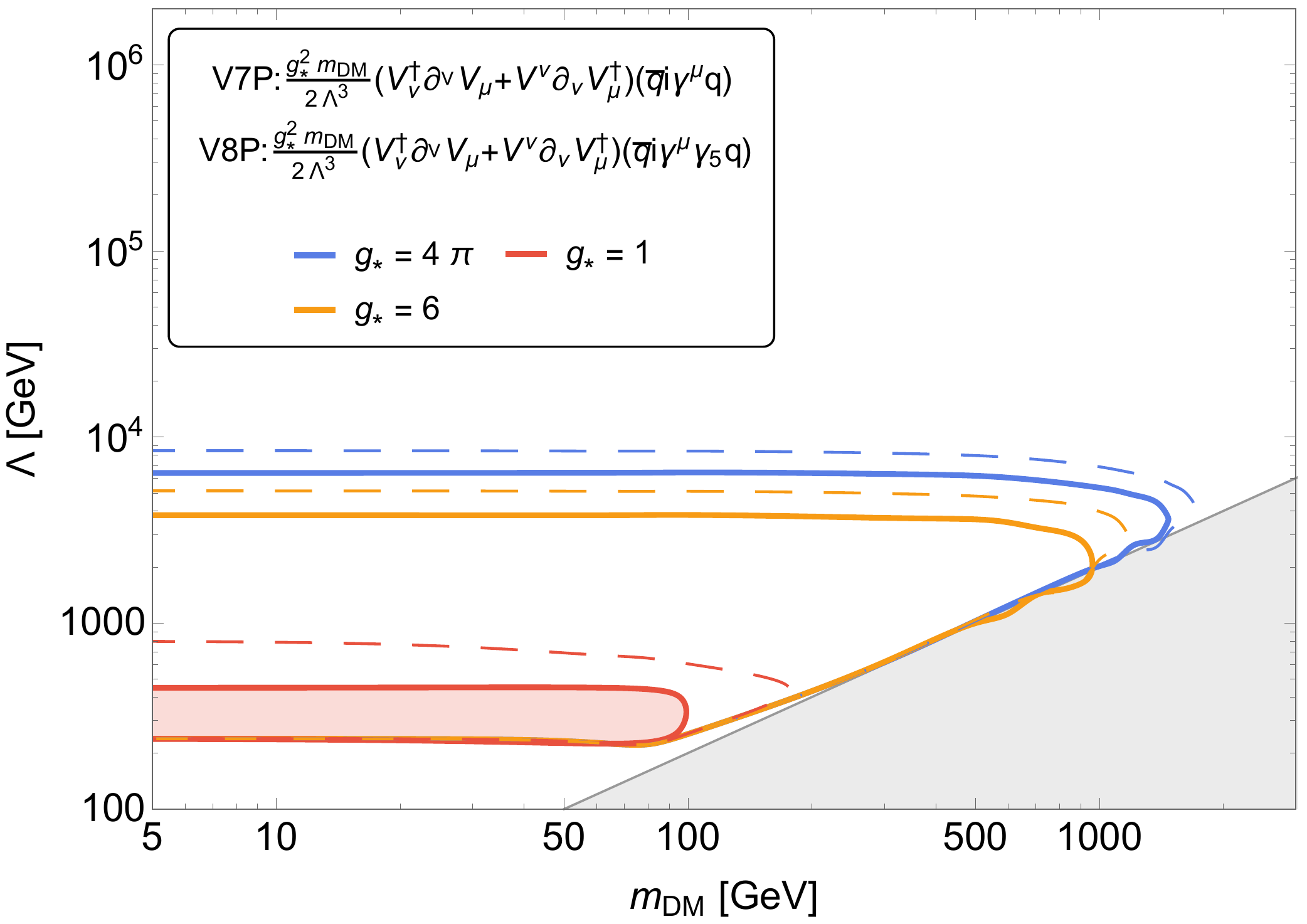}
\includegraphics[width=0.5\textwidth]{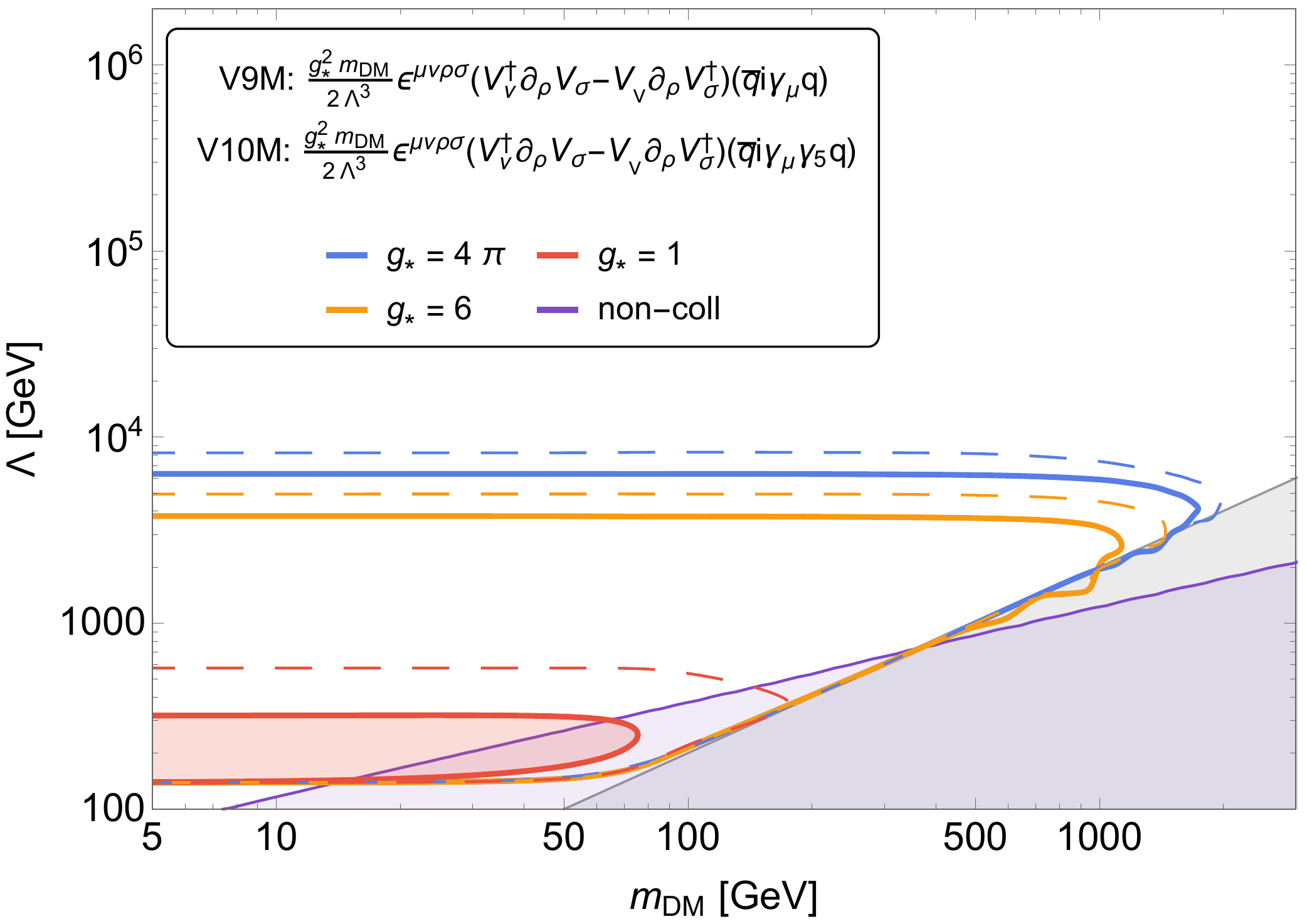} \\
\vskip 0.2cm
(e)\hspace*{0.5\textwidth}\hspace{-0.2cm}(f)\\\vspace*{-2cm}\\\\
\includegraphics[width=0.5\textwidth]{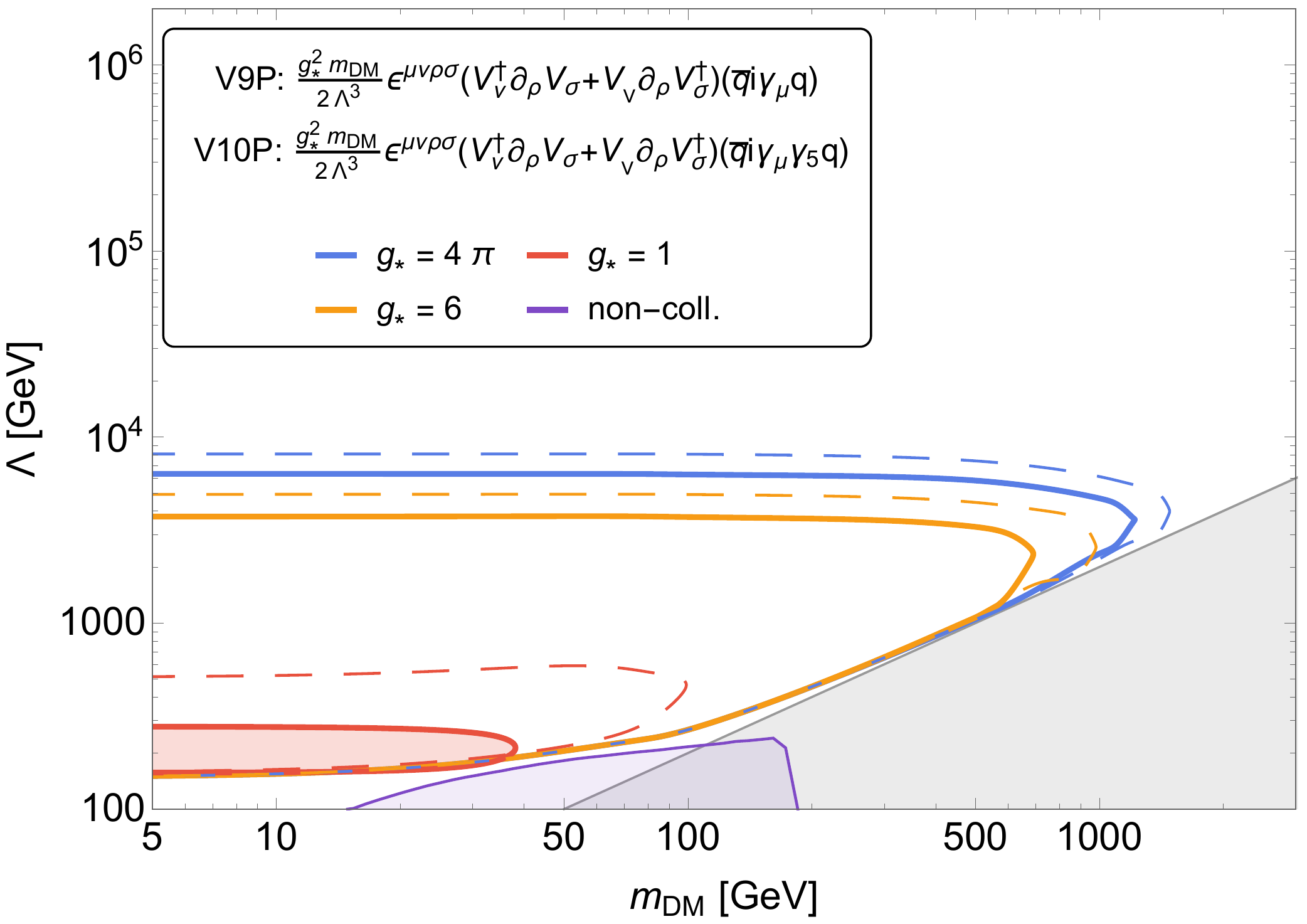}
\includegraphics[width=0.5\textwidth]{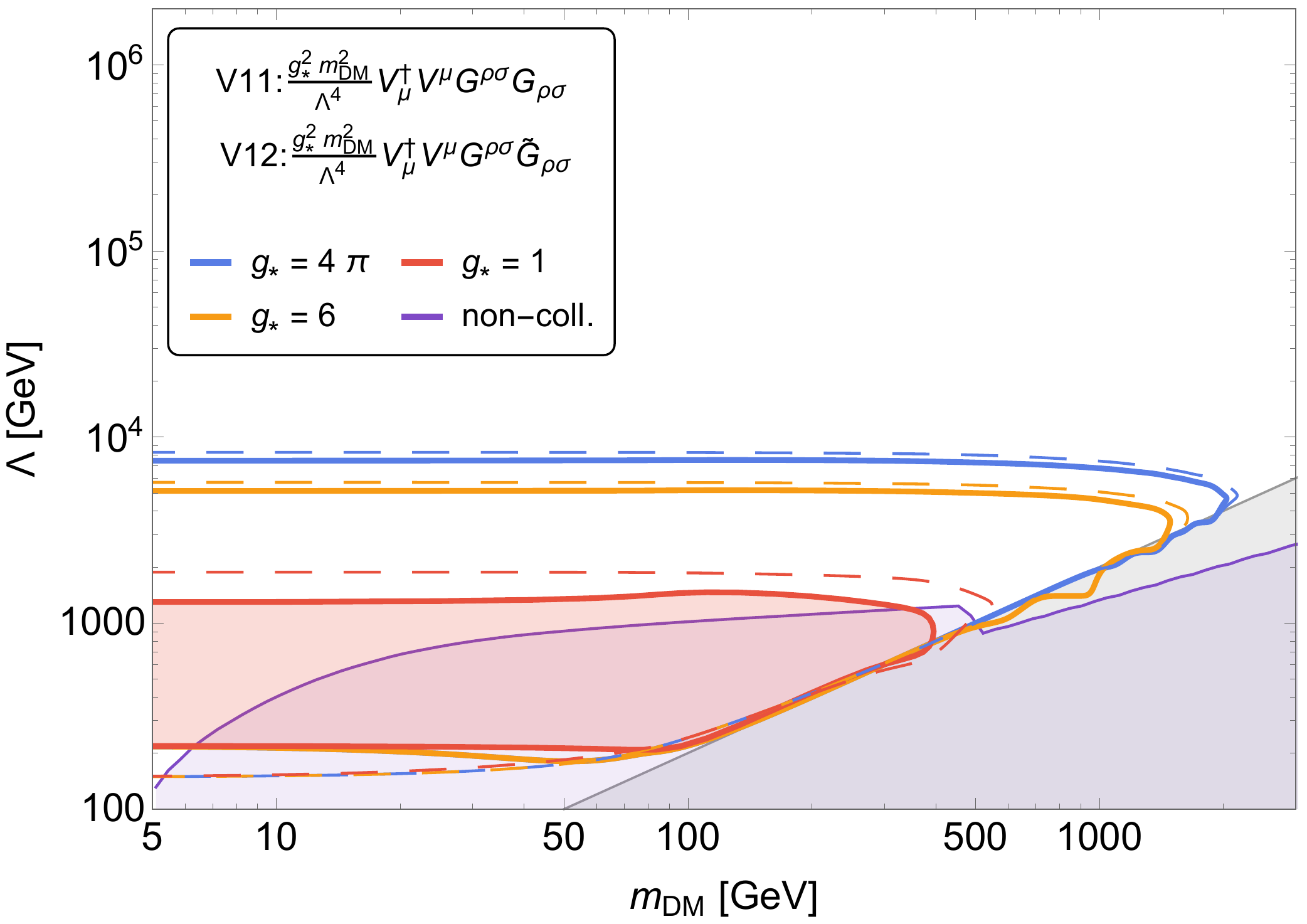}%
\caption{\label{fig:coll-vector2} Constraints from LHC monojet
  searches on the EFT operators V5, V6, V7M, V8M, V7P, V8P, V9M, V10M,
  V9P, V10P, V11, and V12 that exhibit vector DM. The conventions
  are as in Fig.~\ref{fig:coll-scal}.  }
\end{figure}

We display in Figures~\ref{fig:coll-vector1}
and~\ref{fig:coll-vector2} the present constraints on vector DM
operators that stem from the LHC monojet searches. Notice that the
LHC cross sections do not change significantly when
we replace a current by the corresponding pseudo one. Consequently, we
do not present the results for the pseudo currents unless their
non-collider constraints are different.

From the top panels of Fig.~\ref{fig:coll-vector1} we learn that the
LHC monojet constraints on the operators V1 and V2 are equal: for
$g_\star=1$, the excluded region is given by $m_{DM} \lesssim 40$ GeV
and $100 \le \Lambda \le 400$ GeV.  As $g_\star$ increases to $4\pi$,
the excluded region becomes $m_{DM} \lesssim 2$ TeV and
$ 100 \le \Lambda \le 9000$ GeV. Notice that the lower limit on
$\Lambda$ does not depend on the coupling $g_\star$.  
Since the V1 contribution to the SI DM DD is unsuppressed the
strongest bound on this operator comes from the DD searches.  On the
other hand, for the operator V2, the collider and non-collider limits are clearly
complementary: the collider limits dominate for DM masses smaller than
40 GeV while the CMB data give rise to the most stringent limits at DM
masses larger than 40 GeV.
The operators V3 and V4 do not lead to any collider limit for $g_\star = 1$
even for a larger integrated luminosity (300 fb$^{-1}$), therefore,
the non-collider limits are the most important ones, showing again the
synergy between low and high energy data.
This lack of sensitivity on these operators for $g_\star=1$ originates
from the shape of the invariant mass distribution of the DM pairs that
is shifted towards high values. Consequently, the EFT validity cut in
Eq.~(\ref{eq:inv-mass-cut}) discards a large fraction of the events in
this case. This reduction of number of signal events is overcome only
for larger values of coupling $g_\star=6$ ($4\pi$) for which the LHC
reach can be $m_{DM} \lesssim 600$ (900) GeV and
$\Lambda \lesssim 2.5$ (4) TeV.

The top left panel of Fig.~\ref{fig:coll-vector2} displays the LHC monojet
constraints on the operators V5 and V6, which are identical. For
$g_\star=1$ the LHC bounds are the most stringent on these operators
for DM masses up to $\simeq 180$ GeV and $\lambda \lesssim 600$ GeV
while the CMB limits play an important role just in a small DM mass
range around $200$ GeV.
On the other hand, we can see from the right top panel of this figure,
that the operators V7M and V8M are not bound at all for
$g_\star=1$. In the collider case the reason is the same behind the
null results for the operators V3 and V4. These operators can only be
probed by the LHC monojet searches for higher values of $g_\star$.

The left middle panel of Fig.~\ref{fig:coll-vector2} contains the
bounds on the operators V7P and V8P that are only constrained by the
LHC monojet searches that exclude the narrow region
$m_{DM} \lesssim 100$ GeV and $250 \lesssim \Lambda \lesssim 450$ GeV
for $g_\star =1$. For higher values of the coupling, {\it e.g.}
$g_\star=4\pi$ this region is substantially expanded to
$m_{DM} \lesssim 1.3$ TeV and $250 \lesssim \Lambda \lesssim 6500$
GeV. Once again the lower limit on $\Lambda$ depends on $g_\star$
and on the integrated luminosity.
The limits on the operators V9M and V10M are equal and are shown in
the right middle panel of Fig.~\ref{fig:coll-vector2}.  In this case
we witness a nice complementarity between high and low energy data for
these operators for $g_\star=1$. On one hand, the LHC monojet data
excludes the region $m_{DM} \lesssim 70$ GeV and
$ 130 \lesssim \Lambda \lesssim 300$ GeV. On the other hand, the CMB
bounds are the strongest ones for $m_{DM} \gtrsim 65$ GeV.

We can see from the left lower panel of Fig.~\ref{fig:coll-vector2}
that the limits on the operators V9P and V10P are rather loose,
however, the non-collider and collider ones are complementary. The LHC
monojet data exclude the area $m_{DM} \lesssim 40$ GeV and
$160 \lesssim \Lambda \lesssim 280$ GeV, while the SD DM DD searches
are more relevant for DM masses in the range 40--100 GeV. Moreover,
the excluded area is increase for larger values of $g_\star$ or for
larger integrated luminosities.
Last, but not least, the limits on the operators V11 and V12 are
presented in the lower right panel of this figure. In this case, the
monojet constraints are more stringent than the direct detection (valid only for the operator V11) and CMB ones, excluding a
large fraction of the parameter space for $g_\star=1$:
$m_{DM} \lesssim 400$ GeV and $200 \lesssim \Lambda \lesssim 1200$
GeV. Notice that these are the only operators that the $\Lambda$ lower
limits changes for higher couplings or integrated luminosities.

In brief, from Figs~\ref{fig:coll-vector1} and \ref{fig:coll-vector2}
we learn that LHC is sensitive to a substantial numbers of operators
containing vector DM, such as V2, V5 V6, V7P, V8P, V9-V12 which can
not be probed at all or are poorly bound with non-collider searches.
The LHC limits strongly depend on the type of the operator ranging
from few hundred GeV (V1, V2, V5,V6, V7P, V8P, V9M, V9P, V10M, V10P)
to about 1.3 TeV (V11,V12).
Moreover, we also assess the impact of the EFT validity cut in
Eq.~(\ref{eq:inv-mass-cut}) for vector DM, a fact that was not explored in previous studies.

In this scenario, the DD and CMB constraints rescale at a fixed DM mass as
$g_\star^{2/3}$ and $g_\star^{1/2}$ for operators suppressed by $\Lambda^3$ and 
$\Lambda^4$ respectively.  For $g_\star = 4 \pi$, the dominance of the CMB bounds
on the operators V2, V5/V6, V9M/V10M, and V11/V12 occurs for DM masses
larger than 920, 1000, 850, and 1500 GeV respectively. 
Moreover, the collider limits on V3 are more stringent than the DD
ones for $m_{DM} \le$ 50 GeV.
For the other vector DM operators, the conclusions remain the same
for the upper bounds on $\Lambda$. Nevertheless, there is a beautiful
complementarity between the non-collider and LHC searches since the
former exclude the region of small $\Lambda$ that the collider
searches do not cover due to the validity of the EFT.


\section{Conclusions}\label{sec:conclusions}

In this work we have presented accurate and up-to-date constraints on the complete set of dimension five and six operators connecting SM quarks and gluons with a DM candidate, which can be a complex scalar, a Dirac fermion or a complex vector.
\smallskip

We have performed a comprehensive analyses of the complementarity between collider and non-collider searches to probe DM parameter space, including LHC mono-jet data, bounds from SI and SD direct searches, relic density limits and CMB indirect constraints due to the injection of energy produced by DM annihilation in the early universe. Since the characteristic energy scale for LHC and direct searches differs by about six orders of magnitude, to correctly evaluate the experimental sensitivity to the DM EFT operators we have taken into account their running and mixing from the TeV scale to the GeV one. This effect is especially important for operators with pseudo-vector SM quark current (C4, D6, D7 and V4 in Table~\ref{tab:EFToperators}) which, mixing into operators with vector SM quark current, develop a SI cross section. Another important point has been to take into account the realistic uncertainty in the DM DD searches limits due to the uncertainty on the local DM density. In some cases, the uncertainty can quantitatively shift the bounds by one order of magnitude, an important result that must be taken into account to properly address the excluded parameter space. As for indirect searches, we have chosen to present only the bounds coming from CMB data, since they are not plagued by uncertainties on the DM distribution. These bounds are particularly important when the direct detection rates are suppressed. We have also found  that the EFT validity criteria $M_{DM,DM}< \Lambda$ plays an important role for the case of vector DM, when the invariant mass of the DM pair $M_{DM,DM}$ is typically larger than in the case of scalar and fermion DM, and there is no collider bound for relatively small values of $\Lambda$.
\smallskip

Our updated bounds are summarized in Figs.~\ref{fig:coll-scal}-\ref{fig:coll-vector2} for all the operators listed in Table~\ref{tab:EFToperators}.
In the case of scalar DM we have found an important synergy between the different experiments: while for the operators C1 and C3 (with scalar and vector quark currents, respectively) the most stringent bounds stem from SI searches, the best limits on the C5 and C6 operators (describing DM-DM-Gluon-Gluon interactions) come from the LHC mono-jet data, since DD rates of DM scattering on gluon component of the nucleons are suppressed. At the same time, the most important constraints on the C2 operator come from the CMB data, while the limits on C4 are dominated by the LHC searches at low $m_{DM}$ and by SI bounds for large DM mass values.
\smallskip

The overall picture for the Dirac fermion DM is similar to the one of scalar DM: there is a synergy between the collider and non-collider searches. The strongest limits on the D1 and D5 operators are due to the SI searches, while the LHC mono-jet data put the strongest constraints on the D3, D6 and D8 operators. On the other hand, the operators D2, D4 and D9 are bounded by the LHC mono-jet searches at small DM masses, while at larger masses the CMB data dominate the limits. At the same time, the best limits on the D7 and D10 operators
 are defined by the interplay of the LHC searches, direct detection data and CMB constraints, depending on the DM mass.
 \smallskip

In the case of vector DM we can also see an important complementarity between the different experiments to probe DM parameter space. The most stringent limits on the V1, V3 and V4 Wilson coefficients are set by  SI searches. On the other hand, the LHC mono-jet data provides the strongest bounds on the V5, V6, V7P, V8P, V11, and V12 operators for any value of the DM mass. Furthermore, there is a synergy between the LHC and CMB data in probing the V2, V9M and V10M operators in different  DM mass regions. One should also note  the interplay between the LHC mono-jet searches and the SD bounds in bounding the V9P and V10P operators.
We have also found that notwithstanding of combination of  all data  there are no limits on 
the operators V7M and V8M for
$g_\star=1$. 
There is one more important point to stress about the 
interplay of different data in probing vector DM operators:
the lack of the LHC sensitivity to small values of $\Lambda$
discussed above in section~\ref{sec:coll-vector-dm} is complemented  by the potential of CMB data to probe this region. This complementarity is manifest for  many operators (V1, V2, V3, V4, V9M, V10M, V9P, V10P, V11 and V12), for which CMB data partly or completely cover the lower $\Lambda$ region 
which LHC is unable to probe.
\smallskip

As a general remark, in our analysis we have assumed that just one operator is non-vanishing at a time, which may or may not be the case depending on the underlying theory realized  in nature. Nevertheless, our studies can be easily adapted to some specific scenario where the integration of heavy mediators can lead to more than one non-vanishing Wilson coefficient. 
For instance, a $t$-channel scalar mediator coupled to fermion DM would generate a $(\bar{\chi}q) (\bar{q} \chi)$ effective operator.  After a Fierz transformation, this operator can be written as a combination of the fermion operators D1, D4, D5, D8, and D9~\cite{Belyaev:2016pxe}. According to Fig.~\ref{fig:coll-dirac2}, it is clear that in this case the most stringent bounds would be set by SI searches, so that at least in first approximation the bound on $(\bar{\chi}q) (\bar{q} \chi)$ coincides with the bound on D1. Of course, the validity of such a procedure must be analyzed case by case.

\section*{Acknowledgements}
AB acknowledges partial  support from the STFC grant ST/L000296/1.
AB also thanks the NExT Institute, Royal Society Leverhulme Trust Senior Research Fellowship LT140094, 
Royal Society International Exchange grant IE150682 and
Soton-FAPESP grant.
AB acknowledges partial support from the InvisiblesPlus RISE from the European
Union Horizon 2020 research and innovation programme under the Marie Sklodowska-Curie grant
agreement No 690575. EB acknowledges support from the Funda\c{c}\~ao de Amp\'{a}ro \`{a} Pesquisa (FAPESP) under grant 2015/25884-4. OE is supported in part by Conselho
Nacional de Desenvolvimento Cient\'{\i}fico e Tecnol\'ogico (CNPq) and
by FAPESP. GGdC  is supported by FAPESP process 2016/17041-0 and thanks the University of Southampton, the CERN Theoretical Physics Department, the Humboldt University, SISSA, Sapienza and University of Padova for hospitality at various stages of this work. 
F.~I.~acknowledges support from the Simons Foundation and FAPESP process 2014/11070-2.

\bibliography{bib}

\providecommand{\href}[2]{#2}\begingroup\raggedright\begin{thebibliography}{10}

\bibitem{Ade:2015xua}
{\scshape Planck} collaboration, P.~A.~R. Ade et~al., \emph{{Planck 2015
  results. XIII. Cosmological parameters}},
  \href{http://dx.doi.org/10.1051/0004-6361/201525830}{\emph{Astron.
  Astrophys.} {\bf 594} (2016) A13},
  [\href{http://arxiv.org/abs/1502.01589}{{\tt 1502.01589}}].

\bibitem{Blumenthal:1984bp}
G.~R. Blumenthal, S.~M. Faber, J.~R. Primack and M.~J. Rees, \emph{{Formation
  of Galaxies and Large Scale Structure with Cold Dark Matter}},
  \href{http://dx.doi.org/10.1038/311517a0}{\emph{Nature} {\bf 311} (1984)
  517--525}.

\bibitem{Bullock:1999he}
J.~S. Bullock, T.~S. Kolatt, Y.~Sigad, R.~S. Somerville, A.~V. Kravtsov, A.~A.
  Klypin et~al., \emph{{Profiles of dark haloes. Evolution, scatter, and
  environment}},
  \href{http://dx.doi.org/10.1046/j.1365-8711.2001.04068.x}{\emph{Mon. Not.
  Roy. Astron. Soc.} {\bf 321} (2001) 559--575},
  [\href{http://arxiv.org/abs/astro-ph/9908159}{{\tt astro-ph/9908159}}].

\bibitem{Aaboud:2017phn}
{\scshape ATLAS} collaboration, M.~Aaboud et~al., \emph{{Search for dark matter
  and other new phenomena in events with an energetic jet and large missing
  transverse momentum using the ATLAS detector}},
  \href{http://dx.doi.org/10.1007/JHEP01(2018)126}{\emph{JHEP} {\bf 01} (2018)
  126}, [\href{http://arxiv.org/abs/1711.03301}{{\tt 1711.03301}}].

\bibitem{CMS:2017tbk}
{\scshape CMS} collaboration, C.~Collaboration, \emph{{Search for new physics
  in final states with an energetic jet or a hadronically decaying W or Z boson
  using $35.9~\mathrm{fb}^{-1}$ of data at $\sqrt{s} = 13~\mathrm{TeV}$}}, .

\bibitem{ATLAS-CONF-2018-005}
{\scshape ATLAS} collaboration, \emph{{Search for dark matter in events with a
  hadronically decaying vector boson and missing transverse momentum in $pp$
  collisions at $\sqrt{s} = 13$ TeV with the ATLAS detector}},  Tech. Rep.
  ATLAS-CONF-2018-005, CERN, Geneva, Apr, 2018.

\bibitem{Aaboud:2017dor}
{\scshape ATLAS} collaboration, M.~Aaboud et~al., \emph{{Search for dark matter
  at $\sqrt{s}=13$ TeV in final states containing an energetic photon and large
  missing transverse momentum with the ATLAS detector}},
  \href{http://dx.doi.org/10.1140/epjc/s10052-017-4965-8}{\emph{Eur. Phys. J.}
  {\bf C77} (2017) 393}, [\href{http://arxiv.org/abs/1704.03848}{{\tt
  1704.03848}}].

\bibitem{Hinshaw:2012aka}
{\scshape WMAP} collaboration, G.~Hinshaw et~al., \emph{{Nine-Year Wilkinson
  Microwave Anisotropy Probe (WMAP) Observations: Cosmological Parameter
  Results}},
  \href{http://dx.doi.org/10.1088/0067-0049/208/2/19}{\emph{Astrophys. J.
  Suppl.} {\bf 208} (2013) 19}, [\href{http://arxiv.org/abs/1212.5226}{{\tt
  1212.5226}}].

\bibitem{Goodman:1984dc}
M.~W. Goodman and E.~Witten, \emph{{Detectability of Certain Dark Matter
  Candidates}}, \href{http://dx.doi.org/10.1103/PhysRevD.31.3059}{\emph{Phys.
  Rev.} {\bf D31} (1985) 3059}.

\bibitem{Aprile:2017iyp}
{\scshape XENON} collaboration, E.~Aprile et~al., \emph{{First Dark Matter
  Search Results from the XENON1T Experiment}},
  \href{http://dx.doi.org/10.1103/PhysRevLett.119.181301}{\emph{Phys. Rev.
  Lett.} {\bf 119} (2017) 181301}, [\href{http://arxiv.org/abs/1705.06655}{{\tt
  1705.06655}}].

\bibitem{Akerib:2016vxi}
{\scshape LUX} collaboration, D.~S. Akerib et~al., \emph{{Results from a search
  for dark matter in the complete LUX exposure}},
  \href{http://dx.doi.org/10.1103/PhysRevLett.118.021303}{\emph{Phys. Rev.
  Lett.} {\bf 118} (2017) 021303}, [\href{http://arxiv.org/abs/1608.07648}{{\tt
  1608.07648}}].

\bibitem{Fu:2016ega}
{\scshape PandaX-II} collaboration, C.~Fu et~al., \emph{{Spin-Dependent
  Weakly-Interacting-Massive-Particle--Nucleon Cross Section Limits from First
  Data of PandaX-II Experiment}},
  \href{http://dx.doi.org/10.1103/PhysRevLett.120.049902,
  10.1103/PhysRevLett.118.071301}{\emph{Phys. Rev. Lett.} {\bf 118} (2017)
  071301}, [\href{http://arxiv.org/abs/1611.06553}{{\tt 1611.06553}}].

\bibitem{Slatyer:2017sev}
T.~R. Slatyer, \emph{{TASI Lectures on Indirect Detection of Dark Matter}},  in
  \emph{{Theoretical Advanced Study Institute in Elementary Particle Physics:
  Anticipating the Next Discoveries in Particle Physics (TASI 2016) Boulder,
  CO, USA, June 6-July 1, 2016}}, 2017.
\newblock \href{http://arxiv.org/abs/1710.05137}{{\tt 1710.05137}}.

\bibitem{Ackermann:2015zua}
{\scshape Fermi-LAT} collaboration, M.~Ackermann et~al., \emph{{Searching for
  Dark Matter Annihilation from Milky Way Dwarf Spheroidal Galaxies with Six
  Years of Fermi Large Area Telescope Data}},
  \href{http://dx.doi.org/10.1103/PhysRevLett.115.231301}{\emph{Phys. Rev.
  Lett.} {\bf 115} (2015) 231301}, [\href{http://arxiv.org/abs/1503.02641}{{\tt
  1503.02641}}].

\bibitem{Zitzer:2015eqa}
{\scshape VERITAS} collaboration, B.~Zitzer, \emph{{A Search for Dark Matter
  from Dwarf Galaxies using VERITAS}}, {\emph{PoS} {\bf ICRC2015} (2016) 1225},
  [\href{http://arxiv.org/abs/1509.01105}{{\tt 1509.01105}}].

\bibitem{Ahnen:2016qkx}
{\scshape Fermi-LAT, MAGIC} collaboration, M.~L. Ahnen et~al., \emph{{Limits to
  dark matter annihilation cross-section from a combined analysis of MAGIC and
  Fermi-LAT observations of dwarf satellite galaxies}},
  \href{http://dx.doi.org/10.1088/1475-7516/2016/02/039}{\emph{JCAP} {\bf 1602}
  (2016) 039}, [\href{http://arxiv.org/abs/1601.06590}{{\tt 1601.06590}}].

\bibitem{Abdallah:2016ygi}
{\scshape H.E.S.S.} collaboration, H.~Abdallah et~al., \emph{{Search for dark
  matter annihilations towards the inner Galactic halo from 10 years of
  observations with H.E.S.S}},
  \href{http://dx.doi.org/10.1103/PhysRevLett.117.111301}{\emph{Phys. Rev.
  Lett.} {\bf 117} (2016) 111301}, [\href{http://arxiv.org/abs/1607.08142}{{\tt
  1607.08142}}].

\bibitem{Abramowski:2013ax}
{\scshape H.E.S.S.} collaboration, A.~Abramowski et~al., \emph{{Search for
  Photon-Linelike Signatures from Dark Matter Annihilations with H.E.S.S.}},
  \href{http://dx.doi.org/10.1103/PhysRevLett.110.041301}{\emph{Phys. Rev.
  Lett.} {\bf 110} (2013) 041301}, [\href{http://arxiv.org/abs/1301.1173}{{\tt
  1301.1173}}].

\bibitem{Galli:2009zc}
S.~Galli, F.~Iocco, G.~Bertone and A.~Melchiorri, \emph{{CMB constraints on
  Dark Matter models with large annihilation cross-section}},
  \href{http://dx.doi.org/10.1103/PhysRevD.80.023505}{\emph{Phys. Rev.} {\bf
  D80} (2009) 023505}, [\href{http://arxiv.org/abs/0905.0003}{{\tt
  0905.0003}}].

\bibitem{Galli:2011rz}
S.~Galli, F.~Iocco, G.~Bertone and A.~Melchiorri, \emph{{Updated CMB
  constraints on Dark Matter annihilation cross-sections}},
  \href{http://dx.doi.org/10.1103/PhysRevD.84.027302}{\emph{Phys. Rev.} {\bf
  D84} (2011) 027302}, [\href{http://arxiv.org/abs/1106.1528}{{\tt
  1106.1528}}].

\bibitem{Contino:2016jqw}
R.~Contino, A.~Falkowski, F.~Goertz, C.~Grojean and F.~Riva, \emph{{On the
  Validity of the Effective Field Theory Approach to SM Precision Tests}},
  \href{http://dx.doi.org/10.1007/JHEP07(2016)144}{\emph{JHEP} {\bf 07} (2016)
  144}, [\href{http://arxiv.org/abs/1604.06444}{{\tt 1604.06444}}].

\bibitem{Belyaev:2016pxe}
A.~Belyaev, L.~Panizzi, A.~Pukhov and M.~Thomas, \emph{{Dark Matter
  characterization at the LHC in the Effective Field Theory approach}},
  \href{http://dx.doi.org/10.1007/JHEP04(2017)110}{\emph{JHEP} {\bf 04} (2017)
  110}, [\href{http://arxiv.org/abs/1610.07545}{{\tt 1610.07545}}].

\bibitem{Goodman:2010ku}
J.~Goodman, M.~Ibe, A.~Rajaraman, W.~Shepherd, T.~M. Tait et~al.,
  \emph{{Constraints on Dark Matter from Colliders}},
  \href{http://dx.doi.org/10.1103/PhysRevD.82.116010}{\emph{Phys.Rev.} {\bf
  D82} (2010) 116010}, [\href{http://arxiv.org/abs/1008.1783}{{\tt
  1008.1783}}].

\bibitem{Kumar:2015wya}
J.~Kumar, D.~Marfatia and D.~Yaylali, \emph{{Vector dark matter at the LHC}},
  \href{http://dx.doi.org/10.1103/PhysRevD.92.095027}{\emph{Phys. Rev.} {\bf
  D92} (2015) 095027}, [\href{http://arxiv.org/abs/1508.04466}{{\tt
  1508.04466}}].

\bibitem{Bertuzzo:2017lwt}
E.~Bertuzzo, C.~J. Caniu~Barros and G.~Grilli~di Cortona, \emph{{MeV Dark
  Matter: Model Independent Bounds}},
  \href{http://dx.doi.org/10.1007/JHEP09(2017)116}{\emph{JHEP} {\bf 09} (2017)
  116}, [\href{http://arxiv.org/abs/1707.00725}{{\tt 1707.00725}}].

\bibitem{Busoni:2013lha}
G.~Busoni, A.~De~Simone, E.~Morgante and A.~Riotto, \emph{{On the Validity of
  the Effective Field Theory for Dark Matter Searches at the LHC}},
  \href{http://dx.doi.org/10.1016/j.physletb.2013.11.069}{\emph{Phys. Lett.}
  {\bf B728} (2014) 412--421}, [\href{http://arxiv.org/abs/1307.2253}{{\tt
  1307.2253}}].

\bibitem{Busoni:2014sya}
G.~Busoni, A.~De~Simone, J.~Gramling, E.~Morgante and A.~Riotto, \emph{{On the
  Validity of the Effective Field Theory for Dark Matter Searches at the LHC,
  Part II: Complete Analysis for the $s$-channel}},
  \href{http://dx.doi.org/10.1088/1475-7516/2014/06/060}{\emph{JCAP} {\bf 1406}
  (2014) 060}, [\href{http://arxiv.org/abs/1402.1275}{{\tt 1402.1275}}].

\bibitem{Busoni:2014haa}
G.~Busoni, A.~De~Simone, T.~Jacques, E.~Morgante and A.~Riotto, \emph{{On the
  Validity of the Effective Field Theory for Dark Matter Searches at the LHC
  Part III: Analysis for the $t$-channel}},
  \href{http://dx.doi.org/10.1088/1475-7516/2014/09/022}{\emph{JCAP} {\bf 1409}
  (2014) 022}, [\href{http://arxiv.org/abs/1405.3101}{{\tt 1405.3101}}].

\bibitem{Racco:2015dxa}
D.~Racco, A.~Wulzer and F.~Zwirner, \emph{{Robust collider limits on
  heavy-mediator Dark Matter}},
  \href{http://dx.doi.org/10.1007/JHEP05(2015)009}{\emph{JHEP} {\bf 1505}
  (2015) 009}, [\href{http://arxiv.org/abs/1502.04701}{{\tt 1502.04701}}].

\bibitem{Hill:2011be}
R.~J. Hill and M.~P. Solon, \emph{{Universal behavior in the scattering of
  heavy, weakly interacting dark matter on nuclear targets}},
  \href{http://dx.doi.org/10.1016/j.physletb.2012.01.013}{\emph{Phys. Lett.}
  {\bf B707} (2012) 539--545}, [\href{http://arxiv.org/abs/1111.0016}{{\tt
  1111.0016}}].

\bibitem{Frandsen:2012db}
M.~T. Frandsen, U.~Haisch, F.~Kahlhoefer, P.~Mertsch and K.~Schmidt-Hoberg,
  \emph{{Loop-induced dark matter direct detection signals from gamma-ray
  lines}}, \href{http://dx.doi.org/10.1088/1475-7516/2012/10/033}{\emph{JCAP}
  {\bf 1210} (2012) 033}, [\href{http://arxiv.org/abs/1207.3971}{{\tt
  1207.3971}}].

\bibitem{Vecchi:2013iza}
L.~Vecchi, \emph{{WIMPs and Un-Naturalness}},
  \href{http://arxiv.org/abs/1312.5695}{{\tt 1312.5695}}.

\bibitem{Crivellin:2014qxa}
A.~Crivellin, F.~D'Eramo and M.~Procura, \emph{{New Constraints on Dark Matter
  Effective Theories from Standard Model Loops}},
  \href{http://dx.doi.org/10.1103/PhysRevLett.112.191304}{\emph{Phys. Rev.
  Lett.} {\bf 112} (2014) 191304}, [\href{http://arxiv.org/abs/1402.1173}{{\tt
  1402.1173}}].

\bibitem{DEramo:2014nmf}
F.~D'Eramo and M.~Procura, \emph{{Connecting Dark Matter UV Complete Models to
  Direct Detection Rates via Effective Field Theory}},
  \href{http://dx.doi.org/10.1007/JHEP04(2015)054}{\emph{JHEP} {\bf 04} (2015)
  054}, [\href{http://arxiv.org/abs/1411.3342}{{\tt 1411.3342}}].

\bibitem{DEramo:2016gos}
F.~D'Eramo, B.~J. Kavanagh and P.~Panci, \emph{{You can hide but you have to
  run: direct detection with vector mediators}},
  \href{http://dx.doi.org/10.1007/JHEP08(2016)111}{\emph{JHEP} {\bf 08} (2016)
  111}, [\href{http://arxiv.org/abs/1605.04917}{{\tt 1605.04917}}].

\bibitem{Bishara:2017pfq}
F.~Bishara, J.~Brod, B.~Grinstein and J.~Zupan, \emph{{From quarks to nucleons
  in dark matter direct detection}},
  \href{http://dx.doi.org/10.1007/JHEP11(2017)059}{\emph{JHEP} {\bf 11} (2017)
  059}, [\href{http://arxiv.org/abs/1707.06998}{{\tt 1707.06998}}].

\bibitem{Iocco:2015xga}
F.~Iocco, M.~Pato and G.~Bertone, \emph{{Evidence for dark matter in the inner
  Milky Way}}, \href{http://dx.doi.org/10.1038/nphys3237}{\emph{Nature Phys.}
  {\bf 11} (2015) 245--248}, [\href{http://arxiv.org/abs/1502.03821}{{\tt
  1502.03821}}].

\bibitem{Pato:2015dua}
M.~Pato, F.~Iocco and G.~Bertone, \emph{{Dynamical constraints on the dark
  matter distribution in the Milky Way}},
  \href{http://dx.doi.org/10.1088/1475-7516/2015/12/001}{\emph{JCAP} {\bf 1512}
  (2015) 001}, [\href{http://arxiv.org/abs/1504.06324}{{\tt 1504.06324}}].

\bibitem{Read:2014qva}
J.~I. Read, \emph{{The Local Dark Matter Density}},
  \href{http://dx.doi.org/10.1088/0954-3899/41/6/063101}{\emph{J. Phys.} {\bf
  G41} (2014) 063101}, [\href{http://arxiv.org/abs/1404.1938}{{\tt
  1404.1938}}].

\bibitem{Green:2010gw}
A.~M. Green, \emph{{Dependence of direct detection signals on the WIMP velocity
  distribution}},
  \href{http://dx.doi.org/10.1088/1475-7516/2010/10/034}{\emph{JCAP} {\bf 1010}
  (2010) 034}, [\href{http://arxiv.org/abs/1009.0916}{{\tt 1009.0916}}].

\bibitem{Green:2011bv}
A.~M. Green, \emph{{Astrophysical uncertainties on direct detection
  experiments}}, \href{http://dx.doi.org/10.1142/S0217732312300042}{\emph{Mod.
  Phys. Lett.} {\bf A27} (2012) 1230004},
  [\href{http://arxiv.org/abs/1112.0524}{{\tt 1112.0524}}].

\bibitem{Bozorgnia:2016ogo}
N.~Bozorgnia, F.~Calore, M.~Schaller, M.~Lovell, G.~Bertone, C.~S. Frenk
  et~al., \emph{{Simulated Milky Way analogues: implications for dark matter
  direct searches}},
  \href{http://dx.doi.org/10.1088/1475-7516/2016/05/024}{\emph{JCAP} {\bf 1605}
  (2016) 024}, [\href{http://arxiv.org/abs/1601.04707}{{\tt 1601.04707}}].

\bibitem{Brod:2018ust}
J.~Brod, B.~Grinstein, E.~Stamou and J.~Zupan, \emph{{Weak mixing below the
  weak scale in dark-matter direct detection}},
  \href{http://dx.doi.org/10.1007/JHEP02(2018)174}{\emph{JHEP} {\bf 02} (2018)
  174}, [\href{http://arxiv.org/abs/1801.04240}{{\tt 1801.04240}}].

\bibitem{Belanger:2008sj}
G.~Belanger, F.~Boudjema, A.~Pukhov and A.~Semenov, \emph{{Dark matter direct
  detection rate in a generic model with micrOMEGAs 2.2}},
  \href{http://dx.doi.org/10.1016/j.cpc.2008.11.019}{\emph{Comput. Phys.
  Commun.} {\bf 180} (2009) 747--767},
  [\href{http://arxiv.org/abs/0803.2360}{{\tt 0803.2360}}].

\bibitem{DelNobile:2013sia}
M.~Cirelli, E.~Del~Nobile and P.~Panci, \emph{{Tools for model-independent
  bounds in direct dark matter searches}},
  \href{http://dx.doi.org/10.1088/1475-7516/2013/10/019}{\emph{JCAP} {\bf 1310}
  (2013) 019}, [\href{http://arxiv.org/abs/1307.5955}{{\tt 1307.5955}}].

\bibitem{Shifman:1978zn}
M.~A. Shifman, A.~I. Vainshtein and V.~I. Zakharov, \emph{{Remarks on Higgs
  Boson Interactions with Nucleons}},
  \href{http://dx.doi.org/10.1016/0370-2693(78)90481-1}{\emph{Phys. Lett.} {\bf
  78B} (1978) 443--446}.

\bibitem{Drees:1993bu}
M.~Drees and M.~Nojiri, \emph{{Neutralino - nucleon scattering revisited}},
  \href{http://dx.doi.org/10.1103/PhysRevD.48.3483}{\emph{Phys. Rev.} {\bf D48}
  (1993) 3483--3501}, [\href{http://arxiv.org/abs/hep-ph/9307208}{{\tt
  hep-ph/9307208}}].

\bibitem{Belanger:2013oya}
G.~Belanger, F.~Boudjema, A.~Pukhov and A.~Semenov, \emph{{$micrOMEGAs_3$: A
  program for calculating dark matter observables}},
  \href{http://dx.doi.org/10.1016/j.cpc.2013.10.016}{\emph{Comput. Phys.
  Commun.} {\bf 185} (2014) 960--985},
  [\href{http://arxiv.org/abs/1305.0237}{{\tt 1305.0237}}].

\bibitem{runDM}
B.~J. Kavanagh, ``bradkav/rundm: Journal release.''
  \url{https://doi.org/10.5281/zenodo.823249}, July, 2017.

\bibitem{Finkbeiner:2011dx}
D.~P. Finkbeiner, S.~Galli, T.~Lin and T.~R. Slatyer, \emph{{Searching for Dark
  Matter in the CMB: A Compact Parameterization of Energy Injection from New
  Physics}}, \href{http://dx.doi.org/10.1103/PhysRevD.85.043522}{\emph{Phys.
  Rev.} {\bf D85} (2012) 043522}, [\href{http://arxiv.org/abs/1109.6322}{{\tt
  1109.6322}}].

\bibitem{Galli:2013dna}
S.~Galli, T.~R. Slatyer, M.~Valdes and F.~Iocco, \emph{{Systematic
  Uncertainties In Constraining Dark Matter Annihilation From The Cosmic
  Microwave Background}},
  \href{http://dx.doi.org/10.1103/PhysRevD.88.063502}{\emph{Phys. Rev.} {\bf
  D88} (2013) 063502}, [\href{http://arxiv.org/abs/1306.0563}{{\tt
  1306.0563}}].

\bibitem{Weniger:2013hja}
C.~Weniger, P.~D. Serpico, F.~Iocco and G.~Bertone, \emph{{CMB bounds on dark
  matter annihilation: Nucleon energy-losses after recombination}},
  \href{http://dx.doi.org/10.1103/PhysRevD.87.123008}{\emph{Phys. Rev.} {\bf
  D87} (2013) 123008}, [\href{http://arxiv.org/abs/1303.0942}{{\tt
  1303.0942}}].

\bibitem{Slatyer:2009yq}
T.~R. Slatyer, N.~Padmanabhan and D.~P. Finkbeiner, \emph{{CMB Constraints on
  WIMP Annihilation: Energy Absorption During the Recombination Epoch}},
  \href{http://dx.doi.org/10.1103/PhysRevD.80.043526}{\emph{Phys. Rev.} {\bf
  D80} (2009) 043526}, [\href{http://arxiv.org/abs/0906.1197}{{\tt
  0906.1197}}].

\bibitem{Henning:2012rm}
B.~Henning and H.~Murayama, \emph{{Constraints on Light Dark Matter from Big
  Bang Nucleosynthesis}},  \href{http://arxiv.org/abs/1205.6479}{{\tt
  1205.6479}}.

\bibitem{Kawasaki:2015yya}
M.~Kawasaki, K.~Kohri, T.~Moroi and Y.~Takaesu, \emph{{Revisiting Big-Bang
  Nucleosynthesis Constraints on Dark-Matter Annihilation}},
  \href{http://dx.doi.org/10.1016/j.physletb.2015.10.048}{\emph{Phys. Lett.}
  {\bf B751} (2015) 246--250}, [\href{http://arxiv.org/abs/1509.03665}{{\tt
  1509.03665}}].

\bibitem{Ibarra:2018yxq}
A.~Ibarra, B.~J. Kavanagh and A.~Rappelt, \emph{{Bracketing the impact of
  astrophysical uncertainties on local dark matter searches}},
  \href{http://arxiv.org/abs/1806.08714}{{\tt 1806.08714}}.

\bibitem{Benito:2016kyp}
M.~Benito, N.~Bernal, N.~Bozorgnia, F.~Calore and F.~Iocco, \emph{{Particle
  Dark Matter Constraints: the Effect of Galactic Uncertainties}},
  \href{http://dx.doi.org/10.1088/1475-7516/2017/02/007}{\emph{JCAP} {\bf 1702}
  (2017) 007}, [\href{http://arxiv.org/abs/1612.02010}{{\tt 1612.02010}}].

\bibitem{Cui:2017nnn}
{\scshape PandaX-II} collaboration, X.~Cui et~al., \emph{{Dark Matter Results
  From 54-Ton-Day Exposure of PandaX-II Experiment}},
  \href{http://dx.doi.org/10.1103/PhysRevLett.119.181302}{\emph{Phys. Rev.
  Lett.} {\bf 119} (2017) 181302}, [\href{http://arxiv.org/abs/1708.06917}{{\tt
  1708.06917}}].

\bibitem{Alloul:2013bka}
A.~Alloul, N.~D. Christensen, C.~Degrande, C.~Duhr and B.~Fuks,
  \emph{{FeynRules 2.0 - A complete toolbox for tree-level phenomenology}},
  \href{http://dx.doi.org/10.1016/j.cpc.2014.04.012}{\emph{Comput. Phys.
  Commun.} {\bf 185} (2014) 2250--2300},
  [\href{http://arxiv.org/abs/1310.1921}{{\tt 1310.1921}}].

\bibitem{Semenov:2010qt}
A.~Semenov, \emph{{LanHEP - a package for automatic generation of Feynman rules
  from the Lagrangian. Updated version 3.1}},
  \href{http://arxiv.org/abs/1005.1909}{{\tt 1005.1909}}.

\bibitem{Alwall:2011uj}
J.~Alwall, M.~Herquet, F.~Maltoni, O.~Mattelaer and T.~Stelzer, \emph{{MadGraph
  5 : Going Beyond}},
  \href{http://dx.doi.org/10.1007/JHEP06(2011)128}{\emph{JHEP} {\bf 1106}
  (2011) 128}, [\href{http://arxiv.org/abs/1106.0522}{{\tt 1106.0522}}].

\bibitem{Sjostrand2006}
T.~Sjostrand, S.~Mrenna and P.~Z. Skands, \emph{{PYTHIA 6.4 Physics and
  Manual}}, \href{http://dx.doi.org/10.1088/1126-6708/2006/05/026}{\emph{JHEP}
  {\bf 0605} (2006) 026}, [\href{http://arxiv.org/abs/hep-ph/0603175}{{\tt
  hep-ph/0603175}}].

\bibitem{Conte:2012fm}
E.~Conte, B.~Fuks and G.~Serret, \emph{{MadAnalysis 5, A User-Friendly
  Framework for Collider Phenomenology}},
  \href{http://dx.doi.org/10.1016/j.cpc.2012.09.009}{\emph{Comput.Phys.Commun.}
  {\bf 184} (2013) 222--256}, [\href{http://arxiv.org/abs/1206.1599}{{\tt
  1206.1599}}].

\bibitem{deFavereau:2013fsa}
{\scshape DELPHES 3} collaboration, J.~de~Favereau, C.~Delaere, P.~Demin,
  A.~Giammanco, V.~Lemaître, A.~Mertens et~al., \emph{{DELPHES 3, A modular
  framework for fast simulation of a generic collider experiment}},
  \href{http://dx.doi.org/10.1007/JHEP02(2014)057}{\emph{JHEP} {\bf 02} (2014)
  057}, [\href{http://arxiv.org/abs/1307.6346}{{\tt 1307.6346}}].

\bibitem{Collaboration:2242860}
{\scshape CMS Collaboration} collaboration, T.~C. Collaboration,
  \emph{{Simplified likelihood for the re-interpretation of public CMS
  results}},  Tech. Rep. CMS-NOTE-2017-001. CERN-CMS-NOTE-2017-001, CERN,
  Geneva, Jan, 2017.

\bibitem{Aaboud:2016tnv}
{\scshape ATLAS} collaboration, M.~Aaboud et~al., \emph{{Search for new
  phenomena in final states with an energetic jet and large missing transverse
  momentum in $pp$ collisions at $\sqrt{s}=13$ TeV using the ATLAS detector}},
  \href{http://arxiv.org/abs/1604.07773}{{\tt 1604.07773}}.

\end{thebibliography}\endgroup
\bibliographystyle{JHEP}
\end{document}